\documentclass[aps,prc,twocolumn,showpacs,superscriptaddress,nofootinbib,floatfix,10pt]{revtex4-1}
\usepackage{bm}
\usepackage{amssymb}
\usepackage{amsmath}
\usepackage{graphicx} 
\usepackage{xcolor}
\usepackage{placeins}
\usepackage[colorlinks,allcolors=blue]{hyperref}

\begin{document}
\title{Sensitivity of the fusion cross section to the \\ density dependence of the symmetry energy}

\author{P.-G. Reinhard}\email{paul-gerhard.reinhard@physik.uni-erlangen.de}
\affiliation{Institut f\"ur Theoretische Physik, Universit\"at Erlangen, D-91054 Erlangen, Germany}
\author{A.S. Umar}\email{umar@compsci.cas.vanderbilt.edu}
\affiliation{Department of Physics and Astronomy, Vanderbilt University, Nashville, TN 37235, USA}
\author{P.~D.~Stevenson}\email{p.stevenson@surrey.ac.uk}
\affiliation{Department of Physics, University of Surrey, Guildford, Surrey, GU2 7XH, United Kingdom}
\author{J. Piekarewicz}\email{jpiekarewicz@fsu.edu}
\affiliation{Department of Physics, Florida State University, Tallahassee, Florida 32306, USA}
\author{V.E. Oberacker}\email{volker.e.oberacker@vanderbilt.edu}
\affiliation{Department of Physics and Astronomy, Vanderbilt University, Nashville, TN 37235, USA}
\author{J.A. Maruhn}\email{maruhn@th.physik.uni-frankfurt.de}
\affiliation{Institut f\"ur Theoretische Physik, Goethe-Universit\"at, D-60438 Frankfurt am Main, Germany}
\date{\today}
\begin{abstract}
\begin{description}
\item[Background]
The study of the nuclear equation of state (EOS) and the behavior
of nuclear matter under extreme conditions is crucial to our understanding
of many nuclear and astrophysical phenomena. Nuclear reactions serve as one
of the means for studying the EOS.
\item[Purpose]
It is the aim of this paper to discuss the impact of nuclear 
fusion on the EOS. This is a timely subject given the expected 
availability of increasingly exotic beams at rare isotope facilities\,\cite{balantekin2014}.
In practice, we focus on $^{48}$Ca+$^{48}$Ca fusion.
\item[Method]

We employ three different approaches to calculate fusion cross-sections for a set of energy density
functionals with systematically varying nuclear matter properties. Fusion calculations are performed
using frozen densities, using a dynamic microscopic method based on density-constrained time-dependent
Hartree-Fock (DC-TDHF) approach, as well as direct TDHF study of above barrier cross-sections.
For these studies, we employ a family of Skyrme parametrizations 
with systematically varied nuclear matter properties.
\item[Results]
The folding-potential model provides a reasonable first
   estimate of cross sections.
   DC-TDHF which includes dynamical polarization reduces the
   fusion barriers and delivers much better cross sections.
   Full TDHF near the barrier agreee nicely with DC-TDHF.
   Most of the Skyrme forces which we used deliver, on the
   average, fusion cross sections in good agreement with the data.
   Trying to read off a trend in the results, we find a slight
   preference for forces which deliver a slope of symmetry
   energy of $L\approx 50$\,MeV that corresponds to a neutron-skin
   thickness of $^{48}$Ca of $R_\mathrm{skin}\!=\!(0.180\!-\!0.210)$\,fm.
\item[Conclusions]
Fusion reactions in the barrier and sub-barrier region can be a tool to study the EOS and
the neutron skin of nuclei. The success of the approach will depend on reduced experimental
uncertainties of fusion data as well as the development of fusion theories that closely
couple to the microscopic structure and dynamics.
\end{description}

\end{abstract}
\smallskip
\maketitle
\raggedbottom

\section{Introduction}
\label{sec:intro}

Nuclear saturation---the existence of an equilibrium density---is  a hallmark of 
nuclear dynamics. The semi-empirical nuclear mass formula of Bethe and 
Weizs\"acker\,\cite{weizsacker1935,bethe1936} regards the atomic nucleus 
as an incompressible liquid drop consisting of two quantum fluids: one electrically 
charged consisting of $Z$ protons and one electrically neutral containing $N$ 
neutrons. The nearly uniform density found in the interior of heavy nuclei and 
the resulting $A^{1/3}$ scaling of the nuclear size with mass number 
$A\!=\!Z\!+\!N$ feature among the most successful predictions of the liquid 
drop model. However, in reality the liquid drop is not incompressible. Thus, it 
is pertinent to ask how does the liquid-drop energy increase as the density 
departs from its equilibrium value\,\cite{piekarewicz2014}. Besides being 
a fundamental nuclear-structure question, the answer to this inquiry is also
vital to our understanding of the nuclear equation of state (EOS). The EOS 
plays a key role in elucidating the structure of exotic nuclei\,\cite{chen2014},
the dynamics of heavy ion collisions\,\cite{danielewicz2002,tsang2009},
the composition of neutron stars\,\cite{haensel1990,chamel2008,horowitz2004,
utama2016}, and the mechanism of core-collapse 
supernovae\,\cite{bonche1981,watanabe2009,shen2011}. As such, the 
EOS provides a powerful bridge between nuclear physics and astrophysics.

The quest for the nuclear EOS also generates a strong and healthy interplay 
between theoretical, experimental, and observational 
research\,\cite{piekarewicz2007}. New measurements drive new theoretical 
efforts, that in turn uncover new puzzles that trigger even newer experiments 
and observations. It is the aim of this paper to discuss the impact of nuclear 
fusion on the EOS. This is a timely subject given the expected 
availability of increasingly exotic beams at rare isotope facilities\,\cite{balantekin2014}. 
This new frontier will enable reactions of exotic beams with large neutron 
excess\,\cite{baran2005} that will provide critical new information on the 
poorly-known isospin dependence of the 
EOS\,\cite{steiner2005,horowitz2014}. The isospin dependence of the EOS is
encoded in the symmetry energy which, in turn, quantifies the energy cost of 
turning symmetric nuclear matter into pure neutron matter. The density 
dependence of the symmetry energy has become an active and vibrant area 
of research as evinced by the recent topical issue devoted exclusively to this 
topic\,\cite{li2014}.

A physical observable that is particularly sensitive to the density dependence
of the symmetry energy is the neutron-skin thickness of 
${}^{208}$Pb\,\cite{brown2000, furnstahl2002, centelles2010,
roca2011}; the neutron skin is defined as the difference between 
the root-mean-square radii of the neutron distribution ($R_{n}$) relative to that 
of the proton ($R_{p}$). Whereas elastic electron scattering experiments have 
provided a very detailed map of the proton distribution throughout the full nuclear 
chart\,\cite{vries1987,fricke1995,angeli2013}, a determination of  the 
neutron density has traditionally relied on hadronic experiments that are 
hindered by large and uncontrolled uncertainties. However, using 
parity-violating electron scattering---a purely electroweak reaction---the Lead 
Radius EXperiment (PREX) has provided the first clear evidence in favor of a 
neutron-rich skin in ${}^{208}$Pb\,\cite{abrahamyan2012,horowitz2012}. 
Given that the weak charge of the neutron is much larger than the corresponding 
weak charge of the proton, parity-violating electron scattering provides an ideal 
tool to map the neutron distribution\,\cite{donnelly1989}. Since the proton 
radius of ${}^{208}$Pb is accurately known\,\cite{angeli2013}, the determination 
of the neutron radius by the PREX collaboration has effectively determined the 
neutron-skin thickness of ${}^{208}$Pb to be\,\cite{abrahamyan2012}:
\begin{equation}
 R_{\rm skin}^{208}\!=\!R_{n}^{208}\!-\!R_{p}^{208}\!=\!{0.33}^{+0.16}_{-0.18}\,{\rm fm}.
 \label{Rskin208}
\end{equation}
Note that experimental error will be reduced by a factor of three (to $\pm0.06{\rm fm}$) 
 in an already approved experiment ``PREX-II''\,\cite{PREX-II}.

An accurate measurement of $R_{\rm skin}^{208}$ will significantly constrain the
slope of the symmetry energy $L$, a fundamental parameter of the EOS that is 
very closely related to the pressure of pure neutron matter at saturation density. 
The strong correlation between $R_{\rm skin}^{208}$ and 
$L$\,\cite{centelles2009,roca2011} emerges from a simple 
question: \emph{where do the excess neutrons go?} Placing the 44 extra 
neutrons in the core is favored by surface tension but disfavored by the 
symmetry energy, which is larger at the center than at the surface. Conversely, 
moving the excess neutrons to the dilute surface increases the surface tension 
but reduces the symmetry energy. In this way, the neutron-skin thickness of 
$R_{\rm skin}^{208}$ develops from a dynamic competition between the surface 
tension and $L$, which represents the difference between the symmetry energy 
at saturation density and at a lower surface density\,\cite{horowitz2014}. In particular, 
for a ``stiff'' symmetry energy, namely one that increases rapidly with density, it is 
energetically advantageous to move most the excess neutrons to the surface 
where the symmetry energy is low; this generates a thick neutron skin. 

Besides PREX-II, the Calcium Radius EXperiment (CREX) has been already 
approved to measure the weak form factor of ${}^{48}$Ca at the Jefferson 
Laboratory via parity-violating electron scattering\,\cite{CREX}. This 
purely electroweak measurement will be carried out a momentum transfer 
of $q\!\simeq\!0.8\,{\rm fm}^{-1}$ that will allow the extraction of the neutron 
radius of ${}^{48}$Ca with a precision of $0.02\,{\rm fm}$. The advantages 
of such an experiment are many. First, it provides a theoretical bridge 
between ab-initio calculations that are now possible in the case of 
${}^{48}$Ca\,\cite{hagen2016} and density functional theory which is 
at present the optimal (and perhaps unique) theoretical tool to describe 
nuclei all over the nuclear chart. Indeed, the strong correlation observed 
between $R_{\rm skin}^{208}$ and $L$ while likely robust, has been 
discussed exclusively in the context of density functional theory. Second,
a precise measurement of the weak form factor of a medium-mass 
nucleus such as ${}^{48}$Ca can be carried out at larger momentum transfer
where the parity-violating asymmetry is larger. Indeed, the situation in 
${}^{48}$Ca is so favorable that the full weak-charge form factor could
be measured using parity-violating electron scattering\,\cite{lin2015}.
Together then, PREX-II and CREX provide a powerful complementary set 
of inputs to nuclear theory. CREX will allow ab-initio models of nuclear 
structure to be tested which, in turn, will inform density functional theory. These improved 
energy density functionals may then be used with confidence to interpret the 
PREX-II measurement and ultimately place stringent constraints on the 
density dependence of the symmetry energy. 

Most studies of the EOS involve infinite or semi-infinite nuclear matter and 
examine the dependence of the EOS on the underlying parametrization of the 
energy density functional (EDF) as well as its relation to the macroscopic and
macroscopic-microscopic models of nuclear 
matter\,\cite{blaizot1995,reinhard2006b,chen2014}. The basic predictions
of all models can be characterized by a small number of nuclear matter properties 
(NMP) commonly defined at the equilibrium state of symmetric matter. Those 
NMP may be constrained by the rich spectrum of nuclear collective excitations 
of diverse spin-isospin character\,\cite{harakeh2001}. Among the more prominent 
ones are the isoscalar giant monopole resonance,\,\cite{blaizot1980}, the isoscalar 
giant quadrupole resonance\,\cite{bartel1982}, and the isovector giant dipole 
resonance\,\cite{reinhard1999}. Lately, the electric dipole polarizability---which is 
proportional to the inverse energy weighted sum of the isovector dipole 
response---has received considerable attention because of its strong connection 
to the density dependence of the symmetry energy\,\cite{reinhard2010,tamii2011,
piekarewicz2012,roca2013,rossi2013,hashimoto2015,
roca2015}. Time dependent Hartree-Fock 
(TDHF) theory in the small amplitude limit and equivalently the random-phase-approximation (RPA)
approaches provide the means 
for a consistent calculation of these collective excitations\,\cite{reinhard1992,vangiai2001,reinhard2007,maruhn2005,nesterenko2006,
nakatsukasa2005,umar2005a,simenel2001,simenel2003,lacroix1998,paar2007,piekarewicz2014}.

Complementing these small-density collective excitations are the large 
amplitude modes associated with fusion and fission, which provide a 
challenging testing ground for theoretical models against a vast amount 
of existing experimental information. For its relevance to CREX, in this 
paper we limit our study to $^{48}$Ca+$^{48}$Ca fusion. The main goal
of such a study is to explore the sensitivity of the fusion cross section to 
the EOS, particularly to the slope of the symmetry energy which has a 
strong impact on the thickness of the neutron-rich skin of $^{48}$Ca. 
Although theoretical models of nuclear fusion may be hindered by hadronic 
uncertainties, sub-barrier fusion is attractive because of the exponential
character of quantum-mechanical tunneling. Indeed, we conjecture that
sub-barrier fusion should be significantly enhanced for those models that
predict a stiff symmetry energy and consequently a large neutron-skin
thickness in ${}^{48}$Ca. To test this conjecture and to explore the 
sensitivity of the fusion cross section to the EOS we use a series of 
Skyrme EDFs with systematically varied NMP\,\cite{kluepfel2009}. 
We note in passing the close connection between the fusion of two 
neutron-rich nuclei and the merger of two neutron stars. Indeed, the 
late inspiral phase of neutron-star mergers probes the nuclear EOS 
through the tidal deformation of the star which is highly sensitive to 
the stellar radius. Both the thickness of the neutron skin and the radius 
of a neutron star have the same origin: the pressure of neutron-rich matter. 
Hence, models with thicker neutron skins often produce neutron stars 
with larger radii\,\cite{horowitz2001a,horowitz2001b,erler2013}. This interesting 
correlations provides a compelling connection between nuclear physics 
and astrophysics.

The paper has been organized as follows. In Sec.~\ref{sec:EDF} we discuss 
the general features of the parametrization of the EOS, with a more detailed 
discussion left to Appendix~\ref{app:NMP}. This is followed in Sec.~\ref{sec:df}
by the calculation of fusion barriers using the double-folding potential approach 
to examine the impact of NMP on fusion cross-sections using a purely static 
approach. The impact of the dynamics is studied utilizing the density constrained 
TDHF and direct TDHF calculations in Secs.~\ref{sec:dctdhf} and \ref{sec:tdhf},
respectively.  A summary 
of our main findings and some concluding remarks are given in Sec.~\ref{sec:summary}.

\section{Energy density functionals}
\label{sec:EDF}

\begin{figure*}[!htb]
\includegraphics*[width=17.2cm]{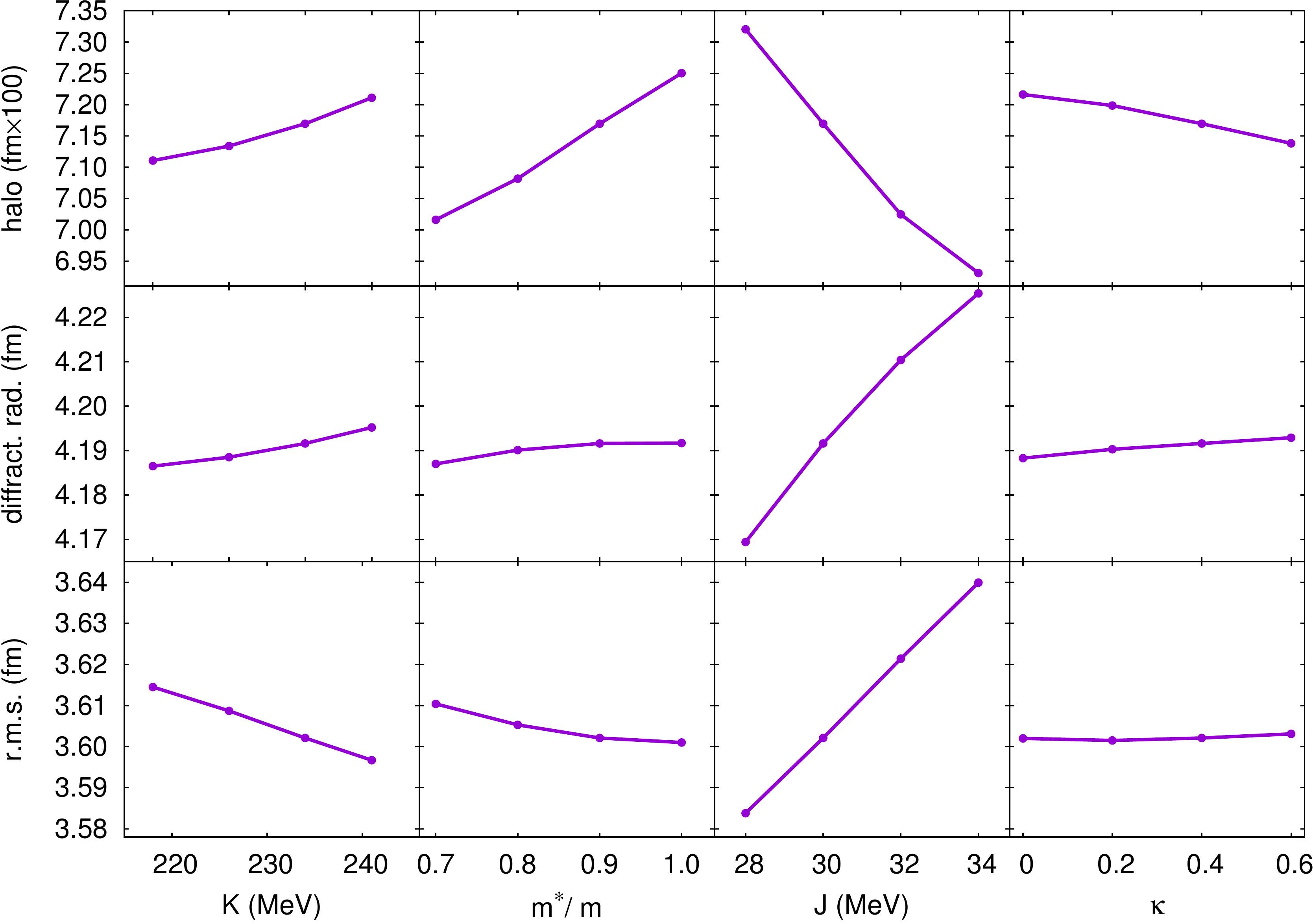}
\caption{\protect (Color online) Neutron root mean-square radius (r.m.s.),
neutron diffraction radius, and neutron halo of $^{48}$Ca obtained from 
a set of Skyrme parametrizations with systematically varied nuclear
matter parameters (incompressibility $K$, effective mass $m^{*}/m$, 
symmetry energy $J$, and $\kappa_\mathrm{TRK}$)~\cite{kluepfel2009} 
for the SV density functionals used in this study}
\label{fig:r_trends}
\end{figure*}

Despite significant progress in nuclear ab-initio calculations 
(for reviews see e.g. Refs.\,\cite{dickhoff1992,pandharipande1997,
epelbaum2009,machleidt2011,hammer2013}), the predictions are 
not yet sophisticated enough to serve as the basis for high-precision 
functionals, especially in the case of medium to heavy nuclei. 
Consequently, most nuclear EDFs are calibrated to experimental data. 
The limited availability of discerning data---especially in the case of the 
isovector sector---as well as approximations and theoretical biases  
leads to residual uncertainties in the fitted parametrizations and 
these propagate to uncertainties in the model predictions; for 
extensive discussions of error propagation and error estimates 
see, e.g., Refs.\,\cite{kortelainen2013,chen2014,dobaczewski2014,erler2014,
reinhard2016,fattoyev2011,piekarewicz2014}). It is our
aim here to explore the sensitivity and variance of model predictions 
for fusion cross sections.  

From the great variety of nuclear EDFs, we concentrate here on the widely 
used Skyrme EDFs as a prototype example. The Skyrme force as an 
effective nuclear interaction dates back to the original work of 
Skyrme~\cite{skyrme1956} which was then utilized in a quantitative 
manner in the pioneering paper of Vautherin\,\cite{vautherin1972}. The 
Skyrme interaction was subsequently developed as to provide an accurate 
description of nuclear structure and dynamics across the full chart of the
nuclides~\cite{bender2003}. For the purpose of the present study, we 
focus on the series of Skyrme parametrizations with systematically 
varied NMP developed in Ref.~\cite{kluepfel2009}. The critical point
exploited here is that there is a one-to-one correspondence between 
NMP and the volume properties of the Skyrme EDF. This allows the 
use of NMP to conveniently characterize each EDF. The calibration
of the EDF results in predictions that determine the NMP with some 
uncertainty. The systematic variation of the NMP is done within this 
uncertainty band. Tracking the trends of an observable along such 
systemically varied parametrizations results in an uncertainty in
the prediction. In turn, such an uncertainty in the prediction 
dictates the experimental precision that would be needed to
better constrain the underlying EDF. 

A detailed description of the Skyrme EDF and its relation to NMP may
be found in Appendix~\ref{app:NMP}. Here, we limit ourselves to summarize 
the strategy involved in generated systematically varied parametrizations. 
In all cases, Skyrme EDFs are fitted to the same pool of bulk properties of 
the nuclear ground state, namely, binding energies, pairing gaps from odd-even 
staggering, a few selected spin-orbit splittings, and bulk properties of the nuclear 
charge distribution, such as the root-mean-square radius, diffraction radius, and
surface thickness; for more details see~\cite{kluepfel2009}. In addition, the fits 
are further constrained by four nuclear matter parameters: (a) the incompressibility 
$K$ of symmetric nuclear matter, (b) the isoscalar effective mass $m^*/m$, 
(c) the symmetry energy at saturation $J$, and (d) the Thomas-Reiche-Kuhn sum 
rule enhancement factor $\kappa_\mathrm{TRK}$ (for more details see 
Appendix~\ref{app:NMP}). The base parametrization is SV-bas with 
$K\!=\!234$ MeV, $m^*/m\!=\!0.9$, $J\!=\!30$ MeV, and 
$\kappa_\mathrm{TRK}\!=\!0.4$. These NMP are chosen such that the underlying
SV-bas parametrization provides a good description of various giant resonances 
and the electric dipole polarizability in $^{208}$Pb. Subsequently, we systematically 
vary each nuclear matter property by holding all others fixed at their optimal value.
Varying $K$ yields SV-K218, SV-K226, and SV-K241, where the corresponding
label indicates the adopted value of $K$. In a similar manner, varying $m^*/m$ 
yields SV-mas10, SV-mas08, and SV-mas07, corresponding to values of the
isoscalar effective mass of $m^*/m=1.0$, 0.8, and 0.7, respectively. Fixing the
symmetry energy at saturation density to $J\!=\!28, 32, 34$\,MeV yields 
SV-sym28, SV-sym32, and SV-sym34. Finally, varying $\kappa_\mathrm{TRK}$ 
yields SV-kap00, SV-kap20, and SV-kap60, corresponding to values of 0.00, 0.20, 
and 0.60, respectively. For further details on the predictions of these 
parametrizations, including the highly uncertain slope of the symmetry energy
$L$, see Table~\ref{tab:nucmat} in Appendix~\ref{app:NMP}.

In order to have a better handle on the variation of the neutron skin as a 
function of the symmetry-energy slope $L$ and its impact on the fusion cross
section, 
we have produced four new forces with dedicated variation of
  $L$. To that end, we keep the four NMP fixed as in SV-bas and in
  addition we set a constraint on $L$. This constraint is then varied
  at the unconstrained value of $L$
  in a range that still produces reasonable fits.
These new forces,
labeled  ``SV-L'', are also tabulated in Table~\ref{tab:nucmat}.

In the next section we investigate the trends that emerge in the fusion barriers and
cross sections for the $^{48}$Ca+$^{48}$Ca reaction as we systematically vary the 
model predictions. Given that we expect that the neutron distribution is the 
most critical piece determining the nuclear attraction at larger distances, we hope to 
gain valuable insights into the density dependence of the symmetry energy. In 
Fig.~\ref{fig:r_trends} we show the emerging trends in neutron radii and halo of 
$^{48}$Ca obtained from the series of systematically varied EDFs. As expected, 
the symmetry energy $J$ has by far the strongest impact on the neutron distribution. 
In contrast, there is some mild sensitivity of the neutron distribution to the isoscalar 
effective mass $m^*/m$ and even less to the incompressibility $K$,
and negligble to the sum rule enhancment $\kappa_\mathrm{TRK}$.

\section{Fusion and NMP}
\label{sec:formalism}

The use of fusion data to elucidate the dependence of cross-sections
on NMP requires a deep understanding of the fusion process and its model 
dependence. Naturally, reducing the highly complex fusion process to the estimation 
of barrier(s) as a function of the distance between the two nuclei $R$ is a
severe simplifying approximation. In order to proceed in this manner, it is critical 
to test this approximation on various energy regimes, as a function of the 
properties of the participating nuclei, and on its sensitivity to the microscopic 
model parameters. For example, in the case of $^{48}$Ca one avoids
complexities associated with deformation~\cite{umar2006a} because of the spherical character
of this nucleus. The height of the ion-ion interaction barrier is largely 
determined by the size of the colliding nuclei and their nuclear skin. As the 
two nuclear surfaces come within the range of the short-range nuclear 
interaction, the barrier starts to deviate from its purely Coulombic form
as a result of the nuclear attraction. In this early stage of the reaction, 
nuclei gradually developed a so-called ``neck'', which subsequently 
evolves into a compound system as the overlapping densities coalesce.
For light and mid-mass systems the barrier is found to be largely insensitive 
to the beam energy\,\cite{keser2012}. At energies at and above the barrier, 
the fusion cross-section displays its highest sensitivity to the neutron-skin 
thickness, which largely determines the height of the barrier. However, at lower 
energies the formation of the neck and density rearrangements are expected 
to be sensitive to the other NMP, such as the incompressibility coefficient and 
the symmetry energy. The $^{48}$Ca+$^{48}$Ca system shows strong fusion 
hindrance at deep sub-barrier energies\,\cite{stefanini2009}. Sophisticated 
coupled-channel calculations using the shallow or compression potential
approach have been done for this system\,\cite{esbensen2010}. These 
calculations are consistent with a neutron-skin thickness in $^{48}$Ca of 0.229\,fm 
based on chosen experimental estimates\,\cite{angeli2013,ray1979}. Note,
that this value is significantly larger than the $R_{\rm skin}\!\lesssim\!0.15\,{\rm fm}$ 
estimate reported recently by Hagen and collaborators using an ab-initio 
approach\,\cite{hagen2016}. Some of the discrepancy may be due to the 
fact that the neutron radius of $^{48}$Ca quoted in Ref.\,\cite{ray1979} 
relies on experiments using elastic proton scattering that are hindered by 
large and uncontrolled hadronic uncertainties. Fortunately, this situation 
will improve greatly by a purely electroweak determination of the neutron 
radius of $^{48}$Ca by the CREX collaboration\,\cite{CREX}.  At this point 
in our analysis it is not possible to make a robust connection of these results 
with the EOS and NMP. In addition, the use of frozen densities in calculating 
the interacting potential is expected to require a slightly larger radius since 
the early stages of the skin-skin attraction and stretching will occur at a 
somewhat smaller distance due to the inability of the densities to dynamically 
stretch.

In this section we employ three different methods to study the dependence of the 
fusion cross section on the various parametrizations of the Skyrme energy density 
functional. First, we use the standard double-folding approach to calculate ion-ion
potentials using frozen densities. Second, we utilize the density-constrained 
time-dependent Hartree-Fock (DC-TDHF) approach to calculate the ion-ion potentials 
directly from TDHF densities. Finally, we perform direct TDHF calculations at above 
barrier energies to compute the fusion cross sections. In what follows we briefly 
outline each method and present the associated results.

\subsection{Double-folding potentials}
\label{sec:df}

In the double-folding approach the nuclear part of the ion-ion interaction potential is 
approximated by the so-called double folding potential $V_F(R)$:
\begin{equation}
 V_F(\mathbf{R})=\int d^3r_1d^3r_2 \rho_1(\mathbf{r}_1)\rho_2(\mathbf{r}_2) 
 V_{NN}(\mathbf{R} + \mathbf{r}_2 - \mathbf{r}_1)\,,
\label{eq:vf}
\end{equation}
where $\rho_1(\mathbf{r}_1)$ and $\rho_2(\mathbf{r}_2)$ are the densities of the two 
nuclei as measured from their respective center of mass and $V_{NN}$ is the effective 
$NN$ interaction. In practice, we have used self-consistent Hartree-Fock (HF) densities 
obtained using a spherical Hartree-Fock code and the M3Y effective NN 
interaction~\cite{satchler1979,bertsch1977,rhoadesbrown1983a,rhoadesbrown1983b}.
The spherical HF code uses a grid spacing of $dr\!=\!0.3$\,fm and maximum radial distance 
of $R_{max}\!=\!10.5$\,fm, which are more than adequate to describe the ground-state 
density of $^{48}$Ca. The double-folding integral is evaluated for the nuclear potential 
together with the Coulomb integral in momentum space~\cite{rhoadesbrown1983b}.
The double-folding potential is expected to be most reliable for distances where the 
nuclei begin to touch and, thus, particularly sensitive to the neutron-skin thickness
of the colliding ions. At smaller distances, the double-folding potential tends to 
overestimate the nuclear interaction as a result of the unphysically large overlap density 
due to the frozen-density approximation.

\begin{figure}[!htb]
	\includegraphics*[width=8.6cm]{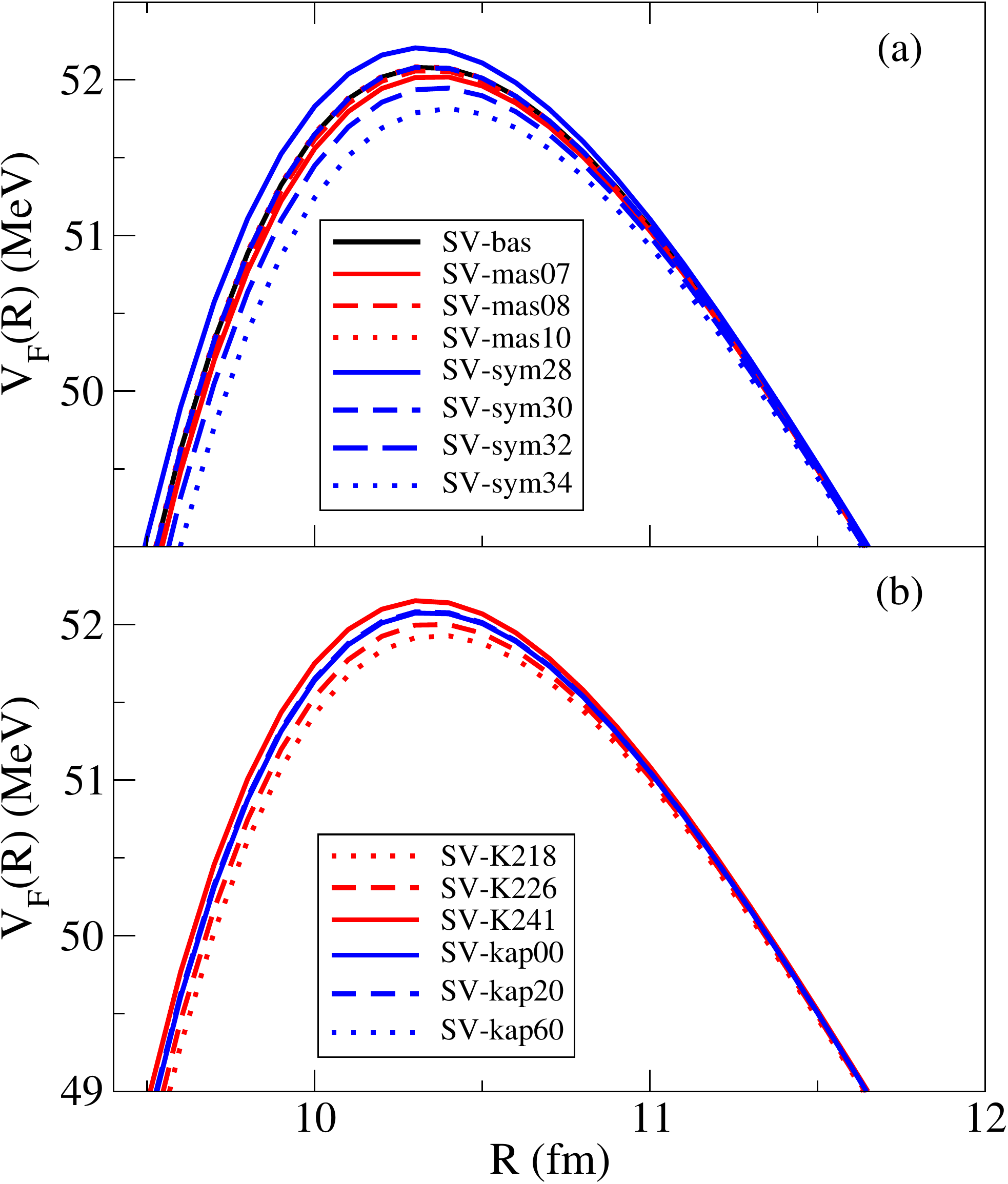}
\caption{\protect
	(Color online) Folding model ion-ion interaction potentials,
	$V_{F}(R)$, for $^{48}$Ca+$^{48}$Ca system using the Skyrme SV
	EDFs; (a) potentials for SV-mas07, SV-mas08, SV-mas10,
	SV-sym28, SV-sym30, SV-sym32, and SV-sym34. Also shown is the
	folding potential corresponding to SV-bas.  (b) potentials for
	SV-K218, SV-K226, SV-K241, SV-kap00, SV-kap20, and
	SV-kap60.}  \label{fig:v_m3y}
\end{figure}

In Fig.~\ref{fig:v_m3y} we show the ion-ion interaction potentials 
defined in Eq.~\ref{eq:vf} for the $^{48}$Ca+$^{48}$Ca system 
using ground-state densities computed for the set of SV forces 
defined in Sec.\ref{sec:EDF}. In particular, Fig.~\ref{fig:v_m3y}(a) 
displays the double folding potential as predicted for the following 
functionals:SV-mas07, SV-mas08, SV-mas10, SV-sym28,
SV-sym30, SV-sym32, and SV-sym34. We observe (red lines) that 
the folding potential is essentially unchanged as the effective mass 
is varied from 0.7 to 1.0. Instead, we find high sensitivity to the 
symmetry energy parameter $J$. The potential peak is largest for 
the smallest value $J\!=\!28$\,MeV and is reduced by as much 
as 0.5~MeV for SV-sym34, which has $J\!=\!34$~MeV. This
finding is consistent with our expectations. A model with a soft 
symmetry energy (such as SV-sym28) predicts a slow increase
in the energy with increasing density and, consequently a thin 
neutron skin; in contrast, a stiff symmetry energy generates a 
thick neutron skin (see Table\,\ref{tab:nucmat}). A thick neutron 
skin enhances the overlap region as the nuclei begin to touch,
thereby resulting in a lower Coulomb barrier. And although half 
an MeV (out of about 50 MeV) may seem as a mild reduction, 
one must recall that nuclear fusion near the Coulomb barrier 
depends exponentially on the height of the barrier.

\begin{figure}[!htb]
	\includegraphics*[width=8.6cm]{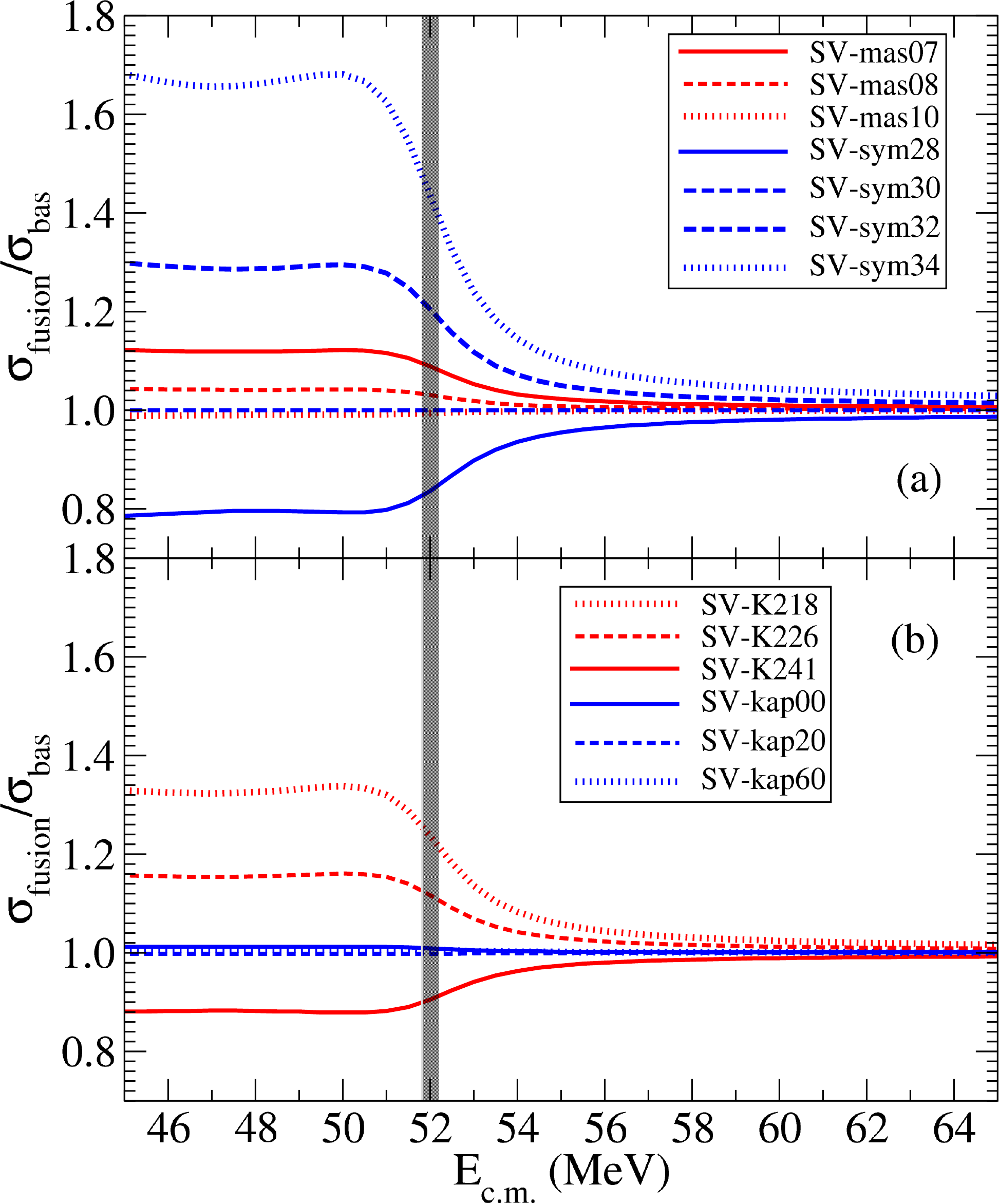} 
\caption{\protect
	(Color online) $^{48}$Ca+$^{48}$Ca fusion cross-sections
	corresponding to the folding model ion-ion interaction
	potentials $V_{F}(R)$ for Skyrme SV EDFs,
	scaled by the fusion cross-section corresponding to the
	SV-bas barrier. 
	(a) cross-sections ratio for SV-mas07, SV-mas08, SV-mas10,
	SV-sym28, SV-sym30, SV-sym32, and SV-sym34. (b) cross-sections ratio for
	SV-K218, SV-K226, SV-K241, SV-kap00, SV-kap20, and SV-kap60.
	The shaded area indicates the range of folding model potential barrier heights
	obtained using the SV forces.}
\label{fig:x_m3y_ratio}
\end{figure}

The folding potentials corresponding to the systematic variation of 
the incompressibility coefficient $K$ and the enhancement factor
$\kappa_\mathrm{TRK}$ are now plotted in Fig.~\ref{fig:v_m3y}(b).  
Here we observe essentially no dependence on $\kappa_\mathrm{TRK}$ 
but some sensitivity, albeit small, to $K$. Larger values of 
the incompressibility result in a higher potential peak, with the total 
difference between the highest and lowest peaks being about 0.25~MeV.  

As already alluded, these results are fully consistent with the variation 
of the ground-state properties of $^{48}$Ca---especially the neutron 
r.m.s. radius shown in Fig.~\ref{fig:r_trends}. Indeed, the bottom row 
of Fig.~\ref{fig:r_trends} clearly indicates that the neutron radius is
highly sensitive to the symmetry energy. Given that the folding 
potential uses frozen densities, the height of the peak 
is highly sensitive to the extent of the nuclear skin. In particular, 
SV-sym28, which predicts the smallest neutron radius in $^{48}$Ca, 
generates the highest peak in the potential since the nuclei can reach 
a smaller ion-ion separation distance $R$ before the outer densities 
start to overlap. That is, a small neutron skin ``delays" the impact of
the nuclear attraction to smaller values of $R$ causing an increase
in the height of the Coulomb barrier. Moreover, a smaller neutron 
skin also manifests itself by having the peak in the potential shift
to a smaller value of $R$ relative to SV-sym34 which, with the stiffest 
symmetry energy, generates the largest neutron skin.


We next examine the fusion cross section for the $^{48}$Ca+$^{48}$Ca
system interacting via the ion-ion potentials shown in Fig.~\ref{fig:v_m3y}.  
When the ion-ion collision is reduced to an ion-ion interaction potential 
dependent on the collective coordinate $R$ the fusion barrier penetrabilities 
$T_L(E_{\mathrm{c.m.}})$ can be obtained by numerical integration of the 
two-body Schr\"odinger equation
\begin{equation}
\left[ \frac{-\hbar^2}{2\mu}\frac{d^2}{dR^2}\!+\!
\frac{L(L\!+\!1)\hbar^2}{2\mu R^2}\!+\!V(R,E)\!-\!E\right]\psi(R)=0\;,
\label{eq:xfus}
\end{equation}
using the {\it incoming wave boundary condition} (IWBC)
method~\cite{hagino1999}.  The potential $V(R,E)$ is the sum of
nuclear and Coulomb potentials.  IWBC assumes that once the 
minimum of the potential is reached fusion will occur. In practice, the
Schr\"odinger equation is integrated from the potential minimum,
$R_\mathrm{min}$, where only an incoming wave is assumed, to a large
asymptotic distance, where it is matched to incoming and outgoing
Coulomb wavefunctions. The barrier penetration factor,
$T_L(E_{\mathrm{c.m.}})$ is the ratio of the incoming flux at 
$R_\mathrm{min}$ to the incoming Coulomb flux at large distances.
Here we implement the IWBC method exactly as it is formulated for the
coupled-channel code CCFULL described in Ref.~\cite{hagino1999}.  This
gives us a consistent way for calculating cross sections at above and
below the barrier via
\begin{equation}
\sigma_f(E_{\mathrm{c.m.}}) = \frac{\pi}{k^2} \sum_{L=0}^{\infty} (2L+1) T_L(E_{\mathrm{c.m.}})\;.
\label{eq:sigfus}
\end{equation}
To elucidate the sensitivity of the $^{48}$Ca+$^{48}$Ca fusion cross section 
to the various EDF parametrizations, we have plotted in Fig.\,\ref{fig:x_m3y_ratio} 
the ratio of the calculated cross sections as a function of center-of-mass energy for 
the potential barriers shown in Fig.\,\ref{fig:v_m3y} relative to the fusion cross section 
obtained from the base parametrization SV-bas. The top and bottom panels of
Fig.\,\ref{fig:x_m3y_ratio} correspond to the potentials displayed in the top and 
bottom panels of Fig.\,\ref{fig:v_m3y}. We have adopted the same vertical scale 
in both panels to emphasize the relative enhancement/quenching factors.
The shaded region indicates the range of barrier heights for the potentials shown
in Fig.~\ref{fig:v_m3y}.
As expected, and fully consistent with our discussion of the potential 
$V_F(R)$, changes in the symmetry energy have by far the largest impact on 
the fusion cross section---particularly below the Coulomb barrier. This is followed 
by the nuclear incompressibility, which displays a significant dependence, and 
finally by the effective mass and TRK enhancement factor both displaying 
minimal impact on the cross section.

\begin{figure}[!htb]
	\includegraphics*[width=8.6cm]{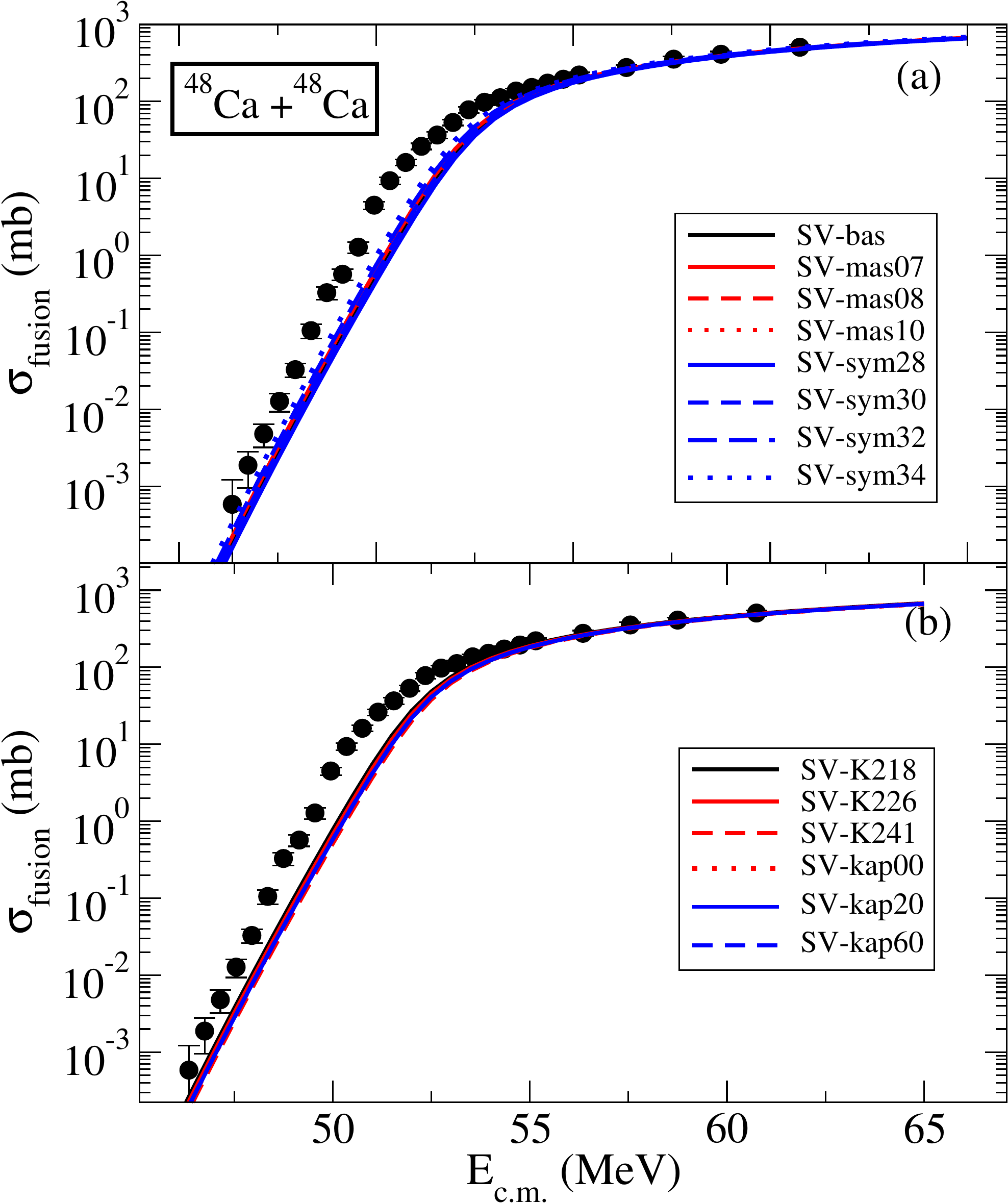} 
\caption{\protect
	(Color online) $^{48}$Ca+$^{48}$Ca fusion cross section
	using an ion-ion interaction $V_{F}(R)$ obtained in the 
	double-folding approximation with densities computed from
	the set of systematically varied Skyrme forces. Results are
	plotted on a logarithmic scale for: (a) SV-mas07, SV-mas08, 
	SV-mas10, SV-sym28, SV-sym30, SV-sym32, and SV-sym34,
	and (b) SV-K218, SV-K226, SV-K241, SV-kap00, SV-kap20, and 
	SV-kap60. Also shown is the cross section corresponding to 
	SV-bas and the experimental data (filled circles) are taken from
	Ref.~\protect\cite{stefanini2009}.}
\label{fig:x_m3y_log}. 
\end{figure}

We close this section by showing in Fig.\,\ref{fig:x_m3y_log} the calculated 
$^{48}$Ca+$^{48}$Ca fusion-excitation function against the experimental
data\,\cite{stefanini2009}. Note that the exponential nature of 
quantum-mechanical tunneling below the Coulomb barrier demands the
use of a logarithmic scale, as the fusion cross section varies over six
orders of magnitude in a 10\,MeV interval in center-of-mass energy. 
Although on such a scale it is difficult to discern significant differences 
among the predictions of the models, some general trends emerge. 
Indeed, whereas the model predictions for the cross section above 
the Coulomb barrier are in reasonable agreement with experiment,
the data below the barrier is consistently underestimated. We attribute 
this discrepancy to the fact that the folding-potential with frozen densities 
is a poor approximation inside the barrier due to an unphysical density 
that is grossly overestimated in the overlap region. We note that for 
c.m. energies above the barrier---where the sensitivity to the neutron-skin 
thickness ceases to be critical---all SV forces produce similar results, with 
the cross sections corresponding to SV-sym34 showing the best overall
agreement. To provide an improved theoretical description of the cross
section and to further examine the sensitivity of the fusion cross section 
to the underlying SV forces, we relax in the next section the double-folding 
approximation in favor of a DC-TDHF approach. 

\subsection{DC-TDHF Method}
\label{sec:dctdhf}
The folding-model calculations ignore dynamical effects such as neck formation, particle exchange, 
pre-equilibrium GDR and other modes during the nuclear overlap phase of the reaction.
During this phase of the collision the primary underlying mechanism is the dynamical change in the 
density along the fusion path which modifies the potential energy.
Obviously, this density change is not instantaneous.  For instance, it was shown in
Ref.~\cite{simenel2013b} that the
development of a neck due to couplings to octupole phonons in $^{40}$Ca+$^{40}$Ca could take 
approximately one zeptosecond.  As a consequence, the dynamical change of the density is most 
significant at low energy (near the barrier-top) where the colliding partners spend enough time 
in the vicinity of each other with little relative kinetic energy.  At high energies, however, the 
nuclei overcome the barrier essentially in their ground-state density.  This energy dependence of
the effect of the couplings on the density evolution was clearly shown in TDHF calculations by 
Washiyama and Lacroix~\cite{washiyama2008}. This naturally translates into an energy
dependence of the nucleus-nucleus potential~\cite{umar2014a,jiang2014}, similar to what was introduced 
phenomenologically in the S\~{a}o-Paulo potential~\cite{chamon2002}.  Consequently, the barrier corresponding
to near barrier-top energies includes dynamical couplings effects and can be referred to as a 
{\it dynamic-adiabatic} barrier, while at high energy the nucleus-nucleus interaction is
determined by a {\it sudden} potential which can be calculated assuming frozen ground-state densities.

In the DC-TDHF approach~\cite{umar2006b} the TDHF
time-evolution takes place without any restrictions. At certain times
during the evolution the instantaneous density is used to perform a
static Hartree-Fock minimization while holding the neutron and proton
densities constrained to be the corresponding instantaneous TDHF
densities~\cite{cusson1985,umar1985}. In essence, this provides us
with the TDHF dynamical path in relation to the multi-dimensional
static energy surface of the combined nuclear system. 
The advantage of DC-TDHF in comparison to the frozen-density
approximation of the previous section is obvious.  The density is, in
fact, not frozen, but polarized in the field of the neighbor fragment
during collision. This is naturally taken into account in the TDHF
propagation and exactly mapped into an ion-ion potential in the
DC-TDHF step. This time-dependent approach automatically incorporates
static and dynamical polarization effects as there are neck formation, 
mass exchange, internal excitations,
deformation effects to all orders, as well as the effect of nuclear
alignment for deformed systems.  In the DC-TDHF method the ion-ion
interaction potential is given by
\begin{equation}
V_{DC}(R)=E_{\mathrm{DC}}(R)-E_{\mathrm{A_{1}}}-E_{\mathrm{A_{2}}}\;,
\label{eq:vr}
\end{equation}
where $E_{\mathrm{DC}}$ is the density-constrained energy at the instantaneous
separation $R(t)$, while $E_{\mathrm{A_{1}}}$ and $E_{\mathrm{A_{2}}}$ are the binding energies of
the two nuclei obtained with the same effective interaction.
In writing Eq.~(\ref{eq:vr}) we have introduced the concept of an adiabatic reference state for
a given TDHF state. The difference between these two energies represents the internal energy.
The adiabatic reference state is the one obtained via the density constraint calculation, which
is the Slater determinant with lowest energy for the given density with vanishing current
and approximates the collective potential energy~\cite{cusson1985}.
We would like to
emphasize again that this procedure does not affect the TDHF time-evolution and
contains no free parameters or normalization.

In addition to the ion-ion potential it is also possible to obtain coordinate
dependent mass parameters. One can compute the ``effective mass'' $M(R)$
using the conservation of energy
\begin{equation}
M(R)=\frac{2[E_{\mathrm{c.m.}}-V_{DC}(R)]}{\dot{R}^{2}}\;,
\label{eq:mr}
\end{equation}
where the collective velocity $\dot{R}$ is directly obtained from the TDHF evolution and the potential
$V_{DC}(R)$ from the density constraint calculations.
In calculating fusion cross-sections this coordinate-dependent mass is used to
obtain a transformed ion-ion potential $V(R)$ with an exact point-transformation as described in Ref.~\cite{umar2009b}.
The coordinate dependent mass only modifies the inner part of the ion-ion potential and its importance was
demonstrated in Ref.~\cite{oberacker2010,umar2014a}.
Recently, the DC-TDHF method was employed for studying fusion reactions relevant for the neutron 
star crust~\cite{umar2012a,simenel2013a}.
\begin{figure}[!htb]
	\includegraphics*[width=8.6cm]{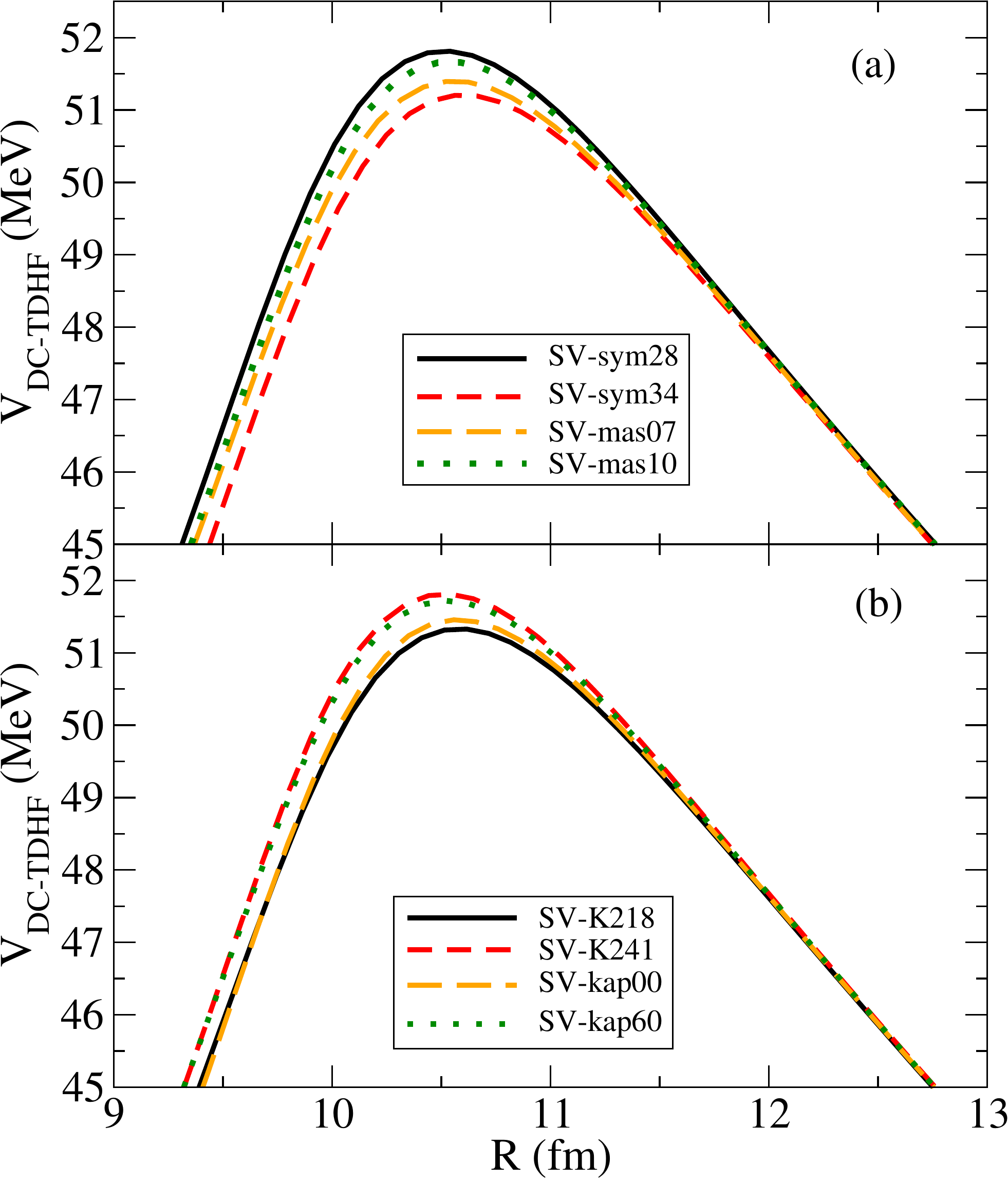}
	\caption{\protect (Color online) DC-TDHF ion-ion interaction potentials $V(R)$ for Skyrme SV EDFs; 
		(a) potentials for SV-mas07, SV-mas10, SV-sym28, and SV-sym34.
		(b) potentials for SV-K218, SV-K241, SV-kap00, and SV-kap60.}
	\label{fig:v_dctdhf}
\end{figure}

TDHF and DC-TDHF calculations were done in a three-dimensional
Cartesian geometry with no symmetry assumptions and using the SV EDFs
discussed above~\cite{umar2006c}.  The three-dimensional Poisson
equation for the Coulomb potential is solved by using Fast-Fourier
Transform (FFT) techniques~\cite{maruhn2014}.  The static HF equations
and the DC-TDHF minimizations are implemented by using the damped
gradient iteration method~\cite{umar1985,umar1989,cusson1985,bottcher1989}.
The box size used for the TDHF calculations in order to utilize the
DC-TDHF method was chosen to be $60\times 30\times 30$~fm$^3$, with a
mesh spacing of $1.0$~fm in all directions.  These values provide very
accurate results due to the employment of sophisticated discretization
techniques~\cite{umar1991a,maruhn2014}.  In Fig.~\ref{fig:v_dctdhf} we
plot the ion-ion potentials for the $^{48}$Ca+$^{48}$Ca system
employing the DC-TDHF method. In this rather expensive fully
time-dependent approach, we have only calculated the potentials
for the SV forces corresponding to the extreme values of the nuclear
matter parameters. In Fig.~\ref{fig:v_dctdhf}(a) we plot the
potentials for EDFs SV-sym28, SV-sym34, SV-mas07, and
SV-mas10. Similar to the folding potentials we observe appreciable
dependence on symmetry energy but little dependence on the effective
mass, although there seems to be a bit more dependence in comparison
to the folding-potential case. Figure~\ref{fig:v_dctdhf}(b) depicts
the DC-TDHF potentials corresponding the EDFs SV-K216, SV-K241,
SV-kap00, and SV-kap60. For these EDFs we see, again, an appreciable
dependence on incompressibility but unlike the folding-potential case
we see a stronger dependence on the value of $\kappa_\mathrm{TRK}$.  This
suggests that the dependence on $\kappa_\mathrm{TRK}$ is more sensitive to the
dynamics such as the formation of a neck and particle transfer which,
in fact, is not so surprising because $\kappa_\mathrm{TRK}$ is the dynamical
isovector response parameter (see Appendix~\ref{app:NMP}).

The calculation of the fusion cross-sections using the DC-TDHF potentials employs the same IWBC approach
used for the fusion cross-section calculations for double-folding potentials.
\begin{figure}[!htb]
	\includegraphics*[width=8.6cm]{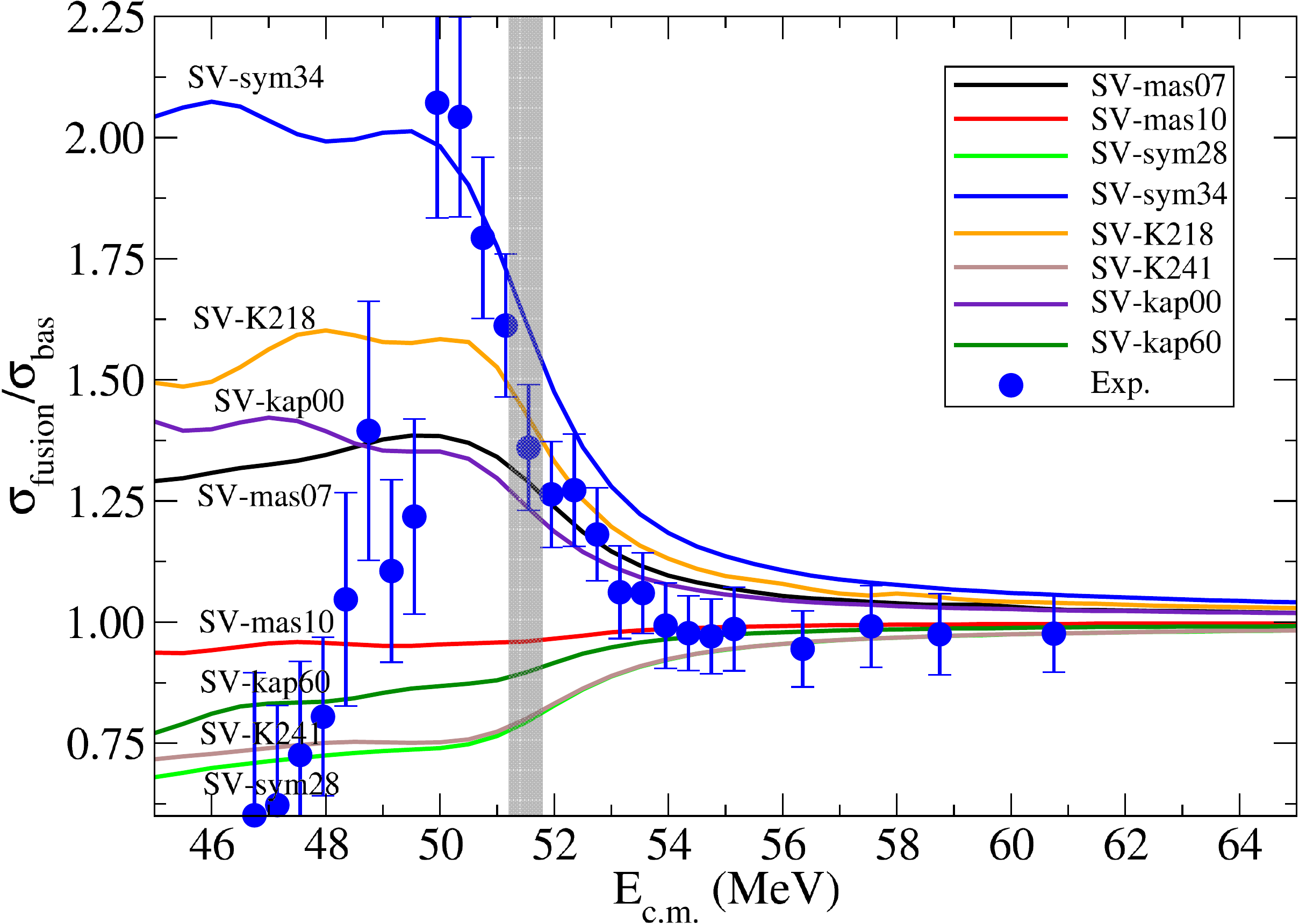}
	\caption{\protect (Color online) Fusion cross-sections corresponding to the DC-TDHF 
		potentials $V(R)$ for Skyrme SV EDFs scaled with the cross-sections corresponding
		to the SV-bas EDF.
		The shaded area indicates the range of DC-TDHF potential barrier heights
	        obtained using the SV forces.
		The experimental data, also scaled with the SV-bas cross-sections,
		(filled circles) are taken from Ref.~\protect\cite{stefanini2009}
		and include statistical as well as $\pm$7\% systematic error.}
	\label{fig:x_dctdhf_scaled}
\end{figure}
Again, to better elucidate the dependence of fusion cross-sections on EDF parametrizations in 
Fig.~\ref{fig:x_dctdhf_scaled} we plot the ratio of the calculated cross-sections
corresponding to the potential barriers shown in Fig.~\ref{fig:v_dctdhf} to the fusion cross-section
of the SV-bas potential barrier as a function of c.m. energy.
The shaded area indicates the range of potential barrier heights for
the potentials used in the study. We note that the dynamical barriers are 300$-$500~keV lower than
the ones obtained from folding potentials.
The experimental data~\cite{stefanini2009} scaled also by the SV-bas cross-sections are shown as
filled circles.
A number of important points may be deduced from Fig.~\ref{fig:x_dctdhf_scaled}.
We again observe that NMP corresponding to symmetry energy has the largest impact on the variation
of the cross-sections. This is followed by incompressibility. A surprising observation is that
the parameter $\kappa$ induces a substantial variation in the cross-sections in this case, whereas
it has practically no impact in the case of the folding potential approach. This indicates that
the dynamics plays an important role in determining the important and correct NMP for nuclear
reactions.
What is also striking is that the parametrizations SV-mas10 and SV-kap60, and to a smaller extent
SV-K241 and SV-sym28 provide an excellent description of the above barrier fusion data.
However, at below barrier energies the dependence on NMP is much more complicated, namely a
strong energy dependence.

\begin{figure}[!htb]
	\includegraphics*[width=8.6cm]{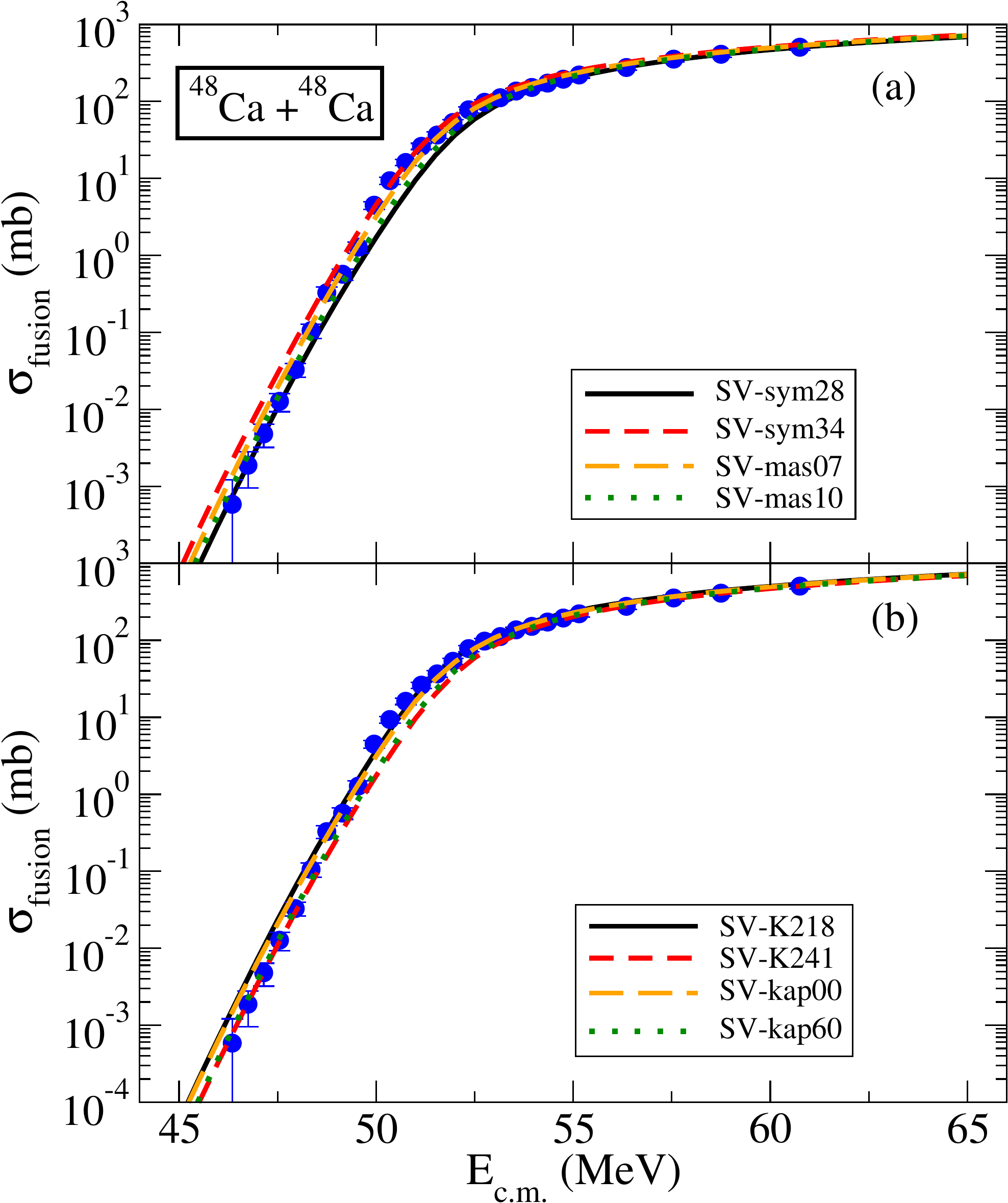}
	\caption{\protect (Color online) Fusion cross-sections corresponding to the DC-TDHF 
		potentials $V(R)$ for Skyrme SV EDFs plot on a $log$ scale;
		(a) cross-sections for SV-sym28, SV-sym34, SV-mas07, and SV-mas10.
		(b) cross-sections for SV-K218, SV-K241, SV-kap00, and SV-kap60.
		The experimental data (filled circles) are taken from Ref.~\protect\cite{stefanini2009}.}
	\label{fig:x_dctdhf_log}
\end{figure}
Again, in Fig.~\ref{fig:x_dctdhf_log} we plot the calculated fusion cross-sections on a logarithmic 
scale. Also shown are the experimental data (circles) from Ref.~\protect\cite{stefanini2009}.
Before focusing on the details we observe that the DC-TDHF ion-ion potentials considerably improve
the agreement with data at sub-barrier energies. This is due to the incorporation of neck dynamics
in the construction of the potential, which modifies the inner part of the barrier. In fact the results 
are surprisingly good since the used EDFs do not include any reaction data for parameter fitting.
One more important observation is that most DC-TDHF calculations to date have used the 
SLy4~\cite{chabanat1998a} forces and one common discrepancy was always identified in 
the energy regime slightly above the barrier peak.
We note that some of the SV forces, particularly SV-sym34 and SV-K218, seem to do a very good job 
in this regime but not as well at deep sub-barrier energies, where SV-sym28 and SV-K241 do better
at these energies.
It would be interesting to see if it is possible to construct an SV force which encompasses both
nuclear matter values to see whether the complete energy regime can be reproduced.
If we examine the DC-TDHF fusion
cross-sections on a linear scale we again observe that the forces
reproducing the barrier peak energy regime
(SV-sym34, SV-mas07, SV-K218, SV-kap00) slightly overestimate the higher
energy cross-sections whereas the forces that reproduce high energy
cross-sections (SV-sym28, SV-mas10, SV-K241, SV-kap60) underestimate the
barrier peak region.

The example shows that (dynamical) rearrangement of the colliding densities
is crucial for a detailed reproduction of the fusion cross
section. But the basic trends are already set by the folding
model which relies exclusively on the neutron density distributions of the
ground states of the colliding nuclei. And these densities are most
sensitive to the density dependence of the symmetry energy.

\begin{figure}[!htb]
\includegraphics*[width=8.6cm]{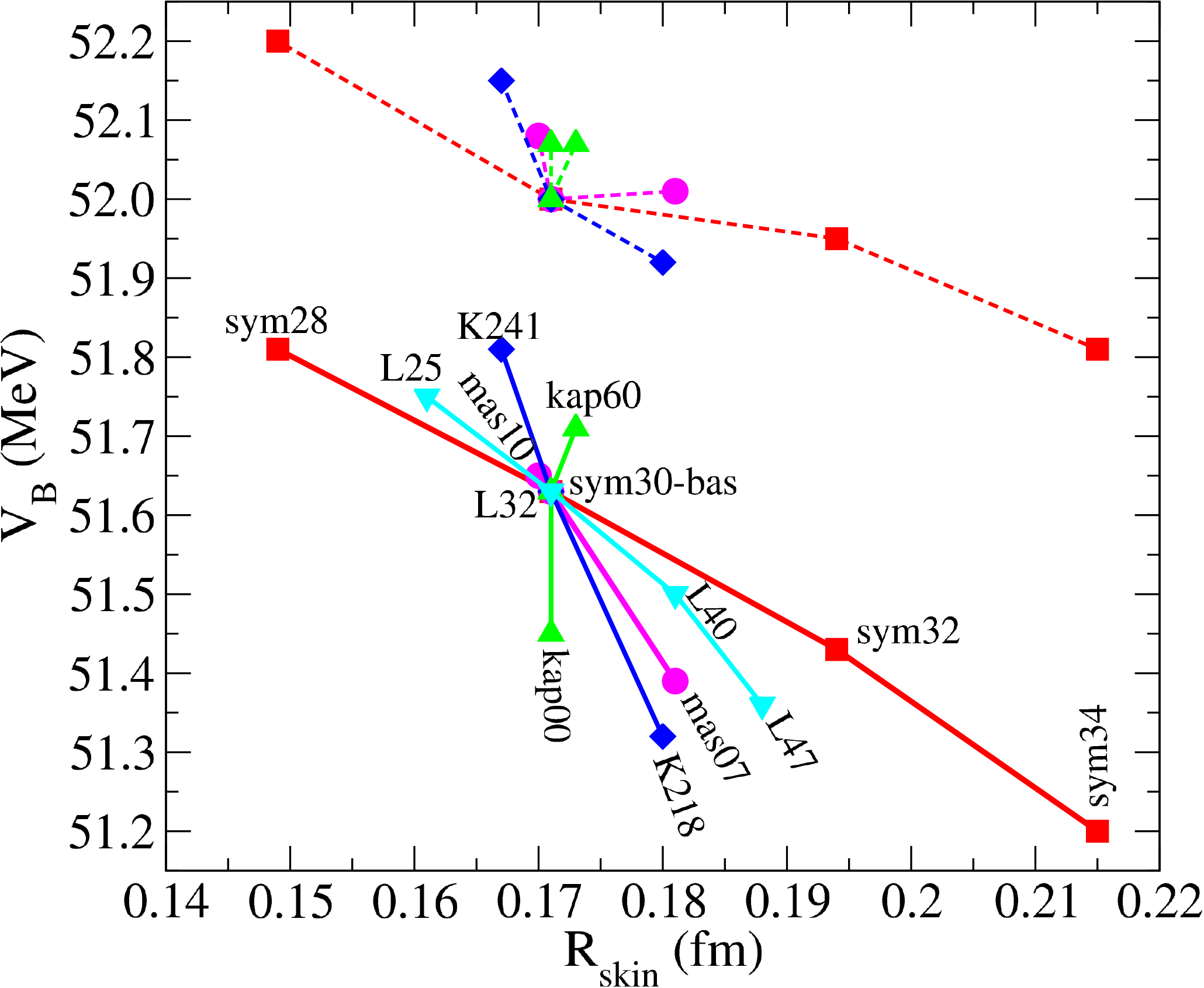}
\caption{\protect (Color online)
The maximum height of the potential energy barrier $V_B$ as a function of 
the neutron skin $R_{\rm skin}$, as predicted by the set of systematically 
varied Skyrme parametrizations used in this paper. The dashed lines show the same
quantities for the folding potentials.}
\label{fig:vb_skin}
\end{figure}

In order to assess in a systematic fashion the sensitivity of the fusion 
cross section to the neutron-skin thickness of ${}^{48}$Ca we have generated 
a set of ``SV-L forces'' that keep all NMP fixed at the SV-bas values and
vary exclusively the slope of the symmetry energy $L$. In Fig.~\ref{fig:vb_skin} we 
plot the potential barrier height for all of the forces used in this study.
For the folding-model potentials we observe an almost linear relationship
between the neutron skin and the barrier height. This is expected since the 
densities are frozen. On the other hand, for the Skyrme parametrizations 
introduced in\,\cite{kluepfel2009}, we observe a more complicated behavior in our 
predictions, although at the extremes there is a clear sensitivity to the 
variations of the symmetry energy. 
We observe that each subset of parametrizations show a nearly linear trend
with the provision that they have differing slopes.
As expected, the systematically varied 
SV-L forces also display an almost linear correlation. 
\begin{figure}[!htb]
\includegraphics*[width=8.6cm]{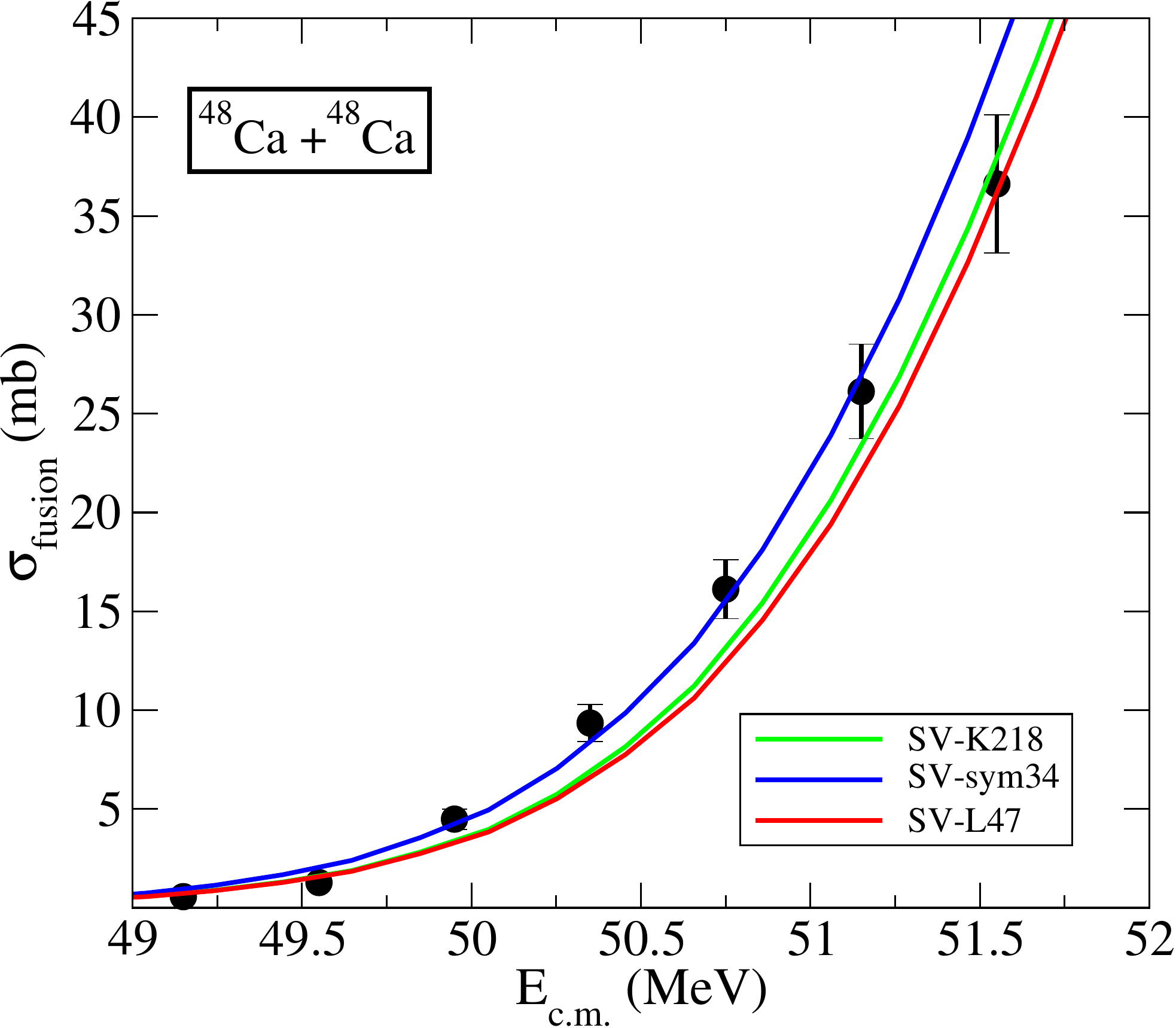}
\caption{\protect (Color online)
Fusion cross-sections in the vicinity of barrier top energies plotted for SV parametrizations
that are most compatible with the fusion data in this energy regime.}
\label{fig:x_barrier}
\end{figure}
In Fig.\,\ref{fig:x_barrier} we plot the SV parametrizations that best reproduce the data
around the barrier top energies. We emphasize that these forces do reproduce the experimental
data to a remarkable accuracy, with errors being on the order of a few millibarns.
While it is not straightforward to pinpoint an unique set of NMP that provide the best
agreement with the data, we observe that there is a tradeoff between having a stiff symmetry
energy and a corresponding large neutron neutron-skin versus a softer symmetry energy and 
a somewhat smaller neutron-skin. Our analysis suggests a value for the symmetry energy closer 
to $J\!=\!30$\,MeV with a relatively small neutron skin. Indeed, a neutron-skin thickness in 
${}^{48}$Ca of $R_{\rm skin}\!=\!(0.180\!-\!0.210)$\,fm provides the optimal reproduction 
of the data.

\subsection{Direct TDHF calculations}
\label{sec:tdhf}
In the case of direct TDHF calculations of fusion cross sections the situation is 
different because they do not include sub-barrier tunneling of the many-body 
wave-function. That is, the fusion probability, $P_{fus.}(L,E_{\mathrm{c.m.}})$, 
for a particular orbital angular momentum $L$ at the center-of-mass energy 
$E_{\mathrm{c.m.}}$ can only be either $P_{fus.}^{TDHF}=0$ or\,$1$.
As a consequence, the quantal expression for the fusion cross-section
\begin{equation}
\sigma_{fus.}(E_{\mathrm{c.m.}}) = \frac{\pi\hbar^2}{2\mu E_{\mathrm{c.m.}}} \sum_{L=0}^\infty (2L+1) P_{fus.}(L,E_{\mathrm{c.m.}})\;,
\label{eq:cs}
\end{equation}
reduces to
\begin{eqnarray}
\sigma_{fus.}(E_{\mathrm{c.m.}}) &=& \frac{\pi\hbar^2}{2\mu E_{\mathrm{c.m.}}} \sum_{L=0}^{L_{max}(E_{\mathrm{c.m.}})} (2L+1) \nonumber \\
&=&  \frac{\pi\hbar^2}{2\mu E_{\mathrm{c.m.}}} [L_{max}(E_{\mathrm{c.m.}})+1]^2\;.
\end{eqnarray}
Here $\mu$ is the reduced mass of the system and $L_{max}$ is the largest value of the 
orbital angular momentum leading to fusion. This is known as the quantum sharp cut-off 
formula~\cite{blair1954}.
\begin{figure}[!htb]
	\includegraphics*[width=8.6cm]{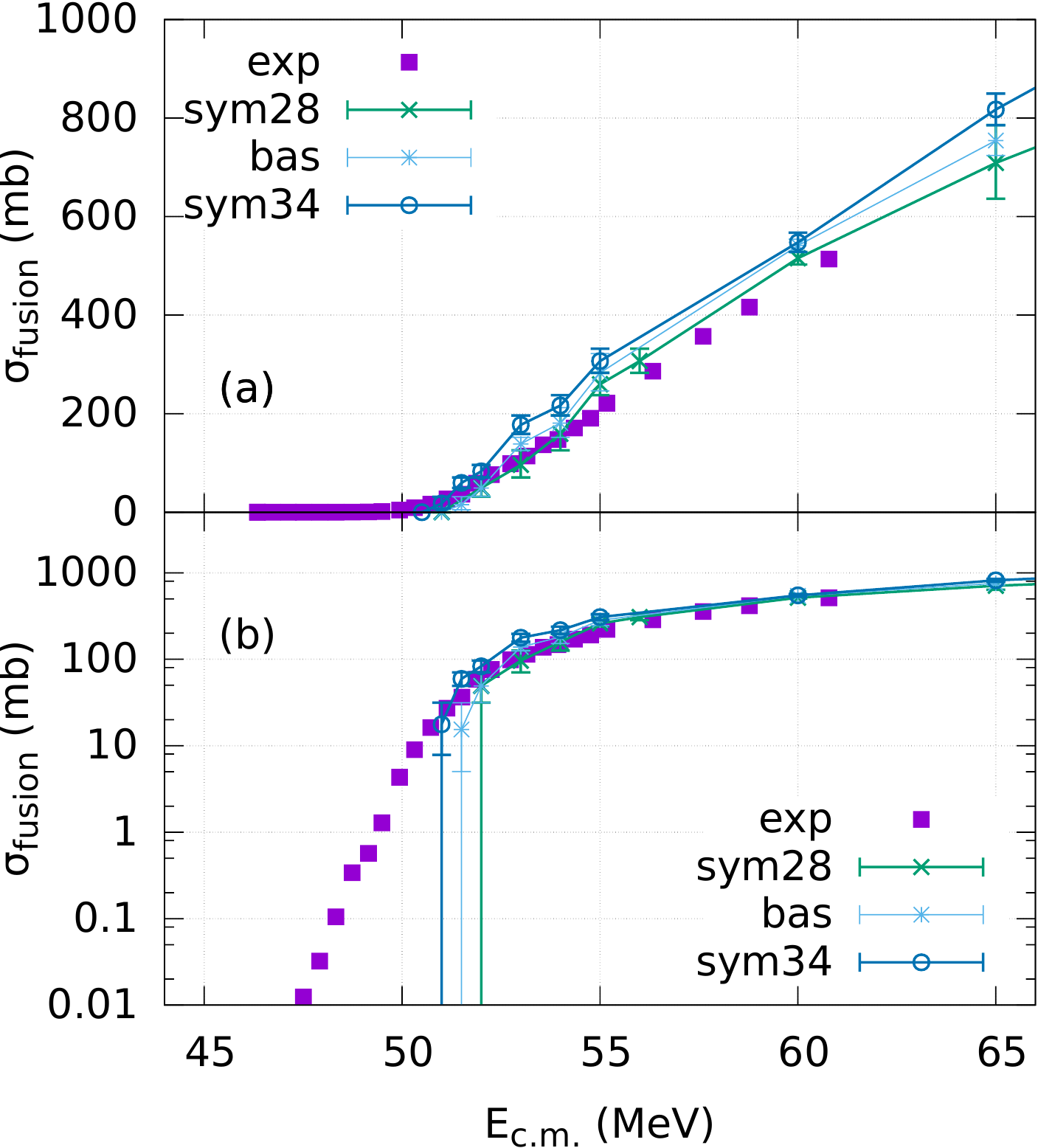}
\caption{\protect
	(Color online) Cross-sections calculated directly from TDHF with SV-bas, SV-sym28, and SV-sym34.  
	The panel (a) shows the data on a plot with a linear $y$--axis and panel (b) shows the same data 
	with a logarithmic $y$--axis.  The experimental data (filled circles) are taken from 
	Ref.~\protect\cite{stefanini2009}.}
\label{fig:tdhf}
\end{figure}

To calculate the fusion cross-section within TDHF it is necessary to make many TDHF calculations to map out the maximum impact parameter for each energy of interest.  We restrict our direct TDHF calculations to explore the systematic variation of the symmetry energy, since these showed the largest dependence in the DC-TDHF calculations.  Figure~\ref{fig:tdhf} shows the calculated and experimental cross-sections for the SV-sym28, SV-bas, and SV-sym34 forces, with symmetry energies 28 MeV, 30 MeV, and 34 MeV respectively.  The error bars associated with the calculations show the accuracy of determination of the critical impact parameter for each energy point indicated.

One notices that the results show the same systematic trends as the DC-TDHF calculations, namely, a softer symmetry energy 
gives a lower cross section. In particular, the force with a symmetry energy of $J\!=\!28$ MeV seems to be slightly favored by 
the experimental data when the collision energy is greater than about 55 MeV, whereas models with a stiffer symmetry energies 
agree better near the barrier.  The inability of direct TDHF calculations to reproduce the fusion cross section below the barrier 
is evident from the logarithmic plot, where the transition to zero cross section is seen for all forces in the region between 51 
and 52 MeV center of mass energy.

\section{Summary and Discussion\label{sec:summary}}
Elucidating the nature of the nuclear force that binds protons and 
neutrons into stable nuclei and rare isotopes is one of the overarching 
questions in nuclear science. In this paper we focused on the role that 
sub-barrier fusion may play in constraining the neutron density of 
${}^{48}$Ca, and in particular its neutron-rich skin. Being dominated by
quantum-mechanical tunneling, the sub-barrier fusion of two ${}^{48}$Ca
nuclei is expected to be dominated by the spatial extent of its neutron 
distribution. This expectation was convincingly validated in this work.
Indeed, models that predict a small neutron skin in ${}^{48}$Ca delay 
the impact of the nuclear attraction to smaller values of the ion-ion 
separation causing an increase in the height of the Coulomb barrier 
relative to models with larger neutron skins. As a result, we have 
established the following correlation: the larger the neutron-skin 
thickness of ${}^{48}$Ca, the larger the ${}^{48}$Ca+${}^{48}$Ca 
sub-barrier fusion cross section. Based on the limited set of systematically 
varied Skyrme forces employed in this work, our results seem to favor a 
slope for the symmetry energy of $L\approx 50$\,MeV and a corresponding
neutron-skin thickness in $^{48}$Ca of 
$R_\mathrm{skin}\!=\!(0.180\!-\!0.210)$\,fm.

The determination of the neutron-skin thickness of a variety of nuclei 
with a significant neutron excess provides critical information on the 
poorly-known isovector sector of the nuclear energy density functional. 
In turn, constraining the isovector sector determines the equation of 
state of neutron-rich matter, particularly the density dependence of the 
symmetry energy. The EOS of neutron-rich matter provides a powerful 
bridge between nuclear physics and astrophysics. Indeed, models that 
predict a large neutron skin in heavy nuclei also tend to predict large 
neutron-star radii. This correlation emerges since  the same pressure 
that pushes against surface tension to generate the neutron skin pushes 
against gravity to shape the stellar radius. Remarkably, this correlation
involves a disparity in length scales of 18 orders of magnitude! In this 
contribution we find another compelling connection between nuclear 
physics and astrophysics. Whereas sub-barrier fusion is hindered by 
the Coulomb barrier, the merger of two neutron stars is hindered by a 
large centrifugal barrier. Yet, the fusion of two neutron-rich nuclei or 
the merger of two neutron stars are both greatly enhanced by a stiff 
symmetry energy that produces large neutron skins and large stellar 
radii. 

Ultimately, the most accurate determination of the neutron-skin thickness 
of ${}^{48}$Ca will come from CREX---a parity-violating experiment 
that relies exclusively on the electro-weak nature of the electron probe. 
Yet, even under the most optimistic scenarios CREX is several years down
the road. In the meantime, it is fruitful to explore complimentary 
approaches that probe similar physics. In this contribution we have 
established---in spite of some inherent hadronic uncertainties---that
${}^{48}$Ca+${}^{48}$Ca fusion near the Coulomb barrier is an attractive 
complement to CREX because of the exponential sensitivity of the 
reaction to the spatial distribution of neutrons.

\section*{Acknowledgments}
This material is based upon work supported by the U.S. Department of 
Energy Office of Science, Office of Nuclear Physics under Award Numbers 
DE-SC0013847 and DE-FD05-92ER40750, by the German BMBF under 
contract No. 05P12RFFTG, and by the UK Science and Technology Facilities 
Council (STFC) under grant numbers ST/J00051/1, ST/J500768/1, and 
ST/M503824/1.

\appendix

\section{Skyrme EDF and nuclear matter parameters (NMP)}
\label{app:NMP}

For simplicity, we only discuss here the part of the Skyrme EDF
containing the time-even couplings. The total energy is composed
from kinetic energy, Skyrme interaction energy, Coulomb energy,
pairing energy, and correlation energy from low-energy collective
modes, usually a center-of-mass and a rotational correlation
\begin{equation*}
  E_\mathrm{total}=
  \int \! d^3 r \; \left\{\mathcal{E}_\mathrm{kin}
  + \mathcal{E}_\mathrm{Sk}\right\}
  + E_\mathrm{Coul}
  + E_\mathrm{pair}
  - E_\mathrm{corr}
  \quad,
\label{eq:Etot}
\end{equation*}
with
\begin{equation}
 \mathcal{E}_\mathrm{kin}=
  \frac{\hbar^2}{2m_p} \tau_p+  \frac{\hbar^2}{2m_n} \tau_n
  \quad,
\end{equation}
and
\begin{equation}
\label{eq:ESkeven}
\begin{array}{rclcl}
 
  \mathcal{E}_\mathrm{Sk}
  &=&
  {C_0^\rho\,\rho_0^2}
  &+&
  {C_1^\rho\,\rho_1^2}
  \\
  &&
  +{C_0^{\rho,\alpha}\,\rho_0^{2+\alpha}}
  &+&
  C_1^{\rho,\alpha}\,\rho_1^2\rho_0^\alpha
  \\
  &&
  +{C_0^{\Delta\rho}\,\rho_0\Delta\rho_0}
  &+&
  C_1^{\Delta\rho}\,\rho_1\Delta\rho_1
  \\
  &&
  +{C_0^{\nabla J}\,\rho_0\nabla\!\cdot\!\vec{J}_0}
  &+&
  C_1^{\nabla J}\,\rho_1\nabla\!\cdot\!\vec{J}_1
  \\[3pt]
  &&
  +C_0^{\tau}\,\rho_0\tau_0
  &+&
  C_1^{\tau}\,\rho_1\tau_1
  \\[3pt]
  &&
  +C_0^{J}\,\vec{J}_0^2
  &+&
  C_1^{J}\,\vec{J}_1^2
\end{array}
\end{equation}
where $\rho_T$ is the local density, $\tau_T$ the kinetic-energy
density, $\vec{J}$ the spin-orbit density, and $T=0,1$ stands for
isospin.
\begin{table}[!htb]
\begin{center}
\begin{tabular}{lrcl}
\hline
\multicolumn{4}{c}{\rule{0pt}{12pt}isoscalar ground state properties}
\\[4pt]
\hline
 equilibrium density:\hspace*{-0.5em}
 &\rule{0pt}{16pt}
 $\rho_\mathrm{eq}$
 &$\leftrightarrow$&
 $\displaystyle\frac{d}{d\rho_0}\frac{E}{A}\Big|_\mathrm{eq}=0$
\\[12pt]
 equilibrium energy:\hspace*{-0.5em}
 &
 $\displaystyle\frac{E}{A}\Big|_\mathrm{eq}$
\\[12pt]
\hline
\multicolumn{4}{c}{\rule{0pt}{12pt}isoscalar response properties}
\\[4pt]
\hline
 incompressibility:\hspace*{-0.5em}
 &
  $  K_\infty$
  &=& 
  $\displaystyle
  9\,\rho_0^2 \, \frac{d^2}{d\rho_0^2} \,
       \frac{{E}}{A}\Big|_\mathrm{eq}
  $
\\[12pt]
  effective mass:\hspace*{-0.5em}
  &
  $\displaystyle
  \frac{\hbar^2}{2m*}
  $
  &=&
  $\displaystyle
   \frac{\hbar^2}{2m}
    + 
    \frac{\partial}{\partial\tau_0} \frac{{E}}{A}\bigg|_\mathrm{eq}
  $
\\[14pt]
\hline
\multicolumn{4}{c}{\rule{0pt}{12pt}isovector response properties}
\\[4pt]
\hline
  symmetry energy:\hspace*{-0.5em}
  &
  $J$
  &=&
  $\displaystyle\rule{0pt}{19pt}
 \frac{1}{2} \frac{d^2}{d\rho_1^2}
  \frac{{E}}{A} \bigg|_\mathrm{eq}
  $
\\[12pt]
  slope of $J$:\hspace*{-0.5em}
  &
  $L$
  &=&
  $\displaystyle
  \frac{3}{2}\rho_0 \frac{d}{d\rho_0}\frac{d^2}{d\rho_1^2}
  \frac{{E}}{A} \bigg|_\mathrm{eq}
  $
\\[12pt]
  TRK sum-rule enh.:\hspace*{-0.5em}
  &
  $\kappa_{\rm TRK}$
  &=& 
  $\displaystyle
  \frac{2m}{\hbar^2}
  \frac{\partial}{\partial\tau_1} 
  \frac{{E}}{A}\bigg|_\mathrm{eq}
  $
\\[12pt]
\hline
\end{tabular}
\end{center}
\caption{\label{tab:nucmatdef}
Definition of the nuclear matter properties (NMP).
All derivatives are to be  taken at the equilibrium point
of  symmetric nuclear matter.
}
\end{table}

Infinite nuclear matter is taken without Coulomb force, pairing, and
correlation correction. The energy per particle becomes
\begin{equation}
  \frac{E}{A}(\rho_0,\rho_1,\tau_0,\tau_1)
  = 
  \frac{\mathcal{E}_\mathrm{kin}+\mathcal{E}_\mathrm{Sk}}{\rho_0}
  \quad,
\end{equation}
where we consider for a while $\rho$ and $\tau$ as independent
variables. Of course, a given system is characterized just
by the densities $\rho_T$ while the kinetic density depends on these
given densities as $\tau_T=\tau_T(\rho_0,\rho_1)$. Thus we have to
distinguish between partial derivatives $\partial/\partial_\tau$ which
take $\tau_T$ as independent and total derivatives $d/d\rho$ which
deal only with $\rho_T$ dependence. The relation is
\begin{equation}
  \frac{d}{d\rho_T}
  =
  \frac{\partial}{\partial\rho_T}
  +
  \sum_{T'}\frac{\partial\tau_{T'}}{\partial\rho_T}
  \frac{\partial}{\partial\tau_{T'}}
  \quad.
\end{equation}
The standard NMP are defined at the equilibrium point
($\rho_0=\rho_\mathrm{eq}$, $\rho_1=0$) of symmetric nuclear matter.
They are summarized in Table~\ref{tab:nucmatdef}.
The enhancement factor for the Thomas-Reiche-Kuhn (TRK) sum
rule~\cite{ring1980} is a widely used way to characterize the isovector
effective mass which is obvious from the given expression involving
derivative with respect to $\tau_1$.  The slope of symmetry energy $L$
characterizes the density dependence of the symmetry energy which
allows to estimate the symmetry energy at half density, i.e. at
surface of finite nuclei.

The NMP in Table~\ref{tab:nucmatdef} can be grouped into four classes:
first, the (isoscalar) ground state properties $\rho_\mathrm{eq}$ and
${E}/{A}\Big|_\mathrm{eq}$, second, isoscalar response properties $K$
and $m^{*}/m$, and third, isovector response properties $J$, $L$,
$\kappa_\mathrm{TRK}$. The response properties determine zero sound in
matter~\cite{thouless1961} and subsequently they are closely related to
giant resonance modes in finite nuclei.  There is
a further category, the surface energies which go already beyond
homogeneous matter and whose definition is rather
involved~\cite{reinhard2006b}. They are not considered here.

Homogeneous matter yields $\Delta\rho=0$ and $\vec{J}=0$ which, in
turn, renders four terms in the functional~(\ref{eq:ESkeven}) inactive.
Thus we have exactly seven interaction parameters ($C_0^{\rho}$,
$C_0^{\rho,\alpha}$, $C_0^{\tau}$, $\alpha$) to determine seven NMP.
The relation is reversible establishing a one-to-one correspondence between
both sets. This allows to consider the NMP equivalently as model
parameters which is, in fact, a more intuitive way to ultimately determine
the model parameters.
\begin{table}[!hbt]
	\begin{tabular}{|l|rcll|rrr|ll|} \hline 
&&&&&&&&\multicolumn{2}{|c|}{Skin} \\
Force
		& \multicolumn{1}{|c}{$K$}
		& \multicolumn{1}{c}{$m^*/m$}
		& \multicolumn{1}{c}{$J$}
		& \multicolumn{1}{c|}{$\kappa$}
		& \multicolumn{1}{c}{$\rho_\mathrm{eq}$}
		& \multicolumn{1}{c}{$E/A$} & \multicolumn{1}{c}{$L$}
		& \multicolumn{1}{|c}{$^{208}$Pb} 
		& \multicolumn{1}{c|}{$^{48}$Ca} 
\\ \hline \hline 
SV-bas   & 234 & 0.9 & 30 & 0.4 & 0.1596 & -15.90 & 32&0.155 &0.171\\ 
\hline \hline 
SV-K218  & 218 & 0.9 & 30 &0.4 & 0.1615 & -15.90 & 35&0.161&0.180\\ 
\hline 
SV-K226  & 226 & 0.9 & 30 & 0.4 & 0.1605 & -15.90 &34&0.159&0.176\\ 
\hline 
SV-K241  & 241 & 0.9 & 30 & 0.4 &0.1588 & -15.91 & 31&0.151&0.167\\ 
\hline \hline 
SV-mas10 &234 & 1.0 & 30 & 0.4 & 0.1594 & -15.91 &28&0.152&0.170\\ 
\hline 
SV-mas08 & 234 & 0.8 & 30 & 0.4 &0.1597 & -15.90 & 40&0.160&0.175\\ 
\hline 
SV-mas07 & 234 &0.7 & 30 & 0.4 & 0.1500 & -15.89 &52&0.152&0.181\\ 
\hline \hline 
SV-sym28 & 234 & 0.9 & 28 &0.4 & 0.1595 & -15.86 &7&0.117&0.149\\ 
\hline 
SV-sym32 & 234	& 0.9 & 32 & 0.4 & 0.1595 & -15.94 &57&0.192&0.194\\ 
\hline
SV-sym34 & 234 & 0.9 & 34 & 0.4 & 0.1592 & -15.97 &81&0.227&0.215\\ 
\hline \hline 
SV-kap00 & 234 & 0.9 & 30 &0.0 & 0.1598 & -15.90 & 40&0.158&0.171\\ 
\hline 
SV-kap20 &234 & 0.9 & 30 & 0.2 & 0.1597 & -15.90 &36&0.155&0.170\\ 
\hline 
SV-kap60 & 234 & 0.9 & 30 & 0.6 &0.1595 & -15.91 &29&0.154&0.173\\ 
\hline \hline 
SV-L25   & 233.5 & 0.9 & 30 & 0.4 &0.1596 & -15.91 &25&0.143&0.161\\
\hline
SV-L32   & 233.3 & 0.9 & 30 & 0.4 &0.1594 & -15.90 &32&0.154&0.171\\
\hline
SV-L40   & 233.3 & 0.9 & 30 & 0.4 &0.1594 & -15.90 &40&0.166&0.181\\
\hline
SV-L47   & 233.4 & 0.9 & 30 & 0.4 &0.1596 & -15.90 &47&0.177&0.188\\
\hline \hline
\end{tabular} 
\caption{\label{tab:nucmat}
		Nuclear matter parameters as defined in
		Ref.~\cite{kluepfel2009} for the various EDF
		parametrizations used in this paper ($K$ in MeV, $J$
		in MeV, $E/A$ in MeV, $\rho_\mathrm{eq}$ in fm$^{-3}$,
		$L$ in MeV, $m^*/m$ dimensionless, and
		$\kappa_\mathrm{TRK}$ dimensionless).  As an
		observable from the finite nucleus, the neutron skin
		$R_n\!-\!R_p$ in $^{208}$Pb and $^{48}$Ca has been added
		in the last column (in units of fm).}
\end{table}

The fact that the NMP describes basic characteristics of a nuclear
EDF suggests that one should explore an EDF by systematic variation of the NMP in
the vicinity of the values for the optimum EDF.  This had been worked
in detail in~\cite{kluepfel2009} delivering a set of systematically
varied Skyrme parametrizations. As explained in Sec.~\ref{sec:EDF},
we are using this set here for exploration of fusion. The
detailed EDF parameters are given in~\cite{kluepfel2009}. Table 
\ref{tab:nucmat} summarizes the NMP for the members of this set.

\FloatBarrier

\bibliography{VU_bibtex_master.bib}

\begin{thebibliography}{106}%
\makeatletter
\providecommand \@ifxundefined [1]{%
 \@ifx{#1\undefined}
}%
\providecommand \@ifnum [1]{%
 \ifnum #1\expandafter \@firstoftwo
 \else \expandafter \@secondoftwo
 \fi
}%
\providecommand \@ifx [1]{%
 \ifx #1\expandafter \@firstoftwo
 \else \expandafter \@secondoftwo
 \fi
}%
\providecommand \natexlab [1]{#1}%
\providecommand \enquote  [1]{``#1''}%
\providecommand \bibnamefont  [1]{#1}%
\providecommand \bibfnamefont [1]{#1}%
\providecommand \citenamefont [1]{#1}%
\providecommand \href@noop [0]{\@secondoftwo}%
\providecommand \href [0]{\begingroup \@sanitize@url \@href}%
\providecommand \@href[1]{\@@startlink{#1}\@@href}%
\providecommand \@@href[1]{\endgroup#1\@@endlink}%
\providecommand \@sanitize@url [0]{\catcode `\\12\catcode `\$12\catcode
  `\&12\catcode `\#12\catcode `\^12\catcode `\_12\catcode `\%12\relax}%
\providecommand \@@startlink[1]{}%
\providecommand \@@endlink[0]{}%
\providecommand \url  [0]{\begingroup\@sanitize@url \@url }%
\providecommand \@url [1]{\endgroup\@href {#1}{\urlprefix }}%
\providecommand \urlprefix  [0]{URL }%
\providecommand \Eprint [0]{\href }%
\providecommand \doibase [0]{http://dx.doi.org/}%
\providecommand \selectlanguage [0]{\@gobble}%
\providecommand \bibinfo  [0]{\@secondoftwo}%
\providecommand \bibfield  [0]{\@secondoftwo}%
\providecommand \translation [1]{[#1]}%
\providecommand \BibitemOpen [0]{}%
\providecommand \bibitemStop [0]{}%
\providecommand \bibitemNoStop [0]{.\EOS\space}%
\providecommand \EOS [0]{\spacefactor3000\relax}%
\providecommand \BibitemShut  [1]{\csname bibitem#1\endcsname}%
\let\auto@bib@innerbib\@empty
\bibitem [{\citenamefont {Balantekin}\ \emph {et~al.}(2014)\citenamefont
  {Balantekin}, \citenamefont {Carlson}, \citenamefont {Dean}, \citenamefont
  {Fuller}, \citenamefont {Furnstahl}, \citenamefont {Hjorth-Jensen},
  \citenamefont {Janssens}, \citenamefont {Li}, \citenamefont {Nazarewicz},
  \citenamefont {Nunes}, \citenamefont {Ormand}, \citenamefont {Reddy},\ and\
  \citenamefont {Sherrill}}]{balantekin2014}%
  \BibitemOpen
  \bibfield  {author} {\bibinfo {author} {\bibfnamefont {A.~B.}\ \bibnamefont
  {Balantekin}}, \bibinfo {author} {\bibfnamefont {J.}~\bibnamefont {Carlson}},
  \bibinfo {author} {\bibfnamefont {D.~J.}\ \bibnamefont {Dean}}, \bibinfo
  {author} {\bibfnamefont {G.~M.}\ \bibnamefont {Fuller}}, \bibinfo {author}
  {\bibfnamefont {R.~J.}\ \bibnamefont {Furnstahl}}, \bibinfo {author}
  {\bibfnamefont {M.}~\bibnamefont {Hjorth-Jensen}}, \bibinfo {author}
  {\bibfnamefont {R.~V.~F.}\ \bibnamefont {Janssens}}, \bibinfo {author}
  {\bibfnamefont {B.-A.}\ \bibnamefont {Li}}, \bibinfo {author} {\bibfnamefont
  {W.}~\bibnamefont {Nazarewicz}}, \bibinfo {author} {\bibfnamefont {F.~M.}\
  \bibnamefont {Nunes}}, \bibinfo {author} {\bibfnamefont {W.~E.}\ \bibnamefont
  {Ormand}}, \bibinfo {author} {\bibfnamefont {S.}~\bibnamefont {Reddy}}, \
  and\ \bibinfo {author} {\bibfnamefont {B.~M.}\ \bibnamefont {Sherrill}},\
  }\href {\doibase 10.1142/S0217732314300109} {\bibfield  {journal} {\bibinfo
  {journal} {Mod. Phys. Lett. A}\ }\textbf {\bibinfo {volume} {29}},\ \bibinfo
  {pages} {1430010} (\bibinfo {year} {2014})}\BibitemShut {NoStop}%
\bibitem [{\citenamefont {{von Weizs\"acker}}(1935)}]{weizsacker1935}%
  \BibitemOpen
  \bibfield  {author} {\bibinfo {author} {\bibfnamefont {C.~F.}\ \bibnamefont
  {{von Weizs\"acker}}},\ }\href
  {http://link.springer.com/article/10.1007%2FBF01337700} {\bibfield  {journal}
  {\bibinfo  {journal} {Z. Phys. A}\ }\textbf {\bibinfo {volume} {96}},\
  \bibinfo {pages} {431} (\bibinfo {year} {1935})}\BibitemShut {NoStop}%
\bibitem [{\citenamefont {Bethe}\ and\ \citenamefont
  {Bacher}(1936)}]{bethe1936}%
  \BibitemOpen
  \bibfield  {author} {\bibinfo {author} {\bibfnamefont {H.~A.}\ \bibnamefont
  {Bethe}}\ and\ \bibinfo {author} {\bibfnamefont {R.~F.}\ \bibnamefont
  {Bacher}},\ }\href {\doibase 10.1103/RevModPhys.8.82} {\bibfield  {journal}
  {\bibinfo  {journal} {Rev. Mod. Phys.}\ }\textbf {\bibinfo {volume} {8}},\
  \bibinfo {pages} {82} (\bibinfo {year} {1936})}\BibitemShut {NoStop}%
\bibitem [{\citenamefont {Piekarewicz}(2014)}]{piekarewicz2014}%
  \BibitemOpen
  \bibfield  {author} {\bibinfo {author} {\bibfnamefont {J.}~\bibnamefont
  {Piekarewicz}},\ }\href {\doibase 10.1140/epja/i2014-14025-x} {\bibfield
  {journal} {\bibinfo  {journal} {Eur. Phys. J. A}\ }\textbf {\bibinfo {volume}
  {50}},\ \bibinfo {pages} {25} (\bibinfo {year} {2014})}\BibitemShut {NoStop}%
\bibitem [{\citenamefont {Chen}\ and\ \citenamefont
  {Piekarewicz}(2014)}]{chen2014}%
  \BibitemOpen
  \bibfield  {author} {\bibinfo {author} {\bibfnamefont {W.-C.}\ \bibnamefont
  {Chen}}\ and\ \bibinfo {author} {\bibfnamefont {J.}~\bibnamefont
  {Piekarewicz}},\ }\href {\doibase 10.1103/PhysRevC.90.044305} {\bibfield
  {journal} {\bibinfo  {journal} {Phys. Rev. C}\ }\textbf {\bibinfo {volume}
  {90}},\ \bibinfo {pages} {044305} (\bibinfo {year} {2014})}\BibitemShut
  {NoStop}%
\bibitem [{\citenamefont {Danielewicz}\ \emph {et~al.}(2002)\citenamefont
  {Danielewicz}, \citenamefont {Lacey},\ and\ \citenamefont
  {Lynch}}]{danielewicz2002}%
  \BibitemOpen
  \bibfield  {author} {\bibinfo {author} {\bibfnamefont {P.}~\bibnamefont
  {Danielewicz}}, \bibinfo {author} {\bibfnamefont {R.}~\bibnamefont {Lacey}},
  \ and\ \bibinfo {author} {\bibfnamefont {W.~G.}\ \bibnamefont {Lynch}},\
  }\href {\doibase 10.1126/science.1078070} {\bibfield  {journal} {\bibinfo
  {journal} {Science}\ }\textbf {\bibinfo {volume} {298}},\ \bibinfo {pages}
  {1592} (\bibinfo {year} {2002})}\BibitemShut {NoStop}%
\bibitem [{\citenamefont {Tsang}\ \emph {et~al.}(2009)\citenamefont {Tsang},
  \citenamefont {Zhang}, \citenamefont {Danielewicz}, \citenamefont {Famiano},
  \citenamefont {Li}, \citenamefont {Lynch},\ and\ \citenamefont
  {Steiner}}]{tsang2009}%
  \BibitemOpen
  \bibfield  {author} {\bibinfo {author} {\bibfnamefont {M.~B.}\ \bibnamefont
  {Tsang}}, \bibinfo {author} {\bibfnamefont {Y.}~\bibnamefont {Zhang}},
  \bibinfo {author} {\bibfnamefont {P.}~\bibnamefont {Danielewicz}}, \bibinfo
  {author} {\bibfnamefont {M.}~\bibnamefont {Famiano}}, \bibinfo {author}
  {\bibfnamefont {Z.}~\bibnamefont {Li}}, \bibinfo {author} {\bibfnamefont
  {W.~G.}\ \bibnamefont {Lynch}}, \ and\ \bibinfo {author} {\bibfnamefont
  {A.~W.}\ \bibnamefont {Steiner}},\ }\href {\doibase
  10.1103/PhysRevLett.102.122701} {\bibfield  {journal} {\bibinfo  {journal}
  {Phys. Rev. Lett.}\ }\textbf {\bibinfo {volume} {102}},\ \bibinfo {pages}
  {122701} (\bibinfo {year} {2009})}\BibitemShut {NoStop}%
\bibitem [{\citenamefont {Haensel}\ and\ \citenamefont
  {Zdunik}(1990)}]{haensel1990}%
  \BibitemOpen
  \bibfield  {author} {\bibinfo {author} {\bibfnamefont {P.}~\bibnamefont
  {Haensel}}\ and\ \bibinfo {author} {\bibfnamefont {J.~L.}\ \bibnamefont
  {Zdunik}},\ }\href
  {http://articles.adsabs.harvard.edu/cgi-bin/nph-iarticle_query?1990A%26A...229..117H&amp;
  data_type=PDF_HIGH&amp; whole_paper=YES&amp; type=PRINTER&amp; filetype=.pdf}
  {\bibfield  {journal} {\bibinfo  {journal} {Astron. Astrophys.}\ }\textbf
  {\bibinfo {volume} {229}},\ \bibinfo {pages} {117} (\bibinfo {year}
  {1990})}\BibitemShut {NoStop}%
\bibitem [{\citenamefont {Chamel}\ and\ \citenamefont
  {Haensel}(2008)}]{chamel2008}%
  \BibitemOpen
  \bibfield  {author} {\bibinfo {author} {\bibfnamefont {N.}~\bibnamefont
  {Chamel}}\ and\ \bibinfo {author} {\bibfnamefont {P.}~\bibnamefont
  {Haensel}},\ }\href {http://www.livingreviews.org/lrr-2008-10} {\bibfield
  {journal} {\bibinfo  {journal} {Living Rev. Relat.}\ }\textbf {\bibinfo
  {volume} {11}} (\bibinfo {year} {2008})}\BibitemShut {NoStop}%
\bibitem [{\citenamefont {Horowitz}\ \emph {et~al.}(2004)\citenamefont
  {Horowitz}, \citenamefont {P\'erez-Garc\'{\i}a},\ and\ \citenamefont
  {Piekarewicz}}]{horowitz2004}%
  \BibitemOpen
  \bibfield  {author} {\bibinfo {author} {\bibfnamefont {C.~J.}\ \bibnamefont
  {Horowitz}}, \bibinfo {author} {\bibfnamefont {M.~A.}\ \bibnamefont
  {P\'erez-Garc\'{\i}a}}, \ and\ \bibinfo {author} {\bibfnamefont
  {J.}~\bibnamefont {Piekarewicz}},\ }\href {\doibase
  10.1103/PhysRevC.69.045804} {\bibfield  {journal} {\bibinfo  {journal} {Phys.
  Rev. C}\ }\textbf {\bibinfo {volume} {69}},\ \bibinfo {pages} {045804}
  (\bibinfo {year} {2004})}\BibitemShut {NoStop}%
\bibitem [{\citenamefont {Utama}\ \emph {et~al.}(2016)\citenamefont {Utama},
  \citenamefont {Piekarewicz},\ and\ \citenamefont {Prosper}}]{utama2016}%
  \BibitemOpen
  \bibfield  {author} {\bibinfo {author} {\bibfnamefont {R.}~\bibnamefont
  {Utama}}, \bibinfo {author} {\bibfnamefont {J.}~\bibnamefont {Piekarewicz}},
  \ and\ \bibinfo {author} {\bibfnamefont {H.~B.}\ \bibnamefont {Prosper}},\
  }\href {\doibase 10.1103/PhysRevC.93.014311} {\bibfield  {journal} {\bibinfo
  {journal} {Phys. Rev. C}\ }\textbf {\bibinfo {volume} {93}},\ \bibinfo
  {pages} {014311} (\bibinfo {year} {2016})}\BibitemShut {NoStop}%
\bibitem [{\citenamefont {Bonche}\ and\ \citenamefont
  {Vautherin}(1981)}]{bonche1981}%
  \BibitemOpen
  \bibfield  {author} {\bibinfo {author} {\bibfnamefont {P.}~\bibnamefont
  {Bonche}}\ and\ \bibinfo {author} {\bibfnamefont {D.}~\bibnamefont
  {Vautherin}},\ }\href {\doibase 10.1016/0375-9474(81)90049-X} {\bibfield
  {journal} {\bibinfo  {journal} {Nucl. Phys. A}\ }\textbf {\bibinfo {volume}
  {372}},\ \bibinfo {pages} {496 } (\bibinfo {year} {1981})}\BibitemShut
  {NoStop}%
\bibitem [{\citenamefont {Watanabe}\ \emph {et~al.}(2009)\citenamefont
  {Watanabe}, \citenamefont {Sonoda}, \citenamefont {Maruyama}, \citenamefont
  {Sato}, \citenamefont {Yasuoka},\ and\ \citenamefont
  {Ebisuzaki}}]{watanabe2009}%
  \BibitemOpen
  \bibfield  {author} {\bibinfo {author} {\bibfnamefont {G.}~\bibnamefont
  {Watanabe}}, \bibinfo {author} {\bibfnamefont {H.}~\bibnamefont {Sonoda}},
  \bibinfo {author} {\bibfnamefont {T.}~\bibnamefont {Maruyama}}, \bibinfo
  {author} {\bibfnamefont {K.}~\bibnamefont {Sato}}, \bibinfo {author}
  {\bibfnamefont {K.}~\bibnamefont {Yasuoka}}, \ and\ \bibinfo {author}
  {\bibfnamefont {T.}~\bibnamefont {Ebisuzaki}},\ }\href {\doibase
  10.1103/PhysRevLett.103.121101} {\bibfield  {journal} {\bibinfo  {journal}
  {Phys. Rev. Lett.}\ }\textbf {\bibinfo {volume} {103}},\ \bibinfo {pages}
  {121101} (\bibinfo {year} {2009})}\BibitemShut {NoStop}%
\bibitem [{\citenamefont {Shen}\ \emph {et~al.}(2011)\citenamefont {Shen},
  \citenamefont {Horowitz},\ and\ \citenamefont {Teige}}]{shen2011}%
  \BibitemOpen
  \bibfield  {author} {\bibinfo {author} {\bibfnamefont {G.}~\bibnamefont
  {Shen}}, \bibinfo {author} {\bibfnamefont {C.~J.}\ \bibnamefont {Horowitz}},
  \ and\ \bibinfo {author} {\bibfnamefont {S.}~\bibnamefont {Teige}},\ }\href
  {\doibase 10.1103/PhysRevC.83.035802} {\bibfield  {journal} {\bibinfo
  {journal} {Phys. Rev. C}\ }\textbf {\bibinfo {volume} {83}},\ \bibinfo
  {pages} {035802} (\bibinfo {year} {2011})}\BibitemShut {NoStop}%
\bibitem [{\citenamefont {Piekarewicz}(2007)}]{piekarewicz2007}%
  \BibitemOpen
  \bibfield  {author} {\bibinfo {author} {\bibfnamefont {J.}~\bibnamefont
  {Piekarewicz}},\ }\href {\doibase 10.1103/PhysRevC.76.064310} {\bibfield
  {journal} {\bibinfo  {journal} {Phys. Rev. C}\ }\textbf {\bibinfo {volume}
  {76}},\ \bibinfo {pages} {064310} (\bibinfo {year} {2007})}\BibitemShut
  {NoStop}%
\bibitem [{\citenamefont {Baran}\ \emph {et~al.}(2005)\citenamefont {Baran},
  \citenamefont {Colonna}, \citenamefont {Greco},\ and\ \citenamefont
  {Di~Toro}}]{baran2005}%
  \BibitemOpen
  \bibfield  {author} {\bibinfo {author} {\bibfnamefont {V.}~\bibnamefont
  {Baran}}, \bibinfo {author} {\bibfnamefont {M.}~\bibnamefont {Colonna}},
  \bibinfo {author} {\bibfnamefont {V.}~\bibnamefont {Greco}}, \ and\ \bibinfo
  {author} {\bibfnamefont {M.}~\bibnamefont {Di~Toro}},\ }\href {\doibase
  10.1016/j.physrep.2004.12.004} {\bibfield  {journal} {\bibinfo  {journal}
  {Phys. Rep.}\ }\textbf {\bibinfo {volume} {410}},\ \bibinfo {pages} {335}
  (\bibinfo {year} {2005})}\BibitemShut {NoStop}%
\bibitem [{\citenamefont {Steiner}\ \emph {et~al.}(2005)\citenamefont
  {Steiner}, \citenamefont {Prakash}, \citenamefont {Lattimer},\ and\
  \citenamefont {Ellis}}]{steiner2005}%
  \BibitemOpen
  \bibfield  {author} {\bibinfo {author} {\bibfnamefont {A.~W.}\ \bibnamefont
  {Steiner}}, \bibinfo {author} {\bibfnamefont {M.}~\bibnamefont {Prakash}},
  \bibinfo {author} {\bibfnamefont {J.~M.}\ \bibnamefont {Lattimer}}, \ and\
  \bibinfo {author} {\bibfnamefont {P.}~\bibnamefont {Ellis}},\ }\href
  {\doibase 10.1016/j.physrep.2005.02.004} {\bibfield  {journal} {\bibinfo
  {journal} {Phys. Rep.}\ }\textbf {\bibinfo {volume} {411}},\ \bibinfo {pages}
  {325} (\bibinfo {year} {2005})}\BibitemShut {NoStop}%
\bibitem [{\citenamefont {Horowitz}\ \emph {et~al.}(2014)\citenamefont
  {Horowitz}, \citenamefont {Brown}, \citenamefont {Kim}, \citenamefont
  {Lynch}, \citenamefont {Michaels}, \citenamefont {Ono}, \citenamefont
  {Piekarewicz}, \citenamefont {Tsang},\ and\ \citenamefont
  {Wolter}}]{horowitz2014}%
  \BibitemOpen
  \bibfield  {author} {\bibinfo {author} {\bibfnamefont {C.~J.}\ \bibnamefont
  {Horowitz}}, \bibinfo {author} {\bibfnamefont {E.~F.}\ \bibnamefont {Brown}},
  \bibinfo {author} {\bibfnamefont {Y.}~\bibnamefont {Kim}}, \bibinfo {author}
  {\bibfnamefont {W.~G.}\ \bibnamefont {Lynch}}, \bibinfo {author}
  {\bibfnamefont {R.}~\bibnamefont {Michaels}}, \bibinfo {author}
  {\bibfnamefont {A.}~\bibnamefont {Ono}}, \bibinfo {author} {\bibfnamefont
  {J.}~\bibnamefont {Piekarewicz}}, \bibinfo {author} {\bibfnamefont {M.~B.}\
  \bibnamefont {Tsang}}, \ and\ \bibinfo {author} {\bibfnamefont {H.~H.}\
  \bibnamefont {Wolter}},\ }\href {\doibase 10.1088/0954-3899/41/9/093001}
  {\bibfield  {journal} {\bibinfo  {journal} {J. Phys. G.}\ }\textbf {\bibinfo
  {volume} {41}},\ \bibinfo {pages} {093001} (\bibinfo {year}
  {2014})}\BibitemShut {NoStop}%
\bibitem [{\citenamefont {{Bao-An Li and \`Angels Ramos and Giuseppe Verde and
  Isaac Vida\~na}}(2014)}]{li2014}%
  \BibitemOpen
  \bibfield  {author} {\bibinfo {author} {\bibnamefont {{Bao-An Li and \`Angels
  Ramos and Giuseppe Verde and Isaac Vida\~na}}},\ }\href
  {http://epja.epj.org/component/toc/?task=topic&id=260} {\bibfield  {journal}
  {\bibinfo  {journal} {Eur. Phys. J. A}\ }\textbf {\bibinfo {volume} {50}},\
  \bibinfo {pages} {1} (\bibinfo {year} {2014})}\BibitemShut {NoStop}%
\bibitem [{\citenamefont {Alex~Brown}(2000)}]{brown2000}%
  \BibitemOpen
  \bibfield  {author} {\bibinfo {author} {\bibfnamefont {B.}~\bibnamefont
  {Alex~Brown}},\ }\href {\doibase 10.1103/PhysRevLett.85.5296} {\bibfield
  {journal} {\bibinfo  {journal} {Phys. Rev. Lett.}\ }\textbf {\bibinfo
  {volume} {85}},\ \bibinfo {pages} {5296} (\bibinfo {year}
  {2000})}\BibitemShut {NoStop}%
\bibitem [{\citenamefont {Furnstahl}(2002)}]{furnstahl2002}%
  \BibitemOpen
  \bibfield  {author} {\bibinfo {author} {\bibfnamefont {R.}~\bibnamefont
  {Furnstahl}},\ }\href {\doibase
  http://dx.doi.org/10.1016/S0375-9474(02)00867-9} {\bibfield  {journal}
  {\bibinfo  {journal} {Nucl. Phys. A}\ }\textbf {\bibinfo {volume} {706}},\
  \bibinfo {pages} {85 } (\bibinfo {year} {2002})}\BibitemShut {NoStop}%
\bibitem [{\citenamefont {Centelles}\ \emph {et~al.}(2010)\citenamefont
  {Centelles}, \citenamefont {Patra}, \citenamefont {Roca-Maza}, \citenamefont
  {Sharma}, \citenamefont {Stevenson},\ and\ \citenamefont
  {Vinas}}]{centelles2010}%
  \BibitemOpen
  \bibfield  {author} {\bibinfo {author} {\bibfnamefont {M.}~\bibnamefont
  {Centelles}}, \bibinfo {author} {\bibfnamefont {S.~K.}\ \bibnamefont
  {Patra}}, \bibinfo {author} {\bibfnamefont {X.}~\bibnamefont {Roca-Maza}},
  \bibinfo {author} {\bibfnamefont {B.~K.}\ \bibnamefont {Sharma}}, \bibinfo
  {author} {\bibfnamefont {P.~D.}\ \bibnamefont {Stevenson}}, \ and\ \bibinfo
  {author} {\bibfnamefont {X.}~\bibnamefont {Vinas}},\ }\href {\doibase
  10.1088/0954-3899/37/7/075107} {\bibfield  {journal} {\bibinfo  {journal} {J.
  Phys. G}\ }\textbf {\bibinfo {volume} {37}},\ \bibinfo {pages} {075107}
  (\bibinfo {year} {2010})}\BibitemShut {NoStop}%
\bibitem [{\citenamefont {Roca-Maza}\ \emph {et~al.}(2011)\citenamefont
  {Roca-Maza}, \citenamefont {Centelles}, \citenamefont {Vi\~nas},\ and\
  \citenamefont {Warda}}]{roca2011}%
  \BibitemOpen
  \bibfield  {author} {\bibinfo {author} {\bibfnamefont {X.}~\bibnamefont
  {Roca-Maza}}, \bibinfo {author} {\bibfnamefont {M.}~\bibnamefont
  {Centelles}}, \bibinfo {author} {\bibfnamefont {X.}~\bibnamefont {Vi\~nas}},
  \ and\ \bibinfo {author} {\bibfnamefont {M.}~\bibnamefont {Warda}},\ }\href
  {\doibase 10.1103/PhysRevLett.106.252501} {\bibfield  {journal} {\bibinfo
  {journal} {Phys. Rev. Lett.}\ }\textbf {\bibinfo {volume} {106}},\ \bibinfo
  {pages} {252501} (\bibinfo {year} {2011})}\BibitemShut {NoStop}%
\bibitem [{\citenamefont {Vries}\ \emph {et~al.}(1987)\citenamefont {Vries},
  \citenamefont {Jager},\ and\ \citenamefont {Vries}}]{vries1987}%
  \BibitemOpen
  \bibfield  {author} {\bibinfo {author} {\bibfnamefont {H.~D.}\ \bibnamefont
  {Vries}}, \bibinfo {author} {\bibfnamefont {C.~D.}\ \bibnamefont {Jager}}, \
  and\ \bibinfo {author} {\bibfnamefont {C.~D.}\ \bibnamefont {Vries}},\ }\href
  {\doibase http://dx.doi.org/10.1016/0092-640X(87)90013-1} {\bibfield
  {journal} {\bibinfo  {journal} {Atom. Data Nucl. Data Tabl.}\ }\textbf
  {\bibinfo {volume} {36}},\ \bibinfo {pages} {495 } (\bibinfo {year}
  {1987})}\BibitemShut {NoStop}%
\bibitem [{\citenamefont {Fricke}\ \emph {et~al.}(1995)\citenamefont {Fricke},
  \citenamefont {Bernhardt}, \citenamefont {Heilig}, \citenamefont {Schaller},
  \citenamefont {Schellenberg}, \citenamefont {Shera},\ and\ \citenamefont
  {Dejager}}]{fricke1995}%
  \BibitemOpen
  \bibfield  {author} {\bibinfo {author} {\bibfnamefont {G.}~\bibnamefont
  {Fricke}}, \bibinfo {author} {\bibfnamefont {C.}~\bibnamefont {Bernhardt}},
  \bibinfo {author} {\bibfnamefont {K.}~\bibnamefont {Heilig}}, \bibinfo
  {author} {\bibfnamefont {L.}~\bibnamefont {Schaller}}, \bibinfo {author}
  {\bibfnamefont {L.}~\bibnamefont {Schellenberg}}, \bibinfo {author}
  {\bibfnamefont {E.}~\bibnamefont {Shera}}, \ and\ \bibinfo {author}
  {\bibfnamefont {C.}~\bibnamefont {Dejager}},\ }\href {\doibase
  http://dx.doi.org/10.1006/adnd.1995.1007} {\bibfield  {journal} {\bibinfo
  {journal} {Atomic Data and Nuclear Data Tables}\ }\textbf {\bibinfo {volume}
  {60}},\ \bibinfo {pages} {177 } (\bibinfo {year} {1995})}\BibitemShut
  {NoStop}%
\bibitem [{\citenamefont {Angeli}\ and\ \citenamefont
  {Marinova}(2013)}]{angeli2013}%
  \BibitemOpen
  \bibfield  {author} {\bibinfo {author} {\bibfnamefont {I.}~\bibnamefont
  {Angeli}}\ and\ \bibinfo {author} {\bibfnamefont {K.}~\bibnamefont
  {Marinova}},\ }\href {\doibase http://dx.doi.org/10.1016/j.adt.2011.12.006}
  {\bibfield  {journal} {\bibinfo  {journal} {At. Data Nucl. Data Tables}\
  }\textbf {\bibinfo {volume} {99}},\ \bibinfo {pages} {69 } (\bibinfo {year}
  {2013})}\BibitemShut {NoStop}%
\bibitem [{\citenamefont {Abrahamyan}\ \emph {et~al.}(2012)\citenamefont
  {Abrahamyan}, \citenamefont {Ahmed}, \citenamefont {Albataineh},
  \citenamefont {Aniol}, \citenamefont {Armstrong}, \citenamefont {Armstrong},
  \citenamefont {Averett}, \citenamefont {Babineau}, \citenamefont {Barbieri},
  \citenamefont {Bellini}, \citenamefont {Beminiwattha}, \citenamefont
  {Benesch}, \citenamefont {Benmokhtar}, \citenamefont {Bielarski},
  \citenamefont {Boeglin}, \citenamefont {Camsonne}, \citenamefont {Canan},
  \citenamefont {Carter}, \citenamefont {Cates}, \citenamefont {Chen},
  \citenamefont {Chen}, \citenamefont {Hen}, \citenamefont {Cusanno},
  \citenamefont {Dalton}, \citenamefont {De~Leo}, \citenamefont {de~Jager},
  \citenamefont {Deconinck}, \citenamefont {Decowski}, \citenamefont {Deng},
  \citenamefont {Deur}, \citenamefont {Dutta}, \citenamefont {Etile},
  \citenamefont {Flay}, \citenamefont {Franklin}, \citenamefont {Friend},
  \citenamefont {Frullani}, \citenamefont {Fuchey}, \citenamefont {Garibaldi},
  \citenamefont {Gasser}, \citenamefont {Gilman}, \citenamefont {Giusa},
  \citenamefont {Glamazdin}, \citenamefont {Gomez}, \citenamefont {Grames},
  \citenamefont {Gu}, \citenamefont {Hansen}, \citenamefont {Hansknecht},
  \citenamefont {Higinbotham}, \citenamefont {Holmes}, \citenamefont
  {Holmstrom}, \citenamefont {Horowitz}, \citenamefont {Hoskins}, \citenamefont
  {Huang}, \citenamefont {Hyde}, \citenamefont {Itard}, \citenamefont {Jen},
  \citenamefont {Jensen}, \citenamefont {Jin}, \citenamefont {Johnston},
  \citenamefont {Kelleher}, \citenamefont {Kliakhandler}, \citenamefont {King},
  \citenamefont {Kowalski}, \citenamefont {Kumar}, \citenamefont {Leacock},
  \citenamefont {Leckey}, \citenamefont {Lee}, \citenamefont {LeRose},
  \citenamefont {Lindgren}, \citenamefont {Liyanage}, \citenamefont {Lubinsky},
  \citenamefont {Mammei}, \citenamefont {Mammoliti}, \citenamefont
  {Margaziotis}, \citenamefont {Markowitz}, \citenamefont {McCreary},
  \citenamefont {McNulty}, \citenamefont {Mercado}, \citenamefont {Meziani},
  \citenamefont {Michaels}, \citenamefont {Mihovilovic}, \citenamefont
  {Muangma}, \citenamefont {Mu\~noz Camacho}, \citenamefont {Nanda},
  \citenamefont {Nelyubin}, \citenamefont {Nuruzzaman}, \citenamefont {Oh},
  \citenamefont {Palmer}, \citenamefont {Parno}, \citenamefont {Paschke},
  \citenamefont {Phillips}, \citenamefont {Poelker}, \citenamefont
  {Pomatsalyuk}, \citenamefont {Posik}, \citenamefont {Puckett}, \citenamefont
  {Quinn}, \citenamefont {Rakhman}, \citenamefont {Reimer}, \citenamefont
  {Riordan}, \citenamefont {Rogan}, \citenamefont {Ron}, \citenamefont {Russo},
  \citenamefont {Saenboonruang}, \citenamefont {Saha}, \citenamefont
  {Sawatzky}, \citenamefont {Shahinyan}, \citenamefont {Silwal}, \citenamefont
  {Sirca}, \citenamefont {Slifer}, \citenamefont {Solvignon}, \citenamefont
  {Souder}, \citenamefont {Sperduto}, \citenamefont {Subedi}, \citenamefont
  {Suleiman}, \citenamefont {Sulkosky}, \citenamefont {Sutera}, \citenamefont
  {Tobias}, \citenamefont {Troth}, \citenamefont {Urciuoli}, \citenamefont
  {Waidyawansa}, \citenamefont {Wang}, \citenamefont {Wexler}, \citenamefont
  {Wilson}, \citenamefont {Wojtsekhowski}, \citenamefont {Yan}, \citenamefont
  {Yao}, \citenamefont {Ye}, \citenamefont {Ye}, \citenamefont {Yim},
  \citenamefont {Zana}, \citenamefont {Zhan}, \citenamefont {Zhang},
  \citenamefont {Zhang}, \citenamefont {Zheng},\ and\ \citenamefont
  {Zhu}}]{abrahamyan2012}%
  \BibitemOpen
  \bibfield  {author} {\bibinfo {author} {\bibfnamefont {S.}~\bibnamefont
  {Abrahamyan}}, \bibinfo {author} {\bibfnamefont {Z.}~\bibnamefont {Ahmed}},
  \bibinfo {author} {\bibfnamefont {H.}~\bibnamefont {Albataineh}}, \bibinfo
  {author} {\bibfnamefont {K.}~\bibnamefont {Aniol}}, \bibinfo {author}
  {\bibfnamefont {D.~S.}\ \bibnamefont {Armstrong}}, \bibinfo {author}
  {\bibfnamefont {W.}~\bibnamefont {Armstrong}}, \bibinfo {author}
  {\bibfnamefont {T.}~\bibnamefont {Averett}}, \bibinfo {author} {\bibfnamefont
  {B.}~\bibnamefont {Babineau}}, \bibinfo {author} {\bibfnamefont
  {A.}~\bibnamefont {Barbieri}}, \bibinfo {author} {\bibfnamefont
  {V.}~\bibnamefont {Bellini}}, \bibinfo {author} {\bibfnamefont
  {R.}~\bibnamefont {Beminiwattha}}, \bibinfo {author} {\bibfnamefont
  {J.}~\bibnamefont {Benesch}}, \bibinfo {author} {\bibfnamefont
  {F.}~\bibnamefont {Benmokhtar}}, \bibinfo {author} {\bibfnamefont
  {T.}~\bibnamefont {Bielarski}}, \bibinfo {author} {\bibfnamefont
  {W.}~\bibnamefont {Boeglin}}, \bibinfo {author} {\bibfnamefont
  {A.}~\bibnamefont {Camsonne}}, \bibinfo {author} {\bibfnamefont
  {M.}~\bibnamefont {Canan}}, \bibinfo {author} {\bibfnamefont
  {P.}~\bibnamefont {Carter}}, \bibinfo {author} {\bibfnamefont {G.~D.}\
  \bibnamefont {Cates}}, \bibinfo {author} {\bibfnamefont {C.}~\bibnamefont
  {Chen}}, \bibinfo {author} {\bibfnamefont {J.-P.}\ \bibnamefont {Chen}},
  \bibinfo {author} {\bibfnamefont {O.}~\bibnamefont {Hen}}, \bibinfo {author}
  {\bibfnamefont {F.}~\bibnamefont {Cusanno}}, \bibinfo {author} {\bibfnamefont
  {M.~M.}\ \bibnamefont {Dalton}}, \bibinfo {author} {\bibfnamefont
  {R.}~\bibnamefont {De~Leo}}, \bibinfo {author} {\bibfnamefont
  {K.}~\bibnamefont {de~Jager}}, \bibinfo {author} {\bibfnamefont
  {W.}~\bibnamefont {Deconinck}}, \bibinfo {author} {\bibfnamefont
  {P.}~\bibnamefont {Decowski}}, \bibinfo {author} {\bibfnamefont
  {X.}~\bibnamefont {Deng}}, \bibinfo {author} {\bibfnamefont {A.}~\bibnamefont
  {Deur}}, \bibinfo {author} {\bibfnamefont {D.}~\bibnamefont {Dutta}},
  \bibinfo {author} {\bibfnamefont {A.}~\bibnamefont {Etile}}, \bibinfo
  {author} {\bibfnamefont {D.}~\bibnamefont {Flay}}, \bibinfo {author}
  {\bibfnamefont {G.~B.}\ \bibnamefont {Franklin}}, \bibinfo {author}
  {\bibfnamefont {M.}~\bibnamefont {Friend}}, \bibinfo {author} {\bibfnamefont
  {S.}~\bibnamefont {Frullani}}, \bibinfo {author} {\bibfnamefont
  {E.}~\bibnamefont {Fuchey}}, \bibinfo {author} {\bibfnamefont
  {F.}~\bibnamefont {Garibaldi}}, \bibinfo {author} {\bibfnamefont
  {E.}~\bibnamefont {Gasser}}, \bibinfo {author} {\bibfnamefont
  {R.}~\bibnamefont {Gilman}}, \bibinfo {author} {\bibfnamefont
  {A.}~\bibnamefont {Giusa}}, \bibinfo {author} {\bibfnamefont
  {A.}~\bibnamefont {Glamazdin}}, \bibinfo {author} {\bibfnamefont
  {J.}~\bibnamefont {Gomez}}, \bibinfo {author} {\bibfnamefont
  {J.}~\bibnamefont {Grames}}, \bibinfo {author} {\bibfnamefont
  {C.}~\bibnamefont {Gu}}, \bibinfo {author} {\bibfnamefont {O.}~\bibnamefont
  {Hansen}}, \bibinfo {author} {\bibfnamefont {J.}~\bibnamefont {Hansknecht}},
  \bibinfo {author} {\bibfnamefont {D.~W.}\ \bibnamefont {Higinbotham}},
  \bibinfo {author} {\bibfnamefont {R.~S.}\ \bibnamefont {Holmes}}, \bibinfo
  {author} {\bibfnamefont {T.}~\bibnamefont {Holmstrom}}, \bibinfo {author}
  {\bibfnamefont {C.~J.}\ \bibnamefont {Horowitz}}, \bibinfo {author}
  {\bibfnamefont {J.}~\bibnamefont {Hoskins}}, \bibinfo {author} {\bibfnamefont
  {J.}~\bibnamefont {Huang}}, \bibinfo {author} {\bibfnamefont {C.~E.}\
  \bibnamefont {Hyde}}, \bibinfo {author} {\bibfnamefont {F.}~\bibnamefont
  {Itard}}, \bibinfo {author} {\bibfnamefont {C.-M.}\ \bibnamefont {Jen}},
  \bibinfo {author} {\bibfnamefont {E.}~\bibnamefont {Jensen}}, \bibinfo
  {author} {\bibfnamefont {G.}~\bibnamefont {Jin}}, \bibinfo {author}
  {\bibfnamefont {S.}~\bibnamefont {Johnston}}, \bibinfo {author}
  {\bibfnamefont {A.}~\bibnamefont {Kelleher}}, \bibinfo {author}
  {\bibfnamefont {K.}~\bibnamefont {Kliakhandler}}, \bibinfo {author}
  {\bibfnamefont {P.~M.}\ \bibnamefont {King}}, \bibinfo {author}
  {\bibfnamefont {S.}~\bibnamefont {Kowalski}}, \bibinfo {author}
  {\bibfnamefont {K.~S.}\ \bibnamefont {Kumar}}, \bibinfo {author}
  {\bibfnamefont {J.}~\bibnamefont {Leacock}}, \bibinfo {author} {\bibfnamefont
  {J.}~\bibnamefont {Leckey}}, \bibinfo {author} {\bibfnamefont {J.~H.}\
  \bibnamefont {Lee}}, \bibinfo {author} {\bibfnamefont {J.~J.}\ \bibnamefont
  {LeRose}}, \bibinfo {author} {\bibfnamefont {R.}~\bibnamefont {Lindgren}},
  \bibinfo {author} {\bibfnamefont {N.}~\bibnamefont {Liyanage}}, \bibinfo
  {author} {\bibfnamefont {N.}~\bibnamefont {Lubinsky}}, \bibinfo {author}
  {\bibfnamefont {J.}~\bibnamefont {Mammei}}, \bibinfo {author} {\bibfnamefont
  {F.}~\bibnamefont {Mammoliti}}, \bibinfo {author} {\bibfnamefont {D.~J.}\
  \bibnamefont {Margaziotis}}, \bibinfo {author} {\bibfnamefont
  {P.}~\bibnamefont {Markowitz}}, \bibinfo {author} {\bibfnamefont
  {A.}~\bibnamefont {McCreary}}, \bibinfo {author} {\bibfnamefont
  {D.}~\bibnamefont {McNulty}}, \bibinfo {author} {\bibfnamefont
  {L.}~\bibnamefont {Mercado}}, \bibinfo {author} {\bibfnamefont {Z.-E.}\
  \bibnamefont {Meziani}}, \bibinfo {author} {\bibfnamefont {R.~W.}\
  \bibnamefont {Michaels}}, \bibinfo {author} {\bibfnamefont {M.}~\bibnamefont
  {Mihovilovic}}, \bibinfo {author} {\bibfnamefont {N.}~\bibnamefont
  {Muangma}}, \bibinfo {author} {\bibfnamefont {C.}~\bibnamefont {Mu\~noz
  Camacho}}, \bibinfo {author} {\bibfnamefont {S.}~\bibnamefont {Nanda}},
  \bibinfo {author} {\bibfnamefont {V.}~\bibnamefont {Nelyubin}}, \bibinfo
  {author} {\bibfnamefont {N.}~\bibnamefont {Nuruzzaman}}, \bibinfo {author}
  {\bibfnamefont {Y.}~\bibnamefont {Oh}}, \bibinfo {author} {\bibfnamefont
  {A.}~\bibnamefont {Palmer}}, \bibinfo {author} {\bibfnamefont
  {D.}~\bibnamefont {Parno}}, \bibinfo {author} {\bibfnamefont {K.~D.}\
  \bibnamefont {Paschke}}, \bibinfo {author} {\bibfnamefont {S.~K.}\
  \bibnamefont {Phillips}}, \bibinfo {author} {\bibfnamefont {B.}~\bibnamefont
  {Poelker}}, \bibinfo {author} {\bibfnamefont {R.}~\bibnamefont
  {Pomatsalyuk}}, \bibinfo {author} {\bibfnamefont {M.}~\bibnamefont {Posik}},
  \bibinfo {author} {\bibfnamefont {A.~J.~R.}\ \bibnamefont {Puckett}},
  \bibinfo {author} {\bibfnamefont {B.}~\bibnamefont {Quinn}}, \bibinfo
  {author} {\bibfnamefont {A.}~\bibnamefont {Rakhman}}, \bibinfo {author}
  {\bibfnamefont {P.~E.}\ \bibnamefont {Reimer}}, \bibinfo {author}
  {\bibfnamefont {S.}~\bibnamefont {Riordan}}, \bibinfo {author} {\bibfnamefont
  {P.}~\bibnamefont {Rogan}}, \bibinfo {author} {\bibfnamefont
  {G.}~\bibnamefont {Ron}}, \bibinfo {author} {\bibfnamefont {G.}~\bibnamefont
  {Russo}}, \bibinfo {author} {\bibfnamefont {K.}~\bibnamefont
  {Saenboonruang}}, \bibinfo {author} {\bibfnamefont {A.}~\bibnamefont {Saha}},
  \bibinfo {author} {\bibfnamefont {B.}~\bibnamefont {Sawatzky}}, \bibinfo
  {author} {\bibfnamefont {A.}~\bibnamefont {Shahinyan}}, \bibinfo {author}
  {\bibfnamefont {R.}~\bibnamefont {Silwal}}, \bibinfo {author} {\bibfnamefont
  {S.}~\bibnamefont {Sirca}}, \bibinfo {author} {\bibfnamefont
  {K.}~\bibnamefont {Slifer}}, \bibinfo {author} {\bibfnamefont
  {P.}~\bibnamefont {Solvignon}}, \bibinfo {author} {\bibfnamefont {P.~A.}\
  \bibnamefont {Souder}}, \bibinfo {author} {\bibfnamefont {M.~L.}\
  \bibnamefont {Sperduto}}, \bibinfo {author} {\bibfnamefont {R.}~\bibnamefont
  {Subedi}}, \bibinfo {author} {\bibfnamefont {R.}~\bibnamefont {Suleiman}},
  \bibinfo {author} {\bibfnamefont {V.}~\bibnamefont {Sulkosky}}, \bibinfo
  {author} {\bibfnamefont {C.~M.}\ \bibnamefont {Sutera}}, \bibinfo {author}
  {\bibfnamefont {W.~A.}\ \bibnamefont {Tobias}}, \bibinfo {author}
  {\bibfnamefont {W.}~\bibnamefont {Troth}}, \bibinfo {author} {\bibfnamefont
  {G.~M.}\ \bibnamefont {Urciuoli}}, \bibinfo {author} {\bibfnamefont
  {B.}~\bibnamefont {Waidyawansa}}, \bibinfo {author} {\bibfnamefont
  {D.}~\bibnamefont {Wang}}, \bibinfo {author} {\bibfnamefont {J.}~\bibnamefont
  {Wexler}}, \bibinfo {author} {\bibfnamefont {R.}~\bibnamefont {Wilson}},
  \bibinfo {author} {\bibfnamefont {B.}~\bibnamefont {Wojtsekhowski}}, \bibinfo
  {author} {\bibfnamefont {X.}~\bibnamefont {Yan}}, \bibinfo {author}
  {\bibfnamefont {H.}~\bibnamefont {Yao}}, \bibinfo {author} {\bibfnamefont
  {Y.}~\bibnamefont {Ye}}, \bibinfo {author} {\bibfnamefont {Z.}~\bibnamefont
  {Ye}}, \bibinfo {author} {\bibfnamefont {V.}~\bibnamefont {Yim}}, \bibinfo
  {author} {\bibfnamefont {L.}~\bibnamefont {Zana}}, \bibinfo {author}
  {\bibfnamefont {X.}~\bibnamefont {Zhan}}, \bibinfo {author} {\bibfnamefont
  {J.}~\bibnamefont {Zhang}}, \bibinfo {author} {\bibfnamefont
  {Y.}~\bibnamefont {Zhang}}, \bibinfo {author} {\bibfnamefont
  {X.}~\bibnamefont {Zheng}}, \ and\ \bibinfo {author} {\bibfnamefont
  {P.}~\bibnamefont {Zhu}} (\bibinfo {collaboration} {PREX Collaboration}),\
  }\href {\doibase 10.1103/PhysRevLett.108.112502} {\bibfield  {journal}
  {\bibinfo  {journal} {Phys. Rev. Lett.}\ }\textbf {\bibinfo {volume} {108}},\
  \bibinfo {pages} {112502} (\bibinfo {year} {2012})}\BibitemShut {NoStop}%
\bibitem [{\citenamefont {Horowitz}\ \emph {et~al.}(2012)\citenamefont
  {Horowitz}, \citenamefont {Ahmed}, \citenamefont {Jen}, \citenamefont
  {Rakhman}, \citenamefont {Souder}, \citenamefont {Dalton}, \citenamefont
  {Liyanage}, \citenamefont {Paschke}, \citenamefont {Saenboonruang},
  \citenamefont {Silwal}, \citenamefont {Franklin}, \citenamefont {Friend},
  \citenamefont {Quinn}, \citenamefont {Kumar}, \citenamefont {McNulty},
  \citenamefont {Mercado}, \citenamefont {Riordan}, \citenamefont {Wexler},
  \citenamefont {Michaels},\ and\ \citenamefont {Urciuoli}}]{horowitz2012}%
  \BibitemOpen
  \bibfield  {author} {\bibinfo {author} {\bibfnamefont {C.~J.}\ \bibnamefont
  {Horowitz}}, \bibinfo {author} {\bibfnamefont {Z.}~\bibnamefont {Ahmed}},
  \bibinfo {author} {\bibfnamefont {C.-M.}\ \bibnamefont {Jen}}, \bibinfo
  {author} {\bibfnamefont {A.}~\bibnamefont {Rakhman}}, \bibinfo {author}
  {\bibfnamefont {P.~A.}\ \bibnamefont {Souder}}, \bibinfo {author}
  {\bibfnamefont {M.~M.}\ \bibnamefont {Dalton}}, \bibinfo {author}
  {\bibfnamefont {N.}~\bibnamefont {Liyanage}}, \bibinfo {author}
  {\bibfnamefont {K.~D.}\ \bibnamefont {Paschke}}, \bibinfo {author}
  {\bibfnamefont {K.}~\bibnamefont {Saenboonruang}}, \bibinfo {author}
  {\bibfnamefont {R.}~\bibnamefont {Silwal}}, \bibinfo {author} {\bibfnamefont
  {G.~B.}\ \bibnamefont {Franklin}}, \bibinfo {author} {\bibfnamefont
  {M.}~\bibnamefont {Friend}}, \bibinfo {author} {\bibfnamefont
  {B.}~\bibnamefont {Quinn}}, \bibinfo {author} {\bibfnamefont {K.~S.}\
  \bibnamefont {Kumar}}, \bibinfo {author} {\bibfnamefont {D.}~\bibnamefont
  {McNulty}}, \bibinfo {author} {\bibfnamefont {L.}~\bibnamefont {Mercado}},
  \bibinfo {author} {\bibfnamefont {S.}~\bibnamefont {Riordan}}, \bibinfo
  {author} {\bibfnamefont {J.}~\bibnamefont {Wexler}}, \bibinfo {author}
  {\bibfnamefont {R.~W.}\ \bibnamefont {Michaels}}, \ and\ \bibinfo {author}
  {\bibfnamefont {G.~M.}\ \bibnamefont {Urciuoli}},\ }\href {\doibase
  10.1103/PhysRevC.85.032501} {\bibfield  {journal} {\bibinfo  {journal} {Phys.
  Rev. C}\ }\textbf {\bibinfo {volume} {85}},\ \bibinfo {pages} {032501}
  (\bibinfo {year} {2012})}\BibitemShut {NoStop}%
\bibitem [{\citenamefont {Donnelly}\ \emph {et~al.}(1989)\citenamefont
  {Donnelly}, \citenamefont {Dubach},\ and\ \citenamefont
  {Sick}}]{donnelly1989}%
  \BibitemOpen
  \bibfield  {author} {\bibinfo {author} {\bibfnamefont {T.}~\bibnamefont
  {Donnelly}}, \bibinfo {author} {\bibfnamefont {J.}~\bibnamefont {Dubach}}, \
  and\ \bibinfo {author} {\bibfnamefont {I.}~\bibnamefont {Sick}},\ }\href
  {\doibase http://dx.doi.org/10.1016/0375-9474(89)90432-6} {\bibfield
  {journal} {\bibinfo  {journal} {Nucl. Phys. A}\ }\textbf {\bibinfo {volume}
  {503}},\ \bibinfo {pages} {589} (\bibinfo {year} {1989})}\BibitemShut
  {NoStop}%
\bibitem [{\citenamefont {Paschke}\ \emph {et~al.}(2012)\citenamefont
  {Paschke}, \citenamefont {Kumar}, \citenamefont {Michaels}, \citenamefont
  {A.Souder},\ and\ \citenamefont {Urciuoli}}]{PREX-II}%
  \BibitemOpen
  \bibfield  {author} {\bibinfo {author} {\bibfnamefont {K.}~\bibnamefont
  {Paschke}}, \bibinfo {author} {\bibfnamefont {K.}~\bibnamefont {Kumar}},
  \bibinfo {author} {\bibfnamefont {R.}~\bibnamefont {Michaels}}, \bibinfo
  {author} {\bibfnamefont {P.}~\bibnamefont {A.Souder}}, \ and\ \bibinfo
  {author} {\bibfnamefont {G.~M.}\ \bibnamefont {Urciuoli}},\ }\href
  {http://hallaweb.jlab.org/parity/prex/prexII.pdf} {\emph {\bibinfo {title}
  {{PREX}--{II}: {P}recision parity-violating measurement of the neutron skin
  of lead}}},\ \bibinfo {type} {Tech. Rep.}\ (\bibinfo  {institution}
  {Jefferson Lab},\ \bibinfo {year} {2012})\BibitemShut {NoStop}%
\bibitem [{\citenamefont {Centelles}\ \emph {et~al.}(2009)\citenamefont
  {Centelles}, \citenamefont {Roca-Maza}, \citenamefont {Vi\~nas},\ and\
  \citenamefont {Warda}}]{centelles2009}%
  \BibitemOpen
  \bibfield  {author} {\bibinfo {author} {\bibfnamefont {M.}~\bibnamefont
  {Centelles}}, \bibinfo {author} {\bibfnamefont {X.}~\bibnamefont
  {Roca-Maza}}, \bibinfo {author} {\bibfnamefont {X.}~\bibnamefont {Vi\~nas}},
  \ and\ \bibinfo {author} {\bibfnamefont {M.}~\bibnamefont {Warda}},\ }\href
  {\doibase 10.1103/PhysRevLett.102.122502} {\bibfield  {journal} {\bibinfo
  {journal} {Phys. Rev. Lett.}\ }\textbf {\bibinfo {volume} {102}},\ \bibinfo
  {pages} {122502} (\bibinfo {year} {2009})}\BibitemShut {NoStop}%
\bibitem [{\citenamefont {Mammei}\ \emph {et~al.}(2013)\citenamefont {Mammei},
  \citenamefont {McNulty}, \citenamefont {Michaels}, \citenamefont {Paschke},
  \citenamefont {Riordan},\ and\ \citenamefont {Souder}}]{CREX}%
  \BibitemOpen
  \bibfield  {author} {\bibinfo {author} {\bibfnamefont {J.}~\bibnamefont
  {Mammei}}, \bibinfo {author} {\bibfnamefont {D.}~\bibnamefont {McNulty}},
  \bibinfo {author} {\bibfnamefont {R.}~\bibnamefont {Michaels}}, \bibinfo
  {author} {\bibfnamefont {K.}~\bibnamefont {Paschke}}, \bibinfo {author}
  {\bibfnamefont {S.}~\bibnamefont {Riordan}}, \ and\ \bibinfo {author}
  {\bibfnamefont {P.~A.}\ \bibnamefont {Souder}},\ }\href
  {http://hallaweb.jlab.org/parity/prex/c-rex2013_v7.pdf} {\emph {\bibinfo
  {title} {{CREX}: {P}arity--violating measurement of the weak charge
  distribution of $^{48}${C}a to 0.02 fm accuracy}}},\ \bibinfo {type} {Tech.
  Rep.}\ (\bibinfo  {institution} {Jefferson Lab},\ \bibinfo {year}
  {2013})\BibitemShut {NoStop}%
\bibitem [{\citenamefont {Hagen}\ \emph {et~al.}(2016)\citenamefont {Hagen},
  \citenamefont {Ekstrom}, \citenamefont {Forssen}, \citenamefont {Jansen},
  \citenamefont {Nazarewicz}, \citenamefont {Papenbrock}, \citenamefont
  {Wendt}, \citenamefont {Bacca}, \citenamefont {Barnea}, \citenamefont
  {Carlsson}, \citenamefont {Drischler}, \citenamefont {Hebeler}, \citenamefont
  {Hjorth-Jensen}, \citenamefont {Miorelli}, \citenamefont {Orlandini},
  \citenamefont {Schwenk},\ and\ \citenamefont {Simonis}}]{hagen2016}%
  \BibitemOpen
  \bibfield  {author} {\bibinfo {author} {\bibfnamefont {G.}~\bibnamefont
  {Hagen}}, \bibinfo {author} {\bibfnamefont {A.}~\bibnamefont {Ekstrom}},
  \bibinfo {author} {\bibfnamefont {C.}~\bibnamefont {Forssen}}, \bibinfo
  {author} {\bibfnamefont {G.~R.}\ \bibnamefont {Jansen}}, \bibinfo {author}
  {\bibfnamefont {W.}~\bibnamefont {Nazarewicz}}, \bibinfo {author}
  {\bibfnamefont {T.}~\bibnamefont {Papenbrock}}, \bibinfo {author}
  {\bibfnamefont {K.~A.}\ \bibnamefont {Wendt}}, \bibinfo {author}
  {\bibfnamefont {S.}~\bibnamefont {Bacca}}, \bibinfo {author} {\bibfnamefont
  {N.}~\bibnamefont {Barnea}}, \bibinfo {author} {\bibfnamefont
  {B.}~\bibnamefont {Carlsson}}, \bibinfo {author} {\bibfnamefont
  {C.}~\bibnamefont {Drischler}}, \bibinfo {author} {\bibfnamefont
  {K.}~\bibnamefont {Hebeler}}, \bibinfo {author} {\bibfnamefont
  {M.}~\bibnamefont {Hjorth-Jensen}}, \bibinfo {author} {\bibfnamefont
  {M.}~\bibnamefont {Miorelli}}, \bibinfo {author} {\bibfnamefont
  {G.}~\bibnamefont {Orlandini}}, \bibinfo {author} {\bibfnamefont
  {A.}~\bibnamefont {Schwenk}}, \ and\ \bibinfo {author} {\bibfnamefont
  {J.}~\bibnamefont {Simonis}},\ }\href {\doibase 10.1038/nphys3529} {\bibfield
   {journal} {\bibinfo  {journal} {Nature}\ }\textbf {\bibinfo {volume} {12}},\
  \bibinfo {pages} {186} (\bibinfo {year} {2016})}\BibitemShut {NoStop}%
\bibitem [{\citenamefont {Lin}\ and\ \citenamefont {Horowitz}(2015)}]{lin2015}%
  \BibitemOpen
  \bibfield  {author} {\bibinfo {author} {\bibfnamefont {Z.}~\bibnamefont
  {Lin}}\ and\ \bibinfo {author} {\bibfnamefont {C.~J.}\ \bibnamefont
  {Horowitz}},\ }\href {\doibase 10.1103/PhysRevC.92.014313} {\bibfield
  {journal} {\bibinfo  {journal} {Phys. Rev. C}\ }\textbf {\bibinfo {volume}
  {92}},\ \bibinfo {pages} {014313} (\bibinfo {year} {2015})}\BibitemShut
  {NoStop}%
\bibitem [{\citenamefont {Blaizot}\ \emph {et~al.}(1995)\citenamefont
  {Blaizot}, \citenamefont {Berger}, \citenamefont {Decharg\'{e}},\ and\
  \citenamefont {Girod}}]{blaizot1995}%
  \BibitemOpen
  \bibfield  {author} {\bibinfo {author} {\bibfnamefont {J.}~\bibnamefont
  {Blaizot}}, \bibinfo {author} {\bibfnamefont {J.}~\bibnamefont {Berger}},
  \bibinfo {author} {\bibfnamefont {J.}~\bibnamefont {Decharg\'{e}}}, \ and\
  \bibinfo {author} {\bibfnamefont {M.}~\bibnamefont {Girod}},\ }\href
  {\doibase 10.1016/0375-9474(95)00294-B} {\bibfield  {journal} {\bibinfo
  {journal} {Nucl. Phys. A}\ }\textbf {\bibinfo {volume} {591}},\ \bibinfo
  {pages} {435} (\bibinfo {year} {1995})}\BibitemShut {NoStop}%
\bibitem [{\citenamefont {Reinhard}\ \emph {et~al.}(2006)\citenamefont
  {Reinhard}, \citenamefont {Bender}, \citenamefont {Nazarewicz},\ and\
  \citenamefont {Vertse}}]{reinhard2006b}%
  \BibitemOpen
  \bibfield  {author} {\bibinfo {author} {\bibfnamefont {P.-G.}\ \bibnamefont
  {Reinhard}}, \bibinfo {author} {\bibfnamefont {M.}~\bibnamefont {Bender}},
  \bibinfo {author} {\bibfnamefont {W.}~\bibnamefont {Nazarewicz}}, \ and\
  \bibinfo {author} {\bibfnamefont {T.}~\bibnamefont {Vertse}},\ }\href
  {\doibase 10.1103/PhysRevC.014309} {\bibfield  {journal} {\bibinfo  {journal}
  {Phys. Rev. C}\ }\textbf {\bibinfo {volume} {73}},\ \bibinfo {pages} {014309}
  (\bibinfo {year} {2006})}\BibitemShut {NoStop}%
\bibitem [{\citenamefont {Harakeh}\ and\ \citenamefont {van~der
  Woude}(2001)}]{harakeh2001}%
  \BibitemOpen
  \bibfield  {author} {\bibinfo {author} {\bibfnamefont {M.~N.}\ \bibnamefont
  {Harakeh}}\ and\ \bibinfo {author} {\bibfnamefont {A.}~\bibnamefont {van~der
  Woude}},\ }\href@noop {} {\emph {\bibinfo {title} {{G}iant {R}esonances :
  {F}undamental {H}igh-frequency {M}odes of {N}uclear {E}xcitation}}}\
  (\bibinfo  {publisher} {Clarendon Press},\ \bibinfo {address} {Oxford},\
  \bibinfo {year} {2001})\BibitemShut {NoStop}%
\bibitem [{\citenamefont {Blaizot}(1980)}]{blaizot1980}%
  \BibitemOpen
  \bibfield  {author} {\bibinfo {author} {\bibfnamefont {J.}~\bibnamefont
  {Blaizot}},\ }\href {\doibase 10.1016/0370-1573(80)90001-0} {\bibfield
  {journal} {\bibinfo  {journal} {Phys. Rep.}\ }\textbf {\bibinfo {volume}
  {64}},\ \bibinfo {pages} {171} (\bibinfo {year} {1980})}\BibitemShut
  {NoStop}%
\bibitem [{\citenamefont {Bartel}\ \emph {et~al.}(1982)\citenamefont {Bartel},
  \citenamefont {Quentin}, \citenamefont {Brack}, \citenamefont {Guet},\ and\
  \citenamefont {H\aa{a}kansson}}]{bartel1982}%
  \BibitemOpen
  \bibfield  {author} {\bibinfo {author} {\bibfnamefont {J.}~\bibnamefont
  {Bartel}}, \bibinfo {author} {\bibfnamefont {P.}~\bibnamefont {Quentin}},
  \bibinfo {author} {\bibfnamefont {M.}~\bibnamefont {Brack}}, \bibinfo
  {author} {\bibfnamefont {C.}~\bibnamefont {Guet}}, \ and\ \bibinfo {author}
  {\bibfnamefont {H.-B.}\ \bibnamefont {H\aa{a}kansson}},\ }\href {\doibase
  10.1016/0375-9474(82)90403-1} {\bibfield  {journal} {\bibinfo  {journal}
  {Nucl. Phys. A}\ }\textbf {\bibinfo {volume} {386}},\ \bibinfo {pages} {79}
  (\bibinfo {year} {1982})}\BibitemShut {NoStop}%
\bibitem [{\citenamefont {Reinhard}(1999)}]{reinhard1999}%
  \BibitemOpen
  \bibfield  {author} {\bibinfo {author} {\bibfnamefont {P.-G.}\ \bibnamefont
  {Reinhard}},\ }\href {\doibase 10.1016/S0375-9474(99)00076-7} {\bibfield
  {journal} {\bibinfo  {journal} {Nucl. Phys. A}\ }\textbf {\bibinfo {volume}
  {649}},\ \bibinfo {pages} {305c} (\bibinfo {year} {1999})}\BibitemShut
  {NoStop}%
\bibitem [{\citenamefont {Reinhard}\ and\ \citenamefont
  {Nazarewicz}(2010)}]{reinhard2010}%
  \BibitemOpen
  \bibfield  {author} {\bibinfo {author} {\bibfnamefont {P.-G.}\ \bibnamefont
  {Reinhard}}\ and\ \bibinfo {author} {\bibfnamefont {W.}~\bibnamefont
  {Nazarewicz}},\ }\href {\doibase 10.1103/PhysRevC.81.051303} {\bibfield
  {journal} {\bibinfo  {journal} {Phys. Rev. C}\ }\textbf {\bibinfo {volume}
  {81}},\ \bibinfo {pages} {051303} (\bibinfo {year} {2010})}\BibitemShut
  {NoStop}%
\bibitem [{\citenamefont {Tamii}\ \emph {et~al.}(2011)\citenamefont {Tamii},
  \citenamefont {Poltoratska}, \citenamefont {{von Neumann-Cosel}},
  \citenamefont {Fujita}, \citenamefont {Adachi}, \citenamefont {Bertulani},
  \citenamefont {Carter}, \citenamefont {Dozono}, \citenamefont {Fujita},
  \citenamefont {Fujita}, \citenamefont {Hatanaka}, \citenamefont {Ishikawa},
  \citenamefont {Itoh}, \citenamefont {Kawabata}, \citenamefont {Kalmykov},
  \citenamefont {Krumbholz}, \citenamefont {Litvinova}, \citenamefont
  {Matsubara}, \citenamefont {Nakanishi}, \citenamefont {Neveling},
  \citenamefont {Okamura}, \citenamefont {Ong}, \citenamefont
  {\"Ozel-Tashenov}, \citenamefont {Ponomarev}, \citenamefont {Richter},
  \citenamefont {Rubio}, \citenamefont {Sakaguchi}, \citenamefont {Sakemi},
  \citenamefont {Sasamoto}, \citenamefont {Shimbara}, \citenamefont {Shimizu},
  \citenamefont {Smit}, \citenamefont {Suzuki}, \citenamefont {Tameshige},
  \citenamefont {Wambach}, \citenamefont {Yamada}, \citenamefont {Yosoi},\ and\
  \citenamefont {Zenihiro}}]{tamii2011}%
  \BibitemOpen
  \bibfield  {author} {\bibinfo {author} {\bibfnamefont {A.}~\bibnamefont
  {Tamii}}, \bibinfo {author} {\bibfnamefont {I.}~\bibnamefont {Poltoratska}},
  \bibinfo {author} {\bibfnamefont {P.}~\bibnamefont {{von Neumann-Cosel}}},
  \bibinfo {author} {\bibfnamefont {Y.}~\bibnamefont {Fujita}}, \bibinfo
  {author} {\bibfnamefont {T.}~\bibnamefont {Adachi}}, \bibinfo {author}
  {\bibfnamefont {C.~A.}\ \bibnamefont {Bertulani}}, \bibinfo {author}
  {\bibfnamefont {J.}~\bibnamefont {Carter}}, \bibinfo {author} {\bibfnamefont
  {M.}~\bibnamefont {Dozono}}, \bibinfo {author} {\bibfnamefont
  {H.}~\bibnamefont {Fujita}}, \bibinfo {author} {\bibfnamefont
  {K.}~\bibnamefont {Fujita}}, \bibinfo {author} {\bibfnamefont
  {K.}~\bibnamefont {Hatanaka}}, \bibinfo {author} {\bibfnamefont
  {D.}~\bibnamefont {Ishikawa}}, \bibinfo {author} {\bibfnamefont
  {M.}~\bibnamefont {Itoh}}, \bibinfo {author} {\bibfnamefont {T.}~\bibnamefont
  {Kawabata}}, \bibinfo {author} {\bibfnamefont {Y.}~\bibnamefont {Kalmykov}},
  \bibinfo {author} {\bibfnamefont {A.~M.}\ \bibnamefont {Krumbholz}}, \bibinfo
  {author} {\bibfnamefont {E.}~\bibnamefont {Litvinova}}, \bibinfo {author}
  {\bibfnamefont {H.}~\bibnamefont {Matsubara}}, \bibinfo {author}
  {\bibfnamefont {K.}~\bibnamefont {Nakanishi}}, \bibinfo {author}
  {\bibfnamefont {R.}~\bibnamefont {Neveling}}, \bibinfo {author}
  {\bibfnamefont {H.}~\bibnamefont {Okamura}}, \bibinfo {author} {\bibfnamefont
  {H.~J.}\ \bibnamefont {Ong}}, \bibinfo {author} {\bibfnamefont
  {B.}~\bibnamefont {\"Ozel-Tashenov}}, \bibinfo {author} {\bibfnamefont
  {V.~Y.}\ \bibnamefont {Ponomarev}}, \bibinfo {author} {\bibfnamefont
  {A.}~\bibnamefont {Richter}}, \bibinfo {author} {\bibfnamefont
  {B.}~\bibnamefont {Rubio}}, \bibinfo {author} {\bibfnamefont
  {H.}~\bibnamefont {Sakaguchi}}, \bibinfo {author} {\bibfnamefont
  {Y.}~\bibnamefont {Sakemi}}, \bibinfo {author} {\bibfnamefont
  {Y.}~\bibnamefont {Sasamoto}}, \bibinfo {author} {\bibfnamefont
  {Y.}~\bibnamefont {Shimbara}}, \bibinfo {author} {\bibfnamefont
  {Y.}~\bibnamefont {Shimizu}}, \bibinfo {author} {\bibfnamefont {F.~D.}\
  \bibnamefont {Smit}}, \bibinfo {author} {\bibfnamefont {T.}~\bibnamefont
  {Suzuki}}, \bibinfo {author} {\bibfnamefont {Y.}~\bibnamefont {Tameshige}},
  \bibinfo {author} {\bibfnamefont {J.}~\bibnamefont {Wambach}}, \bibinfo
  {author} {\bibfnamefont {R.}~\bibnamefont {Yamada}}, \bibinfo {author}
  {\bibfnamefont {M.}~\bibnamefont {Yosoi}}, \ and\ \bibinfo {author}
  {\bibfnamefont {J.}~\bibnamefont {Zenihiro}},\ }\href {\doibase
  10.1103/PhysRevLett.107.062502} {\bibfield  {journal} {\bibinfo  {journal}
  {Phys. Rev. Lett.}\ }\textbf {\bibinfo {volume} {107}},\ \bibinfo {pages}
  {062502} (\bibinfo {year} {2011})}\BibitemShut {NoStop}%
\bibitem [{\citenamefont {Piekarewicz}\ \emph {et~al.}(2012)\citenamefont
  {Piekarewicz}, \citenamefont {Agrawal}, \citenamefont {Col\`o}, \citenamefont
  {Nazarewicz}, \citenamefont {Paar}, \citenamefont {Reinhard}, \citenamefont
  {Roca-Maza},\ and\ \citenamefont {Vretenar}}]{piekarewicz2012}%
  \BibitemOpen
  \bibfield  {author} {\bibinfo {author} {\bibfnamefont {J.}~\bibnamefont
  {Piekarewicz}}, \bibinfo {author} {\bibfnamefont {B.~K.}\ \bibnamefont
  {Agrawal}}, \bibinfo {author} {\bibfnamefont {G.}~\bibnamefont {Col\`o}},
  \bibinfo {author} {\bibfnamefont {W.}~\bibnamefont {Nazarewicz}}, \bibinfo
  {author} {\bibfnamefont {N.}~\bibnamefont {Paar}}, \bibinfo {author}
  {\bibfnamefont {P.-G.}\ \bibnamefont {Reinhard}}, \bibinfo {author}
  {\bibfnamefont {X.}~\bibnamefont {Roca-Maza}}, \ and\ \bibinfo {author}
  {\bibfnamefont {D.}~\bibnamefont {Vretenar}},\ }\href {\doibase
  10.1103/PhysRevC.85.041302} {\bibfield  {journal} {\bibinfo  {journal} {Phys.
  Rev. C}\ }\textbf {\bibinfo {volume} {85}},\ \bibinfo {pages} {041302}
  (\bibinfo {year} {2012})}\BibitemShut {NoStop}%
\bibitem [{\citenamefont {Roca-Maza}\ \emph {et~al.}(2013)\citenamefont
  {Roca-Maza}, \citenamefont {Brenna}, \citenamefont {Col\`o}, \citenamefont
  {Centelles}, \citenamefont {Vi\~nas}, \citenamefont {Agrawal}, \citenamefont
  {Paar}, \citenamefont {Vretenar},\ and\ \citenamefont
  {Piekarewicz}}]{roca2013}%
  \BibitemOpen
  \bibfield  {author} {\bibinfo {author} {\bibfnamefont {X.}~\bibnamefont
  {Roca-Maza}}, \bibinfo {author} {\bibfnamefont {M.}~\bibnamefont {Brenna}},
  \bibinfo {author} {\bibfnamefont {G.}~\bibnamefont {Col\`o}}, \bibinfo
  {author} {\bibfnamefont {M.}~\bibnamefont {Centelles}}, \bibinfo {author}
  {\bibfnamefont {X.}~\bibnamefont {Vi\~nas}}, \bibinfo {author} {\bibfnamefont
  {B.~K.}\ \bibnamefont {Agrawal}}, \bibinfo {author} {\bibfnamefont
  {N.}~\bibnamefont {Paar}}, \bibinfo {author} {\bibfnamefont {D.}~\bibnamefont
  {Vretenar}}, \ and\ \bibinfo {author} {\bibfnamefont {J.}~\bibnamefont
  {Piekarewicz}},\ }\href {\doibase 10.1103/PhysRevC.88.024316} {\bibfield
  {journal} {\bibinfo  {journal} {Phys. Rev. C}\ }\textbf {\bibinfo {volume}
  {88}},\ \bibinfo {pages} {024316} (\bibinfo {year} {2013})}\BibitemShut
  {NoStop}%
\bibitem [{\citenamefont {Rossi}\ \emph {et~al.}(2013)\citenamefont {Rossi},
  \citenamefont {Adrich}, \citenamefont {Aksouh}, \citenamefont {Alvarez-Pol},
  \citenamefont {Aumann}, \citenamefont {Benlliure}, \citenamefont {B\"ohmer},
  \citenamefont {Boretzky}, \citenamefont {Casarejos}, \citenamefont
  {Chartier}, \citenamefont {Chatillon}, \citenamefont {Cortina-Gil},
  \citenamefont {Datta~Pramanik}, \citenamefont {Emling}, \citenamefont
  {Ershova}, \citenamefont {Fernandez-Dominguez}, \citenamefont {Geissel},
  \citenamefont {Gorska}, \citenamefont {Heil}, \citenamefont {Johansson},
  \citenamefont {Junghans}, \citenamefont {Kelic-Heil}, \citenamefont
  {Kiselev}, \citenamefont {Klimkiewicz}, \citenamefont {Kratz}, \citenamefont
  {Kr\"ucken}, \citenamefont {Kurz}, \citenamefont {Labiche}, \citenamefont
  {Le~Bleis}, \citenamefont {Lemmon}, \citenamefont {Litvinov}, \citenamefont
  {Mahata}, \citenamefont {Maierbeck}, \citenamefont {Movsesyan}, \citenamefont
  {Nilsson}, \citenamefont {Nociforo}, \citenamefont {Palit}, \citenamefont
  {Paschalis}, \citenamefont {Plag}, \citenamefont {Reifarth}, \citenamefont
  {Savran}, \citenamefont {Scheit}, \citenamefont {Simon}, \citenamefont
  {S\"ummerer}, \citenamefont {Wagner}, \citenamefont
  {Walu\ifmmode~\acute{s}\else \'{s}\fi{}}, \citenamefont {Weick},\ and\
  \citenamefont {Winkler}}]{rossi2013}%
  \BibitemOpen
  \bibfield  {author} {\bibinfo {author} {\bibfnamefont {D.~M.}\ \bibnamefont
  {Rossi}}, \bibinfo {author} {\bibfnamefont {P.}~\bibnamefont {Adrich}},
  \bibinfo {author} {\bibfnamefont {F.}~\bibnamefont {Aksouh}}, \bibinfo
  {author} {\bibfnamefont {H.}~\bibnamefont {Alvarez-Pol}}, \bibinfo {author}
  {\bibfnamefont {T.}~\bibnamefont {Aumann}}, \bibinfo {author} {\bibfnamefont
  {J.}~\bibnamefont {Benlliure}}, \bibinfo {author} {\bibfnamefont
  {M.}~\bibnamefont {B\"ohmer}}, \bibinfo {author} {\bibfnamefont
  {K.}~\bibnamefont {Boretzky}}, \bibinfo {author} {\bibfnamefont
  {E.}~\bibnamefont {Casarejos}}, \bibinfo {author} {\bibfnamefont
  {M.}~\bibnamefont {Chartier}}, \bibinfo {author} {\bibfnamefont
  {A.}~\bibnamefont {Chatillon}}, \bibinfo {author} {\bibfnamefont
  {D.}~\bibnamefont {Cortina-Gil}}, \bibinfo {author} {\bibfnamefont
  {U.}~\bibnamefont {Datta~Pramanik}}, \bibinfo {author} {\bibfnamefont
  {H.}~\bibnamefont {Emling}}, \bibinfo {author} {\bibfnamefont
  {O.}~\bibnamefont {Ershova}}, \bibinfo {author} {\bibfnamefont
  {B.}~\bibnamefont {Fernandez-Dominguez}}, \bibinfo {author} {\bibfnamefont
  {H.}~\bibnamefont {Geissel}}, \bibinfo {author} {\bibfnamefont
  {M.}~\bibnamefont {Gorska}}, \bibinfo {author} {\bibfnamefont
  {M.}~\bibnamefont {Heil}}, \bibinfo {author} {\bibfnamefont {H.~T.}\
  \bibnamefont {Johansson}}, \bibinfo {author} {\bibfnamefont {A.}~\bibnamefont
  {Junghans}}, \bibinfo {author} {\bibfnamefont {A.}~\bibnamefont
  {Kelic-Heil}}, \bibinfo {author} {\bibfnamefont {O.}~\bibnamefont {Kiselev}},
  \bibinfo {author} {\bibfnamefont {A.}~\bibnamefont {Klimkiewicz}}, \bibinfo
  {author} {\bibfnamefont {J.~V.}\ \bibnamefont {Kratz}}, \bibinfo {author}
  {\bibfnamefont {R.}~\bibnamefont {Kr\"ucken}}, \bibinfo {author}
  {\bibfnamefont {N.}~\bibnamefont {Kurz}}, \bibinfo {author} {\bibfnamefont
  {M.}~\bibnamefont {Labiche}}, \bibinfo {author} {\bibfnamefont
  {T.}~\bibnamefont {Le~Bleis}}, \bibinfo {author} {\bibfnamefont
  {R.}~\bibnamefont {Lemmon}}, \bibinfo {author} {\bibfnamefont {Y.~A.}\
  \bibnamefont {Litvinov}}, \bibinfo {author} {\bibfnamefont {K.}~\bibnamefont
  {Mahata}}, \bibinfo {author} {\bibfnamefont {P.}~\bibnamefont {Maierbeck}},
  \bibinfo {author} {\bibfnamefont {A.}~\bibnamefont {Movsesyan}}, \bibinfo
  {author} {\bibfnamefont {T.}~\bibnamefont {Nilsson}}, \bibinfo {author}
  {\bibfnamefont {C.}~\bibnamefont {Nociforo}}, \bibinfo {author}
  {\bibfnamefont {R.}~\bibnamefont {Palit}}, \bibinfo {author} {\bibfnamefont
  {S.}~\bibnamefont {Paschalis}}, \bibinfo {author} {\bibfnamefont
  {R.}~\bibnamefont {Plag}}, \bibinfo {author} {\bibfnamefont {R.}~\bibnamefont
  {Reifarth}}, \bibinfo {author} {\bibfnamefont {D.}~\bibnamefont {Savran}},
  \bibinfo {author} {\bibfnamefont {H.}~\bibnamefont {Scheit}}, \bibinfo
  {author} {\bibfnamefont {H.}~\bibnamefont {Simon}}, \bibinfo {author}
  {\bibfnamefont {K.}~\bibnamefont {S\"ummerer}}, \bibinfo {author}
  {\bibfnamefont {A.}~\bibnamefont {Wagner}}, \bibinfo {author} {\bibfnamefont
  {W.}~\bibnamefont {Walu\ifmmode~\acute{s}\else \'{s}\fi{}}}, \bibinfo
  {author} {\bibfnamefont {H.}~\bibnamefont {Weick}}, \ and\ \bibinfo {author}
  {\bibfnamefont {M.}~\bibnamefont {Winkler}},\ }\href {\doibase
  10.1103/PhysRevLett.111.242503} {\bibfield  {journal} {\bibinfo  {journal}
  {Phys. Rev. Lett.}\ }\textbf {\bibinfo {volume} {111}},\ \bibinfo {pages}
  {242503} (\bibinfo {year} {2013})}\BibitemShut {NoStop}%
\bibitem [{\citenamefont {Hashimoto}\ \emph {et~al.}(2015)\citenamefont
  {Hashimoto}, \citenamefont {Krumbholz}, \citenamefont {Reinhard},
  \citenamefont {Tamii}, \citenamefont {{von Neumann-Cosel}}, \citenamefont
  {Adachi}, \citenamefont {Aoi}, \citenamefont {Bertulani}, \citenamefont
  {Fujita}, \citenamefont {Fujita}, \citenamefont {Ganio}, \citenamefont
  {Hatanaka}, \citenamefont {Ideguchi}, \citenamefont {Iwamoto}, \citenamefont
  {Kawabata}, \citenamefont {Khai}, \citenamefont {Krugmann}, \citenamefont
  {Martin}, \citenamefont {Matsubara}, \citenamefont {Miki}, \citenamefont
  {Neveling}, \citenamefont {Okamura}, \citenamefont {Ong}, \citenamefont
  {Poltoratska}, \citenamefont {Ponomarev}, \citenamefont {Richter},
  \citenamefont {Sakaguchi}, \citenamefont {Shimbara}, \citenamefont {Shimizu},
  \citenamefont {Simonis}, \citenamefont {Smit}, \citenamefont {S\"{u}soy},
  \citenamefont {Suzuki}, \citenamefont {Thies}, \citenamefont {Yosoi},\ and\
  \citenamefont {Zenihiro}}]{hashimoto2015}%
  \BibitemOpen
  \bibfield  {author} {\bibinfo {author} {\bibfnamefont {T.}~\bibnamefont
  {Hashimoto}}, \bibinfo {author} {\bibfnamefont {A.~M.}\ \bibnamefont
  {Krumbholz}}, \bibinfo {author} {\bibfnamefont {P.-G.}\ \bibnamefont
  {Reinhard}}, \bibinfo {author} {\bibfnamefont {A.}~\bibnamefont {Tamii}},
  \bibinfo {author} {\bibfnamefont {P.}~\bibnamefont {{von Neumann-Cosel}}},
  \bibinfo {author} {\bibfnamefont {T.}~\bibnamefont {Adachi}}, \bibinfo
  {author} {\bibfnamefont {N.}~\bibnamefont {Aoi}}, \bibinfo {author}
  {\bibfnamefont {C.~A.}\ \bibnamefont {Bertulani}}, \bibinfo {author}
  {\bibfnamefont {H.}~\bibnamefont {Fujita}}, \bibinfo {author} {\bibfnamefont
  {Y.}~\bibnamefont {Fujita}}, \bibinfo {author} {\bibfnamefont
  {E.}~\bibnamefont {Ganio}}, \bibinfo {author} {\bibfnamefont
  {K.}~\bibnamefont {Hatanaka}}, \bibinfo {author} {\bibfnamefont
  {E.}~\bibnamefont {Ideguchi}}, \bibinfo {author} {\bibfnamefont
  {C.}~\bibnamefont {Iwamoto}}, \bibinfo {author} {\bibfnamefont
  {T.}~\bibnamefont {Kawabata}}, \bibinfo {author} {\bibfnamefont {N.~T.}\
  \bibnamefont {Khai}}, \bibinfo {author} {\bibfnamefont {A.}~\bibnamefont
  {Krugmann}}, \bibinfo {author} {\bibfnamefont {D.}~\bibnamefont {Martin}},
  \bibinfo {author} {\bibfnamefont {H.}~\bibnamefont {Matsubara}}, \bibinfo
  {author} {\bibfnamefont {K.}~\bibnamefont {Miki}}, \bibinfo {author}
  {\bibfnamefont {R.}~\bibnamefont {Neveling}}, \bibinfo {author}
  {\bibfnamefont {H.}~\bibnamefont {Okamura}}, \bibinfo {author} {\bibfnamefont
  {H.~J.}\ \bibnamefont {Ong}}, \bibinfo {author} {\bibfnamefont
  {I.}~\bibnamefont {Poltoratska}}, \bibinfo {author} {\bibfnamefont {V.~Y.}\
  \bibnamefont {Ponomarev}}, \bibinfo {author} {\bibfnamefont {A.}~\bibnamefont
  {Richter}}, \bibinfo {author} {\bibfnamefont {H.}~\bibnamefont {Sakaguchi}},
  \bibinfo {author} {\bibfnamefont {Y.}~\bibnamefont {Shimbara}}, \bibinfo
  {author} {\bibfnamefont {Y.}~\bibnamefont {Shimizu}}, \bibinfo {author}
  {\bibfnamefont {J.}~\bibnamefont {Simonis}}, \bibinfo {author} {\bibfnamefont
  {F.~D.}\ \bibnamefont {Smit}}, \bibinfo {author} {\bibfnamefont
  {G.}~\bibnamefont {S\"{u}soy}}, \bibinfo {author} {\bibfnamefont
  {T.}~\bibnamefont {Suzuki}}, \bibinfo {author} {\bibfnamefont {J.~H.}\
  \bibnamefont {Thies}}, \bibinfo {author} {\bibfnamefont {M.}~\bibnamefont
  {Yosoi}}, \ and\ \bibinfo {author} {\bibfnamefont {J.}~\bibnamefont
  {Zenihiro}},\ }\href {\doibase 10.1103/PhysRevC.92.031305} {\bibfield
  {journal} {\bibinfo  {journal} {Phys. Rev. C}\ }\textbf {\bibinfo {volume}
  {92}},\ \bibinfo {pages} {031305(R)} (\bibinfo {year} {2015})}\BibitemShut
  {NoStop}%
\bibitem [{\citenamefont {Roca-Maza}\ \emph {et~al.}(2015)\citenamefont
  {Roca-Maza}, \citenamefont {Vi\~nas}, \citenamefont {Centelles},
  \citenamefont {Agrawal}, \citenamefont {Col\`o}, \citenamefont {Paar},
  \citenamefont {Piekarewicz},\ and\ \citenamefont {Vretenar}}]{roca2015}%
  \BibitemOpen
  \bibfield  {author} {\bibinfo {author} {\bibfnamefont {X.}~\bibnamefont
  {Roca-Maza}}, \bibinfo {author} {\bibfnamefont {X.}~\bibnamefont {Vi\~nas}},
  \bibinfo {author} {\bibfnamefont {M.}~\bibnamefont {Centelles}}, \bibinfo
  {author} {\bibfnamefont {B.~K.}\ \bibnamefont {Agrawal}}, \bibinfo {author}
  {\bibfnamefont {G.}~\bibnamefont {Col\`o}}, \bibinfo {author} {\bibfnamefont
  {N.}~\bibnamefont {Paar}}, \bibinfo {author} {\bibfnamefont {J.}~\bibnamefont
  {Piekarewicz}}, \ and\ \bibinfo {author} {\bibfnamefont {D.}~\bibnamefont
  {Vretenar}},\ }\href {\doibase 10.1103/PhysRevC.92.064304} {\bibfield
  {journal} {\bibinfo  {journal} {Phys. Rev. C}\ }\textbf {\bibinfo {volume}
  {92}},\ \bibinfo {pages} {064304} (\bibinfo {year} {2015})}\BibitemShut
  {NoStop}%
\bibitem [{\citenamefont {Reinhard}(1992)}]{reinhard1992}%
  \BibitemOpen
  \bibfield  {author} {\bibinfo {author} {\bibfnamefont {P.-G.}\ \bibnamefont
  {Reinhard}},\ }\href@noop {} {\bibfield  {journal} {\bibinfo  {journal} {Ann.
  Phys. (Leipzig)}\ }\textbf {\bibinfo {volume} {504}},\ \bibinfo {pages} {632}
  (\bibinfo {year} {1992})}\BibitemShut {NoStop}%
\bibitem [{\citenamefont {{Van Giai}}\ \emph {et~al.}(2001)\citenamefont {{Van
  Giai}}, \citenamefont {Bortignon}, \citenamefont {Col\`o}, \citenamefont
  {Ma},\ and\ \citenamefont {Quaglia}}]{vangiai2001}%
  \BibitemOpen
  \bibfield  {author} {\bibinfo {author} {\bibfnamefont {N.}~\bibnamefont {{Van
  Giai}}}, \bibinfo {author} {\bibfnamefont {P.~F.}\ \bibnamefont {Bortignon}},
  \bibinfo {author} {\bibfnamefont {G.}~\bibnamefont {Col\`o}}, \bibinfo
  {author} {\bibfnamefont {Z.-Y.}\ \bibnamefont {Ma}}, \ and\ \bibinfo {author}
  {\bibfnamefont {M.}~\bibnamefont {Quaglia}},\ }\href {\doibase
  http://dx.doi.org/10.1016/S0375-9474(01)00599-1} {\bibfield  {journal}
  {\bibinfo  {journal} {Nuclear Physics A}\ }\textbf {\bibinfo {volume}
  {687}},\ \bibinfo {pages} {44 } (\bibinfo {year} {2001})}\BibitemShut
  {NoStop}%
\bibitem [{\citenamefont {Reinhard}\ \emph {et~al.}(2007)\citenamefont
  {Reinhard}, \citenamefont {{Lu Guo}},\ and\ \citenamefont
  {Maruhn}}]{reinhard2007}%
  \BibitemOpen
  \bibfield  {author} {\bibinfo {author} {\bibfnamefont {P.-G.}\ \bibnamefont
  {Reinhard}}, \bibinfo {author} {\bibnamefont {{Lu Guo}}}, \ and\ \bibinfo
  {author} {\bibfnamefont {J.}~\bibnamefont {Maruhn}},\ }\href {\doibase
  10.1140/epja/i2007-10366-9} {\bibfield  {journal} {\bibinfo  {journal} {Eur.
  Phys. J. A}\ }\textbf {\bibinfo {volume} {32}},\ \bibinfo {pages} {19}
  (\bibinfo {year} {2007})}\BibitemShut {NoStop}%
\bibitem [{\citenamefont {Maruhn}\ \emph {et~al.}(2005)\citenamefont {Maruhn},
  \citenamefont {Reinhard}, \citenamefont {Stevenson}, \citenamefont {{Rikovska
  Stone}},\ and\ \citenamefont {Strayer}}]{maruhn2005}%
  \BibitemOpen
  \bibfield  {author} {\bibinfo {author} {\bibfnamefont {J.~A.}\ \bibnamefont
  {Maruhn}}, \bibinfo {author} {\bibfnamefont {P.~G.}\ \bibnamefont
  {Reinhard}}, \bibinfo {author} {\bibfnamefont {P.~D.}\ \bibnamefont
  {Stevenson}}, \bibinfo {author} {\bibfnamefont {J.}~\bibnamefont {{Rikovska
  Stone}}}, \ and\ \bibinfo {author} {\bibfnamefont {M.~R.}\ \bibnamefont
  {Strayer}},\ }\href {\doibase 10.1103/PhysRevC.71.064328} {\bibfield
  {journal} {\bibinfo  {journal} {Phys. Rev. C}\ }\textbf {\bibinfo {volume}
  {71}},\ \bibinfo {pages} {064328} (\bibinfo {year} {2005})}\BibitemShut
  {NoStop}%
\bibitem [{\citenamefont {Nesterenko}\ \emph {et~al.}(2006)\citenamefont
  {Nesterenko}, \citenamefont {Kleinig}, \citenamefont {Kvasil}, \citenamefont
  {Vesely}, \citenamefont {Reinhard},\ and\ \citenamefont
  {Dolci}}]{nesterenko2006}%
  \BibitemOpen
  \bibfield  {author} {\bibinfo {author} {\bibfnamefont {V.~O.}\ \bibnamefont
  {Nesterenko}}, \bibinfo {author} {\bibfnamefont {W.}~\bibnamefont {Kleinig}},
  \bibinfo {author} {\bibfnamefont {J.}~\bibnamefont {Kvasil}}, \bibinfo
  {author} {\bibfnamefont {P.}~\bibnamefont {Vesely}}, \bibinfo {author}
  {\bibfnamefont {P.-G.}\ \bibnamefont {Reinhard}}, \ and\ \bibinfo {author}
  {\bibfnamefont {D.~S.}\ \bibnamefont {Dolci}},\ }\href {\doibase
  10.1103/PhysRevC.74.064306} {\bibfield  {journal} {\bibinfo  {journal} {Phys.
  Rev. C}\ }\textbf {\bibinfo {volume} {74}},\ \bibinfo {pages} {064306}
  (\bibinfo {year} {2006})}\BibitemShut {NoStop}%
\bibitem [{\citenamefont {Nakatsukasa}\ and\ \citenamefont
  {Yabana}(2005)}]{nakatsukasa2005}%
  \BibitemOpen
  \bibfield  {author} {\bibinfo {author} {\bibfnamefont {T.}~\bibnamefont
  {Nakatsukasa}}\ and\ \bibinfo {author} {\bibfnamefont {K.}~\bibnamefont
  {Yabana}},\ }\href {\doibase 10.1103/PhysRevC.71.024301} {\bibfield
  {journal} {\bibinfo  {journal} {Phys. Rev. C}\ }\textbf {\bibinfo {volume}
  {71}},\ \bibinfo {eid} {024301} (\bibinfo {year} {2005})}\BibitemShut
  {NoStop}%
\bibitem [{\citenamefont {Umar}\ and\ \citenamefont
  {Oberacker}(2005)}]{umar2005a}%
  \BibitemOpen
  \bibfield  {author} {\bibinfo {author} {\bibfnamefont {A.~S.}\ \bibnamefont
  {Umar}}\ and\ \bibinfo {author} {\bibfnamefont {V.~E.}\ \bibnamefont
  {Oberacker}},\ }\href {\doibase 10.1103/PhysRevC.71.034314} {\bibfield
  {journal} {\bibinfo  {journal} {Phys. Rev. C}\ }\textbf {\bibinfo {volume}
  {71}},\ \bibinfo {pages} {034314} (\bibinfo {year} {2005})}\BibitemShut
  {NoStop}%
\bibitem [{\citenamefont {Simenel}\ \emph {et~al.}(2001)\citenamefont
  {Simenel}, \citenamefont {Chomaz},\ and\ \citenamefont {{de
  France}}}]{simenel2001}%
  \BibitemOpen
  \bibfield  {author} {\bibinfo {author} {\bibfnamefont {C.}~\bibnamefont
  {Simenel}}, \bibinfo {author} {\bibfnamefont {P.}~\bibnamefont {Chomaz}}, \
  and\ \bibinfo {author} {\bibfnamefont {G.}~\bibnamefont {{de France}}},\
  }\href {\doibase 10.1103/PhysRevLett.86.2971} {\bibfield  {journal} {\bibinfo
   {journal} {Phys. Rev. Lett.}\ }\textbf {\bibinfo {volume} {86}},\ \bibinfo
  {pages} {2971} (\bibinfo {year} {2001})}\BibitemShut {NoStop}%
\bibitem [{\citenamefont {Simenel}\ and\ \citenamefont
  {Chomaz}(2003)}]{simenel2003}%
  \BibitemOpen
  \bibfield  {author} {\bibinfo {author} {\bibfnamefont {C.}~\bibnamefont
  {Simenel}}\ and\ \bibinfo {author} {\bibfnamefont {P.}~\bibnamefont
  {Chomaz}},\ }\href {\doibase 10.1103/PhysRevC.68.024302} {\bibfield
  {journal} {\bibinfo  {journal} {Phys. Rev. C}\ }\textbf {\bibinfo {volume}
  {68}},\ \bibinfo {pages} {024302} (\bibinfo {year} {2003})}\BibitemShut
  {NoStop}%
\bibitem [{\citenamefont {Lacroix}\ and\ \citenamefont
  {Chomaz}(1998)}]{lacroix1998}%
  \BibitemOpen
  \bibfield  {author} {\bibinfo {author} {\bibfnamefont {D.}~\bibnamefont
  {Lacroix}}\ and\ \bibinfo {author} {\bibfnamefont {P.}~\bibnamefont
  {Chomaz}},\ }\href {\doibase 10.1016/S0375-9474(98)00203-6} {\bibfield
  {journal} {\bibinfo  {journal} {Nucl. Phys. A}\ }\textbf {\bibinfo {volume}
  {636}},\ \bibinfo {pages} {85} (\bibinfo {year} {1998})}\BibitemShut
  {NoStop}%
\bibitem [{\citenamefont {Paar}\ \emph {et~al.}(2007)\citenamefont {Paar},
  \citenamefont {Vretenar}, \citenamefont {Khan},\ and\ \citenamefont
  {Col\`o}}]{paar2007}%
  \BibitemOpen
  \bibfield  {author} {\bibinfo {author} {\bibfnamefont {N.}~\bibnamefont
  {Paar}}, \bibinfo {author} {\bibfnamefont {D.}~\bibnamefont {Vretenar}},
  \bibinfo {author} {\bibfnamefont {E.}~\bibnamefont {Khan}}, \ and\ \bibinfo
  {author} {\bibfnamefont {G.}~\bibnamefont {Col\`o}},\ }\href
  {http://stacks.iop.org/0034-4885/70/i=5/a=R02} {\bibfield  {journal}
  {\bibinfo  {journal} {Rep. Prog. Phys}\ }\textbf {\bibinfo {volume} {70}},\
  \bibinfo {pages} {691} (\bibinfo {year} {2007})}\BibitemShut {NoStop}%
\bibitem [{\citenamefont {Kl\"uepfel}\ \emph {et~al.}(2009)\citenamefont
  {Kl\"uepfel}, \citenamefont {Reinhard}, \citenamefont {B\"urvenich},\ and\
  \citenamefont {Maruhn}}]{kluepfel2009}%
  \BibitemOpen
  \bibfield  {author} {\bibinfo {author} {\bibfnamefont {P.}~\bibnamefont
  {Kl\"uepfel}}, \bibinfo {author} {\bibfnamefont {P.-G.}\ \bibnamefont
  {Reinhard}}, \bibinfo {author} {\bibfnamefont {T.~J.}\ \bibnamefont
  {B\"urvenich}}, \ and\ \bibinfo {author} {\bibfnamefont {J.~A.}\ \bibnamefont
  {Maruhn}},\ }\href {\doibase 10.1103/PhysRevC.79.034310} {\bibfield
  {journal} {\bibinfo  {journal} {Phys. Rev. C}\ }\textbf {\bibinfo {volume}
  {79}},\ \bibinfo {pages} {034310} (\bibinfo {year} {2009})}\BibitemShut
  {NoStop}%
\bibitem [{\citenamefont {Horowitz}\ and\ \citenamefont
  {Piekarewicz}(2001{\natexlab{a}})}]{horowitz2001a}%
  \BibitemOpen
  \bibfield  {author} {\bibinfo {author} {\bibfnamefont {C.~J.}\ \bibnamefont
  {Horowitz}}\ and\ \bibinfo {author} {\bibfnamefont {J.}~\bibnamefont
  {Piekarewicz}},\ }\href {\doibase 10.1103/PhysRevLett.86.5647} {\bibfield
  {journal} {\bibinfo  {journal} {Phys. Rev. Lett.}\ }\textbf {\bibinfo
  {volume} {86}},\ \bibinfo {pages} {5647} (\bibinfo {year}
  {2001}{\natexlab{a}})}\BibitemShut {NoStop}%
\bibitem [{\citenamefont {Horowitz}\ and\ \citenamefont
  {Piekarewicz}(2001{\natexlab{b}})}]{horowitz2001b}%
  \BibitemOpen
  \bibfield  {author} {\bibinfo {author} {\bibfnamefont {C.~J.}\ \bibnamefont
  {Horowitz}}\ and\ \bibinfo {author} {\bibfnamefont {J.}~\bibnamefont
  {Piekarewicz}},\ }\href {\doibase 10.1103/PhysRevC.64.062802} {\bibfield
  {journal} {\bibinfo  {journal} {Phys. Rev. C}\ }\textbf {\bibinfo {volume}
  {64}},\ \bibinfo {pages} {062802} (\bibinfo {year}
  {2001}{\natexlab{b}})}\BibitemShut {NoStop}%
\bibitem [{\citenamefont {Erler}\ \emph {et~al.}(2013)\citenamefont {Erler},
  \citenamefont {Horowitz}, \citenamefont {Nazarewicz}, \citenamefont
  {Rafalski},\ and\ \citenamefont {Reinhard}}]{erler2013}%
  \BibitemOpen
  \bibfield  {author} {\bibinfo {author} {\bibfnamefont {J.}~\bibnamefont
  {Erler}}, \bibinfo {author} {\bibfnamefont {C.~J.}\ \bibnamefont {Horowitz}},
  \bibinfo {author} {\bibfnamefont {W.}~\bibnamefont {Nazarewicz}}, \bibinfo
  {author} {\bibfnamefont {M.}~\bibnamefont {Rafalski}}, \ and\ \bibinfo
  {author} {\bibfnamefont {P.-G.}\ \bibnamefont {Reinhard}},\ }\href {\doibase
  10.1103/PhysRevC.87.044320} {\bibfield  {journal} {\bibinfo  {journal} {Phys.
  Rev. C}\ }\textbf {\bibinfo {volume} {87}},\ \bibinfo {pages} {044320}
  (\bibinfo {year} {2013})}\BibitemShut {NoStop}%
\bibitem [{\citenamefont {Dickhoff}\ and\ \citenamefont
  {M{\"u}ther}(1992)}]{dickhoff1992}%
  \BibitemOpen
  \bibfield  {author} {\bibinfo {author} {\bibfnamefont {W.~H.}\ \bibnamefont
  {Dickhoff}}\ and\ \bibinfo {author} {\bibfnamefont {H.}~\bibnamefont
  {M{\"u}ther}},\ }\href {http://iopscience.iop.org/0034-4885/55/11/002}
  {\bibfield  {journal} {\bibinfo  {journal} {Rep. Prog. Phys.}\ }\textbf
  {\bibinfo {volume} {55}},\ \bibinfo {pages} {1947} (\bibinfo {year}
  {1992})}\BibitemShut {NoStop}%
\bibitem [{\citenamefont {Pandharipande}\ \emph {et~al.}(1997)\citenamefont
  {Pandharipande}, \citenamefont {Sick},\ and\ \citenamefont {deWitt
  Huberts}}]{pandharipande1997}%
  \BibitemOpen
  \bibfield  {author} {\bibinfo {author} {\bibfnamefont {V.~R.}\ \bibnamefont
  {Pandharipande}}, \bibinfo {author} {\bibfnamefont {I.}~\bibnamefont {Sick}},
  \ and\ \bibinfo {author} {\bibfnamefont {P.~K.~A.}\ \bibnamefont {deWitt
  Huberts}},\ }\href {\doibase 10.1103/RevModPhys.69.981} {\bibfield  {journal}
  {\bibinfo  {journal} {Rev. Mod. Phys.}\ }\textbf {\bibinfo {volume} {69}},\
  \bibinfo {pages} {981} (\bibinfo {year} {1997})}\BibitemShut {NoStop}%
\bibitem [{\citenamefont {Epelbaum}\ \emph {et~al.}(2009)\citenamefont
  {Epelbaum}, \citenamefont {Hammer},\ and\ \citenamefont
  {Mei\ss{}ner}}]{epelbaum2009}%
  \BibitemOpen
  \bibfield  {author} {\bibinfo {author} {\bibfnamefont {E.}~\bibnamefont
  {Epelbaum}}, \bibinfo {author} {\bibfnamefont {H.-W.}\ \bibnamefont
  {Hammer}}, \ and\ \bibinfo {author} {\bibfnamefont {U.-G.}\ \bibnamefont
  {Mei\ss{}ner}},\ }\href {\doibase 10.1103/RevModPhys.81.1773} {\bibfield
  {journal} {\bibinfo  {journal} {Rev. Mod. Phys.}\ }\textbf {\bibinfo {volume}
  {81}},\ \bibinfo {pages} {1773} (\bibinfo {year} {2009})}\BibitemShut
  {NoStop}%
\bibitem [{\citenamefont {Machleidt}\ and\ \citenamefont
  {Entem}(2011)}]{machleidt2011}%
  \BibitemOpen
  \bibfield  {author} {\bibinfo {author} {\bibfnamefont {R.}~\bibnamefont
  {Machleidt}}\ and\ \bibinfo {author} {\bibfnamefont {D.~R.}\ \bibnamefont
  {Entem}},\ }\href {\doibase 10.1016/j.physrep.2011.02.001} {\bibfield
  {journal} {\bibinfo  {journal} {Phys. Rep.}\ }\textbf {\bibinfo {volume}
  {503}},\ \bibinfo {pages} {1} (\bibinfo {year} {2011})}\BibitemShut {NoStop}%
\bibitem [{\citenamefont {Hammer}\ \emph {et~al.}(2013)\citenamefont {Hammer},
  \citenamefont {Nogga},\ and\ \citenamefont {Schwenk}}]{hammer2013}%
  \BibitemOpen
  \bibfield  {author} {\bibinfo {author} {\bibfnamefont {H.-W.}\ \bibnamefont
  {Hammer}}, \bibinfo {author} {\bibfnamefont {A.}~\bibnamefont {Nogga}}, \
  and\ \bibinfo {author} {\bibfnamefont {A.}~\bibnamefont {Schwenk}},\ }\href
  {\doibase 10.1103/RevModPhys.85.197} {\bibfield  {journal} {\bibinfo
  {journal} {Rev. Mod. Phys.}\ }\textbf {\bibinfo {volume} {85}},\ \bibinfo
  {pages} {197} (\bibinfo {year} {2013})}\BibitemShut {NoStop}%
\bibitem [{\citenamefont {Kortelainen}\ \emph {et~al.}(2013)\citenamefont
  {Kortelainen}, \citenamefont {Erler}, \citenamefont {Nazarewicz},
  \citenamefont {Birge}, \citenamefont {Gao},\ and\ \citenamefont
  {Olsen}}]{kortelainen2013}%
  \BibitemOpen
  \bibfield  {author} {\bibinfo {author} {\bibfnamefont {M.}~\bibnamefont
  {Kortelainen}}, \bibinfo {author} {\bibfnamefont {J.}~\bibnamefont {Erler}},
  \bibinfo {author} {\bibfnamefont {W.}~\bibnamefont {Nazarewicz}}, \bibinfo
  {author} {\bibfnamefont {N.}~\bibnamefont {Birge}}, \bibinfo {author}
  {\bibfnamefont {Y.}~\bibnamefont {Gao}}, \ and\ \bibinfo {author}
  {\bibfnamefont {E.}~\bibnamefont {Olsen}},\ }\href {\doibase
  10.1103/PhysRevC.88.031305} {\bibfield  {journal} {\bibinfo  {journal} {Phys.
  Rev. C}\ }\textbf {\bibinfo {volume} {88}},\ \bibinfo {pages} {031305}
  (\bibinfo {year} {2013})}\BibitemShut {NoStop}%
\bibitem [{\citenamefont {Dobaczewski}\ \emph {et~al.}(2014)\citenamefont
  {Dobaczewski}, \citenamefont {Nazarewicz},\ and\ \citenamefont
  {Reinhard}}]{dobaczewski2014}%
  \BibitemOpen
  \bibfield  {author} {\bibinfo {author} {\bibfnamefont {J.}~\bibnamefont
  {Dobaczewski}}, \bibinfo {author} {\bibfnamefont {W.}~\bibnamefont
  {Nazarewicz}}, \ and\ \bibinfo {author} {\bibfnamefont {P.-G.}\ \bibnamefont
  {Reinhard}},\ }\href {http://dx.doi.org/10.1088/0954-3899/41/7/074001}
  {\bibfield  {journal} {\bibinfo  {journal} {J. Phys. G}\ }\textbf {\bibinfo
  {volume} {41}},\ \bibinfo {pages} {074001} (\bibinfo {year}
  {2014})}\BibitemShut {NoStop}%
\bibitem [{\citenamefont {Erler}\ and\ \citenamefont
  {Reinhard}(2014)}]{erler2014}%
  \BibitemOpen
  \bibfield  {author} {\bibinfo {author} {\bibfnamefont {J.}~\bibnamefont
  {Erler}}\ and\ \bibinfo {author} {\bibfnamefont {P.-G.}\ \bibnamefont
  {Reinhard}},\ }\href {http://dx.doi.org/10.1088/0954-3899/42/3/034026}
  {\bibfield  {journal} {\bibinfo  {journal} {J. Phys. G}\ }\textbf {\bibinfo
  {volume} {42}},\ \bibinfo {pages} {034026} (\bibinfo {year}
  {2014})}\BibitemShut {NoStop}%
\bibitem [{\citenamefont {Reinhard}(2016)}]{reinhard2016}%
  \BibitemOpen
  \bibfield  {author} {\bibinfo {author} {\bibfnamefont {P.-G.}\ \bibnamefont
  {Reinhard}},\ }\href {http://dx.doi.org/10.1088/0031-8949/91/2/023002}
  {\bibfield  {journal} {\bibinfo  {journal} {Phys. Scr.}\ }\textbf {\bibinfo
  {volume} {91}},\ \bibinfo {pages} {023002} (\bibinfo {year}
  {2016})}\BibitemShut {NoStop}%
\bibitem [{\citenamefont {Fattoyev}\ and\ \citenamefont
  {Piekarewicz}(2011)}]{fattoyev2011}%
  \BibitemOpen
  \bibfield  {author} {\bibinfo {author} {\bibfnamefont {F.~J.}\ \bibnamefont
  {Fattoyev}}\ and\ \bibinfo {author} {\bibfnamefont {J.}~\bibnamefont
  {Piekarewicz}},\ }\href {\doibase 10.1103/PhysRevC.84.064302} {\bibfield
  {journal} {\bibinfo  {journal} {Phys. Rev. C}\ }\textbf {\bibinfo {volume}
  {84}},\ \bibinfo {pages} {064302} (\bibinfo {year} {2011})}\BibitemShut
  {NoStop}%
\bibitem [{\citenamefont {Skyrme}(1956)}]{skyrme1956}%
  \BibitemOpen
  \bibfield  {author} {\bibinfo {author} {\bibfnamefont {T.~H.~R.}\
  \bibnamefont {Skyrme}},\ }\href {\doibase 10.1080/14786435608238186}
  {\bibfield  {journal} {\bibinfo  {journal} {Phil. Mag.}\ }\textbf {\bibinfo
  {volume} {1}},\ \bibinfo {pages} {1043} (\bibinfo {year} {1956})}\BibitemShut
  {NoStop}%
\bibitem [{\citenamefont {Vautherin}\ and\ \citenamefont
  {Brink}(1972)}]{vautherin1972}%
  \BibitemOpen
  \bibfield  {author} {\bibinfo {author} {\bibfnamefont {D.}~\bibnamefont
  {Vautherin}}\ and\ \bibinfo {author} {\bibfnamefont {D.~M.}\ \bibnamefont
  {Brink}},\ }\href {\doibase 10.1103/PhysRevC.5.626} {\bibfield  {journal}
  {\bibinfo  {journal} {Phys. Rev. C}\ }\textbf {\bibinfo {volume} {5}},\
  \bibinfo {pages} {626} (\bibinfo {year} {1972})}\BibitemShut {NoStop}%
\bibitem [{\citenamefont {Bender}\ \emph {et~al.}(2003)\citenamefont {Bender},
  \citenamefont {Heenen},\ and\ \citenamefont {Reinhard}}]{bender2003}%
  \BibitemOpen
  \bibfield  {author} {\bibinfo {author} {\bibfnamefont {M.}~\bibnamefont
  {Bender}}, \bibinfo {author} {\bibfnamefont {P.~H.}\ \bibnamefont {Heenen}},
  \ and\ \bibinfo {author} {\bibfnamefont {P.-G.}\ \bibnamefont {Reinhard}},\
  }\href {\doibase 10.1103/RevModPhys.75.121} {\bibfield  {journal} {\bibinfo
  {journal} {Rev. Mod. Phys.}\ }\textbf {\bibinfo {volume} {75}},\ \bibinfo
  {pages} {121} (\bibinfo {year} {2003})}\BibitemShut {NoStop}%
\bibitem [{\citenamefont {Umar}\ and\ \citenamefont
  {Oberacker}(2006{\natexlab{a}})}]{umar2006a}%
  \BibitemOpen
  \bibfield  {author} {\bibinfo {author} {\bibfnamefont {A.~S.}\ \bibnamefont
  {Umar}}\ and\ \bibinfo {author} {\bibfnamefont {V.~E.}\ \bibnamefont
  {Oberacker}},\ }\href {\doibase 10.1103/PhysRevC.74.061601} {\bibfield
  {journal} {\bibinfo  {journal} {Phys. Rev. C}\ }\textbf {\bibinfo {volume}
  {74}},\ \bibinfo {pages} {061601} (\bibinfo {year}
  {2006}{\natexlab{a}})}\BibitemShut {NoStop}%
\bibitem [{\citenamefont {Keser}\ \emph {et~al.}(2012)\citenamefont {Keser},
  \citenamefont {Umar},\ and\ \citenamefont {Oberacker}}]{keser2012}%
  \BibitemOpen
  \bibfield  {author} {\bibinfo {author} {\bibfnamefont {R.}~\bibnamefont
  {Keser}}, \bibinfo {author} {\bibfnamefont {A.~S.}\ \bibnamefont {Umar}}, \
  and\ \bibinfo {author} {\bibfnamefont {V.~E.}\ \bibnamefont {Oberacker}},\
  }\href {\doibase 10.1103/PhysRevC.85.044606} {\bibfield  {journal} {\bibinfo
  {journal} {Phys. Rev. C}\ }\textbf {\bibinfo {volume} {85}},\ \bibinfo
  {pages} {044606} (\bibinfo {year} {2012})}\BibitemShut {NoStop}%
\bibitem [{\citenamefont {Stefanini}\ \emph {et~al.}(2009)\citenamefont
  {Stefanini}, \citenamefont {Montagnoli}, \citenamefont {Silvestri},
  \citenamefont {Corradi}, \citenamefont {Courtin}, \citenamefont {Fioretto},
  \citenamefont {Guiot}, \citenamefont {Haas}, \citenamefont {Lebhertz},
  \citenamefont {Mason}, \citenamefont {Scarlassara},\ and\ \citenamefont
  {Szilner}}]{stefanini2009}%
  \BibitemOpen
  \bibfield  {author} {\bibinfo {author} {\bibfnamefont {A.~M.}\ \bibnamefont
  {Stefanini}}, \bibinfo {author} {\bibfnamefont {G.}~\bibnamefont
  {Montagnoli}}, \bibinfo {author} {\bibfnamefont {R.}~\bibnamefont
  {Silvestri}}, \bibinfo {author} {\bibfnamefont {L.}~\bibnamefont {Corradi}},
  \bibinfo {author} {\bibfnamefont {S.}~\bibnamefont {Courtin}}, \bibinfo
  {author} {\bibfnamefont {E.}~\bibnamefont {Fioretto}}, \bibinfo {author}
  {\bibfnamefont {B.}~\bibnamefont {Guiot}}, \bibinfo {author} {\bibfnamefont
  {F.}~\bibnamefont {Haas}}, \bibinfo {author} {\bibfnamefont {D.}~\bibnamefont
  {Lebhertz}}, \bibinfo {author} {\bibfnamefont {P.}~\bibnamefont {Mason}},
  \bibinfo {author} {\bibfnamefont {F.}~\bibnamefont {Scarlassara}}, \ and\
  \bibinfo {author} {\bibfnamefont {S.}~\bibnamefont {Szilner}},\ }\href
  {\doibase 10.1016/j.physletb.2009.07.017} {\bibfield  {journal} {\bibinfo
  {journal} {Phys. Lett. B}\ }\textbf {\bibinfo {volume} {679}},\ \bibinfo
  {pages} {95} (\bibinfo {year} {2009})}\BibitemShut {NoStop}%
\bibitem [{\citenamefont {Esbensen}\ \emph {et~al.}(2010)\citenamefont
  {Esbensen}, \citenamefont {Jiang},\ and\ \citenamefont
  {Stefanini}}]{esbensen2010}%
  \BibitemOpen
  \bibfield  {author} {\bibinfo {author} {\bibfnamefont {H.}~\bibnamefont
  {Esbensen}}, \bibinfo {author} {\bibfnamefont {C.~L.}\ \bibnamefont {Jiang}},
  \ and\ \bibinfo {author} {\bibfnamefont {A.~M.}\ \bibnamefont {Stefanini}},\
  }\href {\doibase 10.1103/PhysRevC.82.054621} {\bibfield  {journal} {\bibinfo
  {journal} {Phys. Rev. C}\ }\textbf {\bibinfo {volume} {82}},\ \bibinfo
  {pages} {054621} (\bibinfo {year} {2010})}\BibitemShut {NoStop}%
\bibitem [{\citenamefont {Ray}(1979)}]{ray1979}%
  \BibitemOpen
  \bibfield  {author} {\bibinfo {author} {\bibfnamefont {L.}~\bibnamefont
  {Ray}},\ }\href {\doibase 10.1103/PhysRevC.19.1855} {\bibfield  {journal}
  {\bibinfo  {journal} {Phys. Rev. C}\ }\textbf {\bibinfo {volume} {19}},\
  \bibinfo {pages} {1855} (\bibinfo {year} {1979})}\BibitemShut {NoStop}%
\bibitem [{\citenamefont {Satchler}\ and\ \citenamefont
  {Love}(1979)}]{satchler1979}%
  \BibitemOpen
  \bibfield  {author} {\bibinfo {author} {\bibfnamefont {G.~R.}\ \bibnamefont
  {Satchler}}\ and\ \bibinfo {author} {\bibfnamefont {W.~G.}\ \bibnamefont
  {Love}},\ }\href {\doibase 10.1016/0370-1573(79)90081-4} {\bibfield
  {journal} {\bibinfo  {journal} {Phys. Rep.}\ }\textbf {\bibinfo {volume}
  {55}},\ \bibinfo {pages} {183} (\bibinfo {year} {1979})}\BibitemShut
  {NoStop}%
\bibitem [{\citenamefont {Bertsch}\ \emph {et~al.}(1977)\citenamefont
  {Bertsch}, \citenamefont {Borysowicz}, \citenamefont {Mcmanus},\ and\
  \citenamefont {Love}}]{bertsch1977}%
  \BibitemOpen
  \bibfield  {author} {\bibinfo {author} {\bibfnamefont {G.}~\bibnamefont
  {Bertsch}}, \bibinfo {author} {\bibfnamefont {J.}~\bibnamefont {Borysowicz}},
  \bibinfo {author} {\bibfnamefont {H.}~\bibnamefont {Mcmanus}}, \ and\
  \bibinfo {author} {\bibfnamefont {W.~G.}\ \bibnamefont {Love}},\ }\href
  {\doibase 10.1016/0375-9474(77)90392-X} {\bibfield  {journal} {\bibinfo
  {journal} {Nucl. Phys. A}\ }\textbf {\bibinfo {volume} {284}},\ \bibinfo
  {pages} {399} (\bibinfo {year} {1977})}\BibitemShut {NoStop}%
\bibitem [{\citenamefont {Rhoades-Brown}\ and\ \citenamefont
  {Oberacker}(1983)}]{rhoadesbrown1983a}%
  \BibitemOpen
  \bibfield  {author} {\bibinfo {author} {\bibfnamefont {M.~J.}\ \bibnamefont
  {Rhoades-Brown}}\ and\ \bibinfo {author} {\bibfnamefont {V.~E.}\ \bibnamefont
  {Oberacker}},\ }\href {\doibase 10.1103/PhysRevLett.50.1435} {\bibfield
  {journal} {\bibinfo  {journal} {Phys. Rev. Lett.}\ }\textbf {\bibinfo
  {volume} {50}},\ \bibinfo {pages} {1435} (\bibinfo {year}
  {1983})}\BibitemShut {NoStop}%
\bibitem [{\citenamefont {Rhoades-Brown}\ \emph {et~al.}(1983)\citenamefont
  {Rhoades-Brown}, \citenamefont {Oberacker}, \citenamefont {Seiwert},\ and\
  \citenamefont {Greiner}}]{rhoadesbrown1983b}%
  \BibitemOpen
  \bibfield  {author} {\bibinfo {author} {\bibfnamefont {M.~J.}\ \bibnamefont
  {Rhoades-Brown}}, \bibinfo {author} {\bibfnamefont {V.~E.}\ \bibnamefont
  {Oberacker}}, \bibinfo {author} {\bibfnamefont {M.}~\bibnamefont {Seiwert}},
  \ and\ \bibinfo {author} {\bibfnamefont {W.}~\bibnamefont {Greiner}},\ }\href
  {\doibase 10.1007/BF01419514} {\bibfield  {journal} {\bibinfo  {journal} {Z.
  Phys. A}\ }\textbf {\bibinfo {volume} {310}},\ \bibinfo {pages} {287}
  (\bibinfo {year} {1983})}\BibitemShut {NoStop}%
\bibitem [{\citenamefont {Hagino}\ \emph {et~al.}(1999)\citenamefont {Hagino},
  \citenamefont {Rowley},\ and\ \citenamefont {Kruppa}}]{hagino1999}%
  \BibitemOpen
  \bibfield  {author} {\bibinfo {author} {\bibfnamefont {K.}~\bibnamefont
  {Hagino}}, \bibinfo {author} {\bibfnamefont {N.}~\bibnamefont {Rowley}}, \
  and\ \bibinfo {author} {\bibfnamefont {A.~T.}\ \bibnamefont {Kruppa}},\
  }\href {\doibase 10.1016/S0010-4655(99)00243-X} {\bibfield  {journal}
  {\bibinfo  {journal} {Comput. Phys. Commun.}\ }\textbf {\bibinfo {volume}
  {123}},\ \bibinfo {pages} {143} (\bibinfo {year} {1999})}\BibitemShut
  {NoStop}%
\bibitem [{\citenamefont {Simenel}\ \emph
  {et~al.}(2013{\natexlab{a}})\citenamefont {Simenel}, \citenamefont
  {Dasgupta}, \citenamefont {Hinde},\ and\ \citenamefont
  {Williams}}]{simenel2013b}%
  \BibitemOpen
  \bibfield  {author} {\bibinfo {author} {\bibfnamefont {C.}~\bibnamefont
  {Simenel}}, \bibinfo {author} {\bibfnamefont {M.}~\bibnamefont {Dasgupta}},
  \bibinfo {author} {\bibfnamefont {D.~J.}\ \bibnamefont {Hinde}}, \ and\
  \bibinfo {author} {\bibfnamefont {E.}~\bibnamefont {Williams}},\ }\href
  {\doibase 10.1103/PhysRevC.88.064604} {\bibfield  {journal} {\bibinfo
  {journal} {Phys. Rev. C}\ }\textbf {\bibinfo {volume} {88}},\ \bibinfo
  {pages} {064604} (\bibinfo {year} {2013}{\natexlab{a}})}\BibitemShut
  {NoStop}%
\bibitem [{\citenamefont {{Kouhei Washiyama}}\ and\ \citenamefont {{Denis
  Lacroix}}(2008)}]{washiyama2008}%
  \BibitemOpen
  \bibfield  {author} {\bibinfo {author} {\bibnamefont {{Kouhei Washiyama}}}\
  and\ \bibinfo {author} {\bibnamefont {{Denis Lacroix}}},\ }\href {\doibase
  10.1103/PhysRevC.78.024610} {\bibfield  {journal} {\bibinfo  {journal} {Phys.
  Rev. C}\ }\textbf {\bibinfo {volume} {78}},\ \bibinfo {pages} {024610}
  (\bibinfo {year} {2008})}\BibitemShut {NoStop}%
\bibitem [{\citenamefont {Umar}\ \emph {et~al.}(2014)\citenamefont {Umar},
  \citenamefont {Simenel},\ and\ \citenamefont {Oberacker}}]{umar2014a}%
  \BibitemOpen
  \bibfield  {author} {\bibinfo {author} {\bibfnamefont {A.~S.}\ \bibnamefont
  {Umar}}, \bibinfo {author} {\bibfnamefont {C.}~\bibnamefont {Simenel}}, \
  and\ \bibinfo {author} {\bibfnamefont {V.~E.}\ \bibnamefont {Oberacker}},\
  }\href {\doibase 10.1103/PhysRevC.89.034611} {\bibfield  {journal} {\bibinfo
  {journal} {Phys. Rev. C}\ }\textbf {\bibinfo {volume} {89}},\ \bibinfo
  {pages} {034611} (\bibinfo {year} {2014})}\BibitemShut {NoStop}%
\bibitem [{\citenamefont {Jiang}\ \emph {et~al.}(2014)\citenamefont {Jiang},
  \citenamefont {Maruhn},\ and\ \citenamefont {Yan}}]{jiang2014}%
  \BibitemOpen
  \bibfield  {author} {\bibinfo {author} {\bibfnamefont {X.}~\bibnamefont
  {Jiang}}, \bibinfo {author} {\bibfnamefont {J.~A.}\ \bibnamefont {Maruhn}}, \
  and\ \bibinfo {author} {\bibfnamefont {S.}~\bibnamefont {Yan}},\ }\href
  {\doibase 10.1103/PhysRevC.90.064618} {\bibfield  {journal} {\bibinfo
  {journal} {Phys. Rev. C}\ }\textbf {\bibinfo {volume} {90}},\ \bibinfo
  {pages} {064618} (\bibinfo {year} {2014})}\BibitemShut {NoStop}%
\bibitem [{\citenamefont {Chamon}\ \emph {et~al.}(2002)\citenamefont {Chamon},
  \citenamefont {Carlson}, \citenamefont {Gasques}, \citenamefont {Pereira},
  \citenamefont {{De Conti}}, \citenamefont {Alvarez}, \citenamefont {Hussein},
  \citenamefont {Ribeiro}, \citenamefont {Rossi},\ and\ \citenamefont
  {Silva}}]{chamon2002}%
  \BibitemOpen
  \bibfield  {author} {\bibinfo {author} {\bibfnamefont {L.~C.}\ \bibnamefont
  {Chamon}}, \bibinfo {author} {\bibfnamefont {B.~V.}\ \bibnamefont {Carlson}},
  \bibinfo {author} {\bibfnamefont {L.~R.}\ \bibnamefont {Gasques}}, \bibinfo
  {author} {\bibfnamefont {D.}~\bibnamefont {Pereira}}, \bibinfo {author}
  {\bibfnamefont {C.}~\bibnamefont {{De Conti}}}, \bibinfo {author}
  {\bibfnamefont {M.~A.~G.}\ \bibnamefont {Alvarez}}, \bibinfo {author}
  {\bibfnamefont {M.~S.}\ \bibnamefont {Hussein}}, \bibinfo {author}
  {\bibfnamefont {M.~A.~C.}\ \bibnamefont {Ribeiro}}, \bibinfo {author}
  {\bibfnamefont {E.~S.}\ \bibnamefont {Rossi}}, \ and\ \bibinfo {author}
  {\bibfnamefont {C.~P.}\ \bibnamefont {Silva}},\ }\href {\doibase
  10.1103/PhysRevC.66.014610} {\bibfield  {journal} {\bibinfo  {journal} {Phys.
  Rev. C}\ }\textbf {\bibinfo {volume} {66}},\ \bibinfo {pages} {014610}
  (\bibinfo {year} {2002})}\BibitemShut {NoStop}%
\bibitem [{\citenamefont {Umar}\ and\ \citenamefont
  {Oberacker}(2006{\natexlab{b}})}]{umar2006b}%
  \BibitemOpen
  \bibfield  {author} {\bibinfo {author} {\bibfnamefont {A.~S.}\ \bibnamefont
  {Umar}}\ and\ \bibinfo {author} {\bibfnamefont {V.~E.}\ \bibnamefont
  {Oberacker}},\ }\href {\doibase 10.1103/PhysRevC.74.021601} {\bibfield
  {journal} {\bibinfo  {journal} {Phys. Rev. C}\ }\textbf {\bibinfo {volume}
  {74}},\ \bibinfo {pages} {021601} (\bibinfo {year}
  {2006}{\natexlab{b}})}\BibitemShut {NoStop}%
\bibitem [{\citenamefont {Cusson}\ \emph {et~al.}(1985)\citenamefont {Cusson},
  \citenamefont {Reinhard}, \citenamefont {Strayer}, \citenamefont {Maruhn},\
  and\ \citenamefont {Greiner}}]{cusson1985}%
  \BibitemOpen
  \bibfield  {author} {\bibinfo {author} {\bibfnamefont {R.~Y.}\ \bibnamefont
  {Cusson}}, \bibinfo {author} {\bibfnamefont {P.-G.}\ \bibnamefont
  {Reinhard}}, \bibinfo {author} {\bibfnamefont {M.~R.}\ \bibnamefont
  {Strayer}}, \bibinfo {author} {\bibfnamefont {J.~A.}\ \bibnamefont {Maruhn}},
  \ and\ \bibinfo {author} {\bibfnamefont {W.}~\bibnamefont {Greiner}},\ }\href
  {\doibase 10.1007/BF01415725} {\bibfield  {journal} {\bibinfo  {journal} {Z.
  Phys. A}\ }\textbf {\bibinfo {volume} {320}},\ \bibinfo {pages} {475}
  (\bibinfo {year} {1985})}\BibitemShut {NoStop}%
\bibitem [{\citenamefont {Umar}\ \emph {et~al.}(1985)\citenamefont {Umar},
  \citenamefont {Strayer}, \citenamefont {Cusson}, \citenamefont {Reinhard},\
  and\ \citenamefont {Bromley}}]{umar1985}%
  \BibitemOpen
  \bibfield  {author} {\bibinfo {author} {\bibfnamefont {A.~S.}\ \bibnamefont
  {Umar}}, \bibinfo {author} {\bibfnamefont {M.~R.}\ \bibnamefont {Strayer}},
  \bibinfo {author} {\bibfnamefont {R.~Y.}\ \bibnamefont {Cusson}}, \bibinfo
  {author} {\bibfnamefont {P.-G.}\ \bibnamefont {Reinhard}}, \ and\ \bibinfo
  {author} {\bibfnamefont {D.~A.}\ \bibnamefont {Bromley}},\ }\href {\doibase
  10.1103/PhysRevC.32.172} {\bibfield  {journal} {\bibinfo  {journal} {Phys.
  Rev. C}\ }\textbf {\bibinfo {volume} {32}},\ \bibinfo {pages} {172} (\bibinfo
  {year} {1985})}\BibitemShut {NoStop}%
\bibitem [{\citenamefont {Umar}\ and\ \citenamefont
  {Oberacker}(2009)}]{umar2009b}%
  \BibitemOpen
  \bibfield  {author} {\bibinfo {author} {\bibfnamefont {A.~S.}\ \bibnamefont
  {Umar}}\ and\ \bibinfo {author} {\bibfnamefont {V.~E.}\ \bibnamefont
  {Oberacker}},\ }\href {\doibase 10.1140/epja/i2008-10712-5} {\bibfield
  {journal} {\bibinfo  {journal} {Eur. Phys. J. A}\ }\textbf {\bibinfo {volume}
  {39}},\ \bibinfo {pages} {243} (\bibinfo {year} {2009})}\BibitemShut
  {NoStop}%
\bibitem [{\citenamefont {Oberacker}\ \emph {et~al.}(2010)\citenamefont
  {Oberacker}, \citenamefont {Umar}, \citenamefont {Maruhn},\ and\
  \citenamefont {Reinhard}}]{oberacker2010}%
  \BibitemOpen
  \bibfield  {author} {\bibinfo {author} {\bibfnamefont {V.~E.}\ \bibnamefont
  {Oberacker}}, \bibinfo {author} {\bibfnamefont {A.~S.}\ \bibnamefont {Umar}},
  \bibinfo {author} {\bibfnamefont {J.~A.}\ \bibnamefont {Maruhn}}, \ and\
  \bibinfo {author} {\bibfnamefont {P.-G.}\ \bibnamefont {Reinhard}},\ }\href
  {\doibase 10.1103/PhysRevC.82.034603} {\bibfield  {journal} {\bibinfo
  {journal} {Phys. Rev. C}\ }\textbf {\bibinfo {volume} {82}},\ \bibinfo
  {pages} {034603} (\bibinfo {year} {2010})}\BibitemShut {NoStop}%
\bibitem [{\citenamefont {Umar}\ \emph {et~al.}(2012)\citenamefont {Umar},
  \citenamefont {Oberacker},\ and\ \citenamefont {Horowitz}}]{umar2012a}%
  \BibitemOpen
  \bibfield  {author} {\bibinfo {author} {\bibfnamefont {A.~S.}\ \bibnamefont
  {Umar}}, \bibinfo {author} {\bibfnamefont {V.~E.}\ \bibnamefont {Oberacker}},
  \ and\ \bibinfo {author} {\bibfnamefont {C.~J.}\ \bibnamefont {Horowitz}},\
  }\href {\doibase 10.1103/PhysRevC.85.055801} {\bibfield  {journal} {\bibinfo
  {journal} {Phys. Rev. C}\ }\textbf {\bibinfo {volume} {85}},\ \bibinfo
  {pages} {055801} (\bibinfo {year} {2012})}\BibitemShut {NoStop}%
\bibitem [{\citenamefont {Simenel}\ \emph
  {et~al.}(2013{\natexlab{b}})\citenamefont {Simenel}, \citenamefont {Keser},
  \citenamefont {Umar},\ and\ \citenamefont {Oberacker}}]{simenel2013a}%
  \BibitemOpen
  \bibfield  {author} {\bibinfo {author} {\bibfnamefont {C.}~\bibnamefont
  {Simenel}}, \bibinfo {author} {\bibfnamefont {R.}~\bibnamefont {Keser}},
  \bibinfo {author} {\bibfnamefont {A.~S.}\ \bibnamefont {Umar}}, \ and\
  \bibinfo {author} {\bibfnamefont {V.~E.}\ \bibnamefont {Oberacker}},\ }\href
  {\doibase 10.1103/PhysRevC.88.024617} {\bibfield  {journal} {\bibinfo
  {journal} {Phys. Rev. C}\ }\textbf {\bibinfo {volume} {88}},\ \bibinfo
  {pages} {024617} (\bibinfo {year} {2013}{\natexlab{b}})}\BibitemShut
  {NoStop}%
\bibitem [{\citenamefont {Umar}\ and\ \citenamefont
  {Oberacker}(2006{\natexlab{c}})}]{umar2006c}%
  \BibitemOpen
  \bibfield  {author} {\bibinfo {author} {\bibfnamefont {A.~S.}\ \bibnamefont
  {Umar}}\ and\ \bibinfo {author} {\bibfnamefont {V.~E.}\ \bibnamefont
  {Oberacker}},\ }\href {\doibase 10.1103/PhysRevC.73.054607} {\bibfield
  {journal} {\bibinfo  {journal} {Phys. Rev. C}\ }\textbf {\bibinfo {volume}
  {73}},\ \bibinfo {pages} {054607} (\bibinfo {year}
  {2006}{\natexlab{c}})}\BibitemShut {NoStop}%
\bibitem [{\citenamefont {Maruhn}\ \emph {et~al.}(2014)\citenamefont {Maruhn},
  \citenamefont {Reinhard}, \citenamefont {Stevenson},\ and\ \citenamefont
  {Umar}}]{maruhn2014}%
  \BibitemOpen
  \bibfield  {author} {\bibinfo {author} {\bibfnamefont {J.~A.}\ \bibnamefont
  {Maruhn}}, \bibinfo {author} {\bibfnamefont {P.-G.}\ \bibnamefont
  {Reinhard}}, \bibinfo {author} {\bibfnamefont {P.~D.}\ \bibnamefont
  {Stevenson}}, \ and\ \bibinfo {author} {\bibfnamefont {A.~S.}\ \bibnamefont
  {Umar}},\ }\href {\doibase 10.1016/j.cpc.2014.04.008} {\bibfield  {journal}
  {\bibinfo  {journal} {Comp. Phys. Comm.}\ }\textbf {\bibinfo {volume}
  {185}},\ \bibinfo {pages} {2195} (\bibinfo {year} {2014})}\BibitemShut
  {NoStop}%
\bibitem [{\citenamefont {Umar}\ \emph {et~al.}(1989)\citenamefont {Umar},
  \citenamefont {Strayer}, \citenamefont {Reinhard}, \citenamefont {Davies},\
  and\ \citenamefont {Lee}}]{umar1989}%
  \BibitemOpen
  \bibfield  {author} {\bibinfo {author} {\bibfnamefont {A.~S.}\ \bibnamefont
  {Umar}}, \bibinfo {author} {\bibfnamefont {M.~R.}\ \bibnamefont {Strayer}},
  \bibinfo {author} {\bibfnamefont {P.-G.}\ \bibnamefont {Reinhard}}, \bibinfo
  {author} {\bibfnamefont {K.~T.~R.}\ \bibnamefont {Davies}}, \ and\ \bibinfo
  {author} {\bibfnamefont {S.-J.}\ \bibnamefont {Lee}},\ }\href {\doibase
  10.1103/PhysRevC.40.706} {\bibfield  {journal} {\bibinfo  {journal} {Phys.
  Rev. C}\ }\textbf {\bibinfo {volume} {40}},\ \bibinfo {pages} {706} (\bibinfo
  {year} {1989})}\BibitemShut {NoStop}%
\bibitem [{\citenamefont {Bottcher}\ \emph {et~al.}(1989)\citenamefont
  {Bottcher}, \citenamefont {Strayer}, \citenamefont {Umar},\ and\
  \citenamefont {Reinhard}}]{bottcher1989}%
  \BibitemOpen
  \bibfield  {author} {\bibinfo {author} {\bibfnamefont {C.}~\bibnamefont
  {Bottcher}}, \bibinfo {author} {\bibfnamefont {M.~R.}\ \bibnamefont
  {Strayer}}, \bibinfo {author} {\bibfnamefont {A.~S.}\ \bibnamefont {Umar}}, \
  and\ \bibinfo {author} {\bibfnamefont {P.-G.}\ \bibnamefont {Reinhard}},\
  }\href {\doibase 10.1103/PhysRevA.40.4182} {\bibfield  {journal} {\bibinfo
  {journal} {Phys. Rev. A}\ }\textbf {\bibinfo {volume} {40}},\ \bibinfo
  {pages} {4182} (\bibinfo {year} {1989})}\BibitemShut {NoStop}%
\bibitem [{\citenamefont {Umar}\ \emph {et~al.}(1991)\citenamefont {Umar},
  \citenamefont {Strayer}, \citenamefont {Wu}, \citenamefont {Dean},\ and\
  \citenamefont {G\"u\c{c}l\"u}}]{umar1991a}%
  \BibitemOpen
  \bibfield  {author} {\bibinfo {author} {\bibfnamefont {A.~S.}\ \bibnamefont
  {Umar}}, \bibinfo {author} {\bibfnamefont {M.~R.}\ \bibnamefont {Strayer}},
  \bibinfo {author} {\bibfnamefont {J.~S.}\ \bibnamefont {Wu}}, \bibinfo
  {author} {\bibfnamefont {D.~J.}\ \bibnamefont {Dean}}, \ and\ \bibinfo
  {author} {\bibfnamefont {M.~C.}\ \bibnamefont {G\"u\c{c}l\"u}},\ }\href
  {\doibase 10.1103/PhysRevC.44.2512} {\bibfield  {journal} {\bibinfo
  {journal} {Phys. Rev. C}\ }\textbf {\bibinfo {volume} {44}},\ \bibinfo
  {pages} {2512} (\bibinfo {year} {1991})}\BibitemShut {NoStop}%
\bibitem [{\citenamefont {Chabanat}\ \emph {et~al.}(1998)\citenamefont
  {Chabanat}, \citenamefont {Bonche}, \citenamefont {Haensel}, \citenamefont
  {Meyer},\ and\ \citenamefont {Schaeffer}}]{chabanat1998a}%
  \BibitemOpen
  \bibfield  {author} {\bibinfo {author} {\bibfnamefont {E.}~\bibnamefont
  {Chabanat}}, \bibinfo {author} {\bibfnamefont {P.}~\bibnamefont {Bonche}},
  \bibinfo {author} {\bibfnamefont {P.}~\bibnamefont {Haensel}}, \bibinfo
  {author} {\bibfnamefont {J.}~\bibnamefont {Meyer}}, \ and\ \bibinfo {author}
  {\bibfnamefont {R.}~\bibnamefont {Schaeffer}},\ }\href {\doibase
  10.1016/S0375-9474(98)00180-8} {\bibfield  {journal} {\bibinfo  {journal}
  {Nucl. Phys. A}\ }\textbf {\bibinfo {volume} {635}},\ \bibinfo {pages} {231}
  (\bibinfo {year} {1998})}\BibitemShut {NoStop}%
\bibitem [{\citenamefont {Blair}(1954)}]{blair1954}%
  \BibitemOpen
  \bibfield  {author} {\bibinfo {author} {\bibfnamefont {J.~S.}\ \bibnamefont
  {Blair}},\ }\href {\doibase 10.1103/PhysRev.95.1218} {\bibfield  {journal}
  {\bibinfo  {journal} {Phys. Rev.}\ }\textbf {\bibinfo {volume} {95}},\
  \bibinfo {pages} {1218} (\bibinfo {year} {1954})}\BibitemShut {NoStop}%
\bibitem [{\citenamefont {Ring}\ and\ \citenamefont {Schuck}(1980)}]{ring1980}%
  \BibitemOpen
  \bibfield  {author} {\bibinfo {author} {\bibfnamefont {P.}~\bibnamefont
  {Ring}}\ and\ \bibinfo {author} {\bibfnamefont {P.}~\bibnamefont {Schuck}},\
  }\href@noop {} {\emph {\bibinfo {title} {{T}he {N}uclear {M}any--{B}ody
  {P}roblem}}}\ (\bibinfo  {publisher} {Springer--Verl.},\ \bibinfo {address}
  {New York, Heidelberg, Berlin},\ \bibinfo {year} {1980})\BibitemShut
  {NoStop}%
\bibitem [{\citenamefont {Thouless}(1961)}]{thouless1961}%
  \BibitemOpen
  \bibfield  {author} {\bibinfo {author} {\bibfnamefont {D.~J.}\ \bibnamefont
  {Thouless}},\ }\href@noop {} {\emph {\bibinfo {title} {{T}he {Q}uantum
  {M}echanics of {M}any--{B}ody {S}ystems}}}\ (\bibinfo  {publisher} {Academic
  Press},\ \bibinfo {address} {New York},\ \bibinfo {year} {1961})\BibitemShut
  {NoStop}%
\end{thebibliography}%

\end{document}